\documentclass[aoas]{imsart}

\RequirePackage{amsthm,amsmath,amsfonts,amssymb}
\RequirePackage[authoryear]{natbib}
\RequirePackage[colorlinks,citecolor=blue,urlcolor=blue]{hyperref}%% uncomment this for coloring bibliography citations and linked URLs
\RequirePackage{graphicx}%% uncomment this for including figures
\usepackage{caption}
\usepackage{subcaption}
\usepackage{multirow}
 \usepackage{booktabs} 

\startlocaldefs
\numberwithin{equation}{section}
\theoremstyle{plain}

\newtheorem{theorem}{Theorem}[section]
\newtheorem{lemma}[theorem]{Lemma}

\newtheorem{algorithm}[theorem]{Algorithm}
\theoremstyle{remark}

\newcommand{\R}{\mathbb{R}}

\renewcommand{\tilde}{\widetilde}
\renewcommand{\hat}{\widehat}

\allowdisplaybreaks

\endlocaldefs

\begin{document}

\begin{frontmatter}
\title{A semi-parametric approach for estimating consumer valuation distributions using second price auctions}
%\title{A sample article title with some additional note\thanksref{T1}}
\runtitle{Estimating Valuation distributions using Second Price Auctions}
%\thankstext{T1}{A sample of additional note to the title.}

\begin{aug}
%%%%%%%%%%%%%%%%%%%%%%%%%%%%%%%%%%%%%%%%%%%%%%%
%% Only one address is permitted per author. %%
%% Only division, organization and e-mail is %%
%% included in the address.                  %%
%% Additional information can be included in %%
\author[A]{\fnms{Sourav} \snm{Mukherjee}\thanksref{coau}\ead[label=e2]{sm37sds@gmail.com}},
\author[A]{\fnms{Ziqian} \snm{Yang}\thanksref{coau}\ead[label=e4]{zi.yang@ufl.edu}},
\author[B]{\fnms{Rohit K} \snm{Patra}\ead[label=e1]{rkumarpatra@gmail.com}},
\and
\author[A]{\fnms{Kshitij} \snm{Khare}\ead[label=e3]{kdkhare@stat.ufl.edu}}
\address[A]{Department of Statistics, University of Florida}\printead{e2,e4,e3}

\address[B]{LinkedIn Inc}\printead{e1}

\thankstext{coau}{co-first authors}
%%%%%%%%%%%%%%%%%%%%%%%%%%%%%%%%%%%%%%%%%%%%%%
%% Addresses                                %%
%%%%%%%%%%%%%%%%%%%%%%%%%%%%%%%%%%%%%%%%%%%%%%

%\address[B]{???, \printead{e2,e3}}
\runauthor{Mukherjee, Yang, Patra, and Khare}
\end{aug}

\begin{abstract}
We focus on online second price auctions, where bids are made sequentially, and the winning bidder pays the maximum of the second-highest bid and a seller specified starting price. For many such auctions, the seller does not see all the bids or the total number of bidders accessing the auction, and only observes the current selling prices throughout the course of the auction. We develop a novel semi-parametric approach to estimate the underlying consumer valuation distribution based on this data. Previous semi-parametric or non-parametric approaches in the literature only use the final selling price and assume knowledge of the total number of bidders. The resulting estimate, in particular, can be used by the seller to compute the optimal profit-maximizing price for the product. Our approach is free of tuning parameters, and we demonstrate its computational and statistical efficiency in a variety of simulation settings, and also on an Xbox 7-day auction dataset on eBay.
\end{abstract}


\end{frontmatter}
%%%%%%%%%%%%%%%%%%%%%%%%%%%%%%%%%%%%%%%%%%%%%%
%% Please use \tableofcontents for articles %%
%% with 50 pages and more                   %%
%%%%%%%%%%%%%%%%%%%%%%%%%%%%%%%%%%%%%%%%%%%%%%
%\tableofcontents

%%%%%%%%%%%%%%%%%%%%%%%%%%%%%%%%%%%%%%%%%%%%%%
%%%% Main text entry area:

\section{Introduction}\label{sec:Introduction}

%\begin{itemize}
%\item IPV auctions
%\item Second price auctions
%\item Identifiability issues due to not observing total number of bidders (Song, 2004)
%\item Using selling price over the entire duration of the auctions, and not only final selling price
%\item Simplifications: bidders can only bid once, constant or two-tiered arrival rate
%\item Comment on the last paragraph in Section 2.2 of George and Hui (2012) - second highest current price is not the third 
%highest valuation, so not easy to interpret or incorporate in their framework without making very restrictive assumptions 
%about order of bidding. But current price information is still useful, and carries information about $F$. 
%\end{itemize}

\noindent
In a second price auction {\color{blue} with starting price}, the product on sale is awarded to the highest bidder if the corresponding bid is higher than a seller-specified starting price $r$. The price paid by the winner is however, the maximum of the starting price and the second highest bid. These auctions have been the industry standard for a long time, and are attractive to sellers as they induce the bidders to bid their true ``private value"
for the product, i.e., the maximum price they wish to pay for it. While some platforms have recently moved to first price auctions, the second price auction is still widely used on E-commerce platforms such as {\it eBay}, {\it Rokt}, and online ad exchanges such as {\it Xandr}. The analysis of data obtained from these auctions presents unique challenges. For a clear understanding of these challenges, we first discuss in detail the auction framework, the observed data and the quantity of interest that we want to estimate/extract. 

{\it Auction framework}: We consider an auction setting where a 
single product is on sale for a fixed time window $[0, \tau]$. 
The seller sets the starting price $r$, which is used as the 
current selling price at time $0$. Any bidder who arrives 
subsequently is \emph{eligible} to place a bid only if his/her bid value 
is higher than the current selling price at that time. If the bid is 
placed, the current selling price is updated to the {\it 
second-highest bid value} among the set of all placed bids up to 
that time, including this latest bid (the starting price is also treated as a placed bid).~\footnote{Typically, a small increment (e.g. $\$0.01$) is also added to the second highest bid, but this insignificant increment is unlikely to influence bidder's behaviour, and we ignore it in our analysis for ease of exposition.} For example, suppose the current selling price at a given time is $\$ 4$ and the highest placed bid value up to that time is $\$ 5$. If a bidder comes and bids $\$ 3$, the bid will not be placed. If the bidder were to bid $\$ 4.5$, the bid would be placed and the current selling price would be updated to $\$ 4.5$ (since now the second largest placed bid is $\$4.5$). If the bidder were to bid $\$ 6$, the bid would be placed and the current selling price would be updated to $\$ 5$. At the end of the auction period, if no bid above the starting price is placed, the item goes unsold, otherwise it is sold to the highest bidder at the selling price at time $\tau$. This final selling price is the second highest placed bid (including the starting price) throughout the course of the auction. 

\begin{figure}
    \centering
    \includegraphics[width = \linewidth]{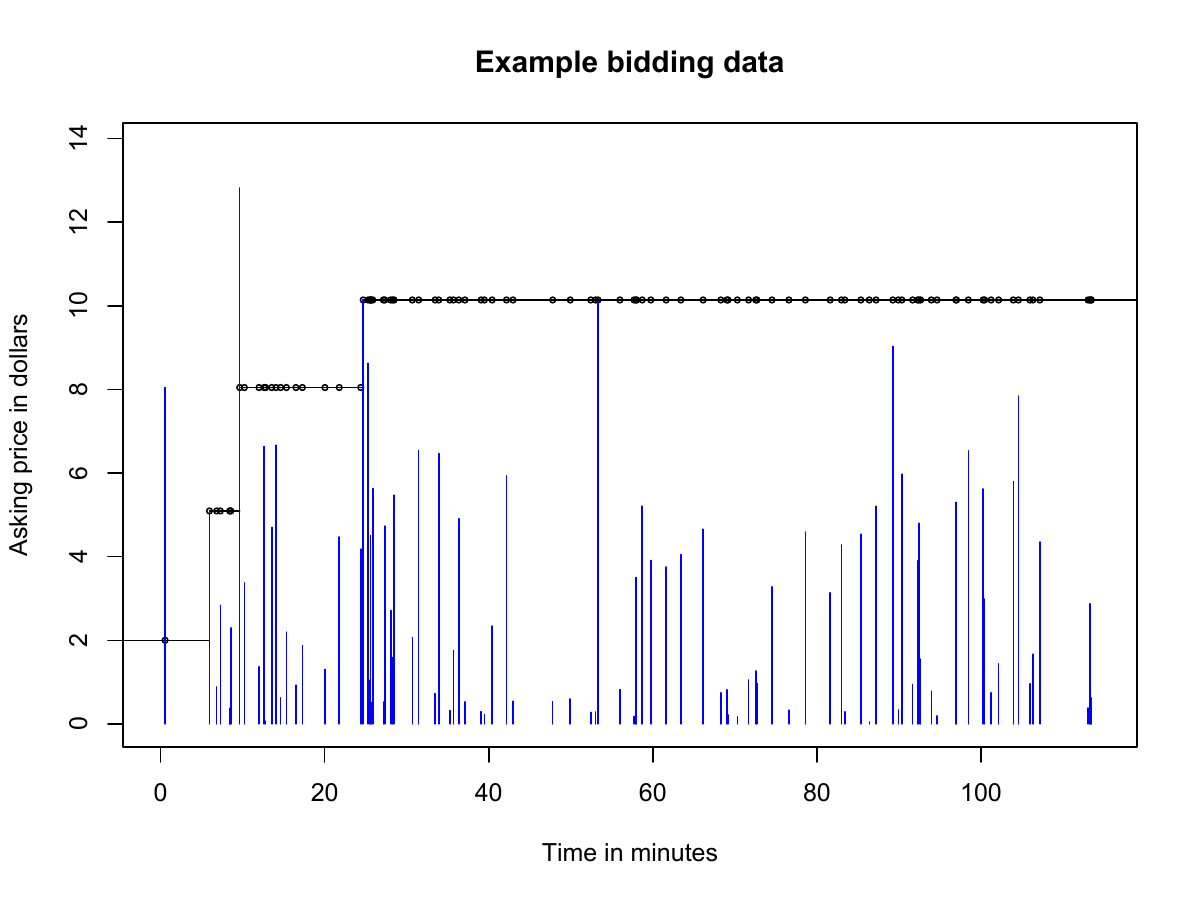}
    \caption{\textit{An illustration of a single second price auction. True bid values are generated from a Pareto distribution and starting price is at $\$2$. Blue vertical lines are the bid values, black horizontal lines are the current selling prices, and black dots are the time points when the bids are made.}}
    \label{fig:example.single.plot}
\end{figure}

{\it An illustration of a single second price auction}: Figure~\ref{fig:example.single.plot} provides a concrete illustration of how a second price auction works. The data for this auction has been simulated from a setting where the bids follow a Pareto distribution with location parameter $3$ and dispersion parameter $100$. The waiting times between bids are generated from an exponential distribution with rate parameter $1$. In total, $100$ bids are generated in a time period $\tau$ of around $115$ minutes. The starting price for the auction is $\$2$. In Figure~\ref{fig:example.single.plot}, the bid values and the current selling prices during the course of the auction are represented by the blue vertical lines and the black horizontal lines, respectively. The black dots on the black horizontal lines represent the time points (in minutes) of the $100$ bids. As can be seen from Figure~\ref{fig:example.single.plot}, the initial selling price is equal to the starting price ($\$2$). We have the first bid of around $\$8.05$ at $0.55$ minutes. Since it's higher than the starting price of $\$2$, this bid will be placed, but the starting price still remains the current selling price at $0.55$ minutes. Recall that arriving consumers only see the current selling price, and hence a consumer with a $\$5.09$ valuation would be both willing and eligible to place their bid when the current selling price is $\$2$. Hence, the second bid of $\$5.09$ occurring at $5.96$ minutes ($5.41$ minutes after the first bid's occurrence) is placed, and subsequently the current selling price jumps to $\$5.09$ as it's the second highest value among the starting price ($\$2$) and the two existing placed bids ($\$8.05$ and $\$5.09$).  
We don't observe any jumps in the current selling price for the next few bids as they all are less than the current selling price of $\$5.09$ (and hence are not eligible to be placed). Then we see another jump at $9.65$ minutes, where a bid of $\$12.82$ which makes the current selling price jump to $\$8.05$. The next few bids again happen to be less than the current selling price ($\$8.05$). At around $24$ minutes, we see a last jump in the current selling price to $\$10.14$ (based on a new bid of $\$10.14$ which exceeds $\$8.05$). The subsequent bids are all less than $\$10.14$, it remains the current selling price throughout the rest of the auction period. This can be observed through the flat horizontal black line at $\$10.14$ in the time period of $(24.7, 114.43)$ minutes. The final selling price for the auction is therefore $\$10.14$. The observed data for the above auction is the sequence 
of current selling prices given by $(\$2, \$5.09, \$8.05, \$10.14)$ 
and the sequence of times at which there was a change in the current selling price, given by $(5.96, 9.65, 24)$.

{\it Observed data}: The observed data is the sequence of current
selling price values (also sometimes referred to as the standing price) throughout the course of the auction, and the times at which there is a change in the selling/standing price. Typically, such data is available for multiple auctions of the same product. For example, in 
Section \ref{sec:Empirical_application}, we analyze data with current selling 
prices for $93$ different eBay $7$-day auctions for Xbox. A key observation to 
make here is that consumers who access the auction but have bids which are 
less than the current selling price (standing price) are not allowed to place 
their bids. In other words, instead of observing the bids of all the customers 
who access the auction, we only observe the running second largest value of such bids. 

\begin{figure}
    \centering
    \includegraphics[width = \linewidth]{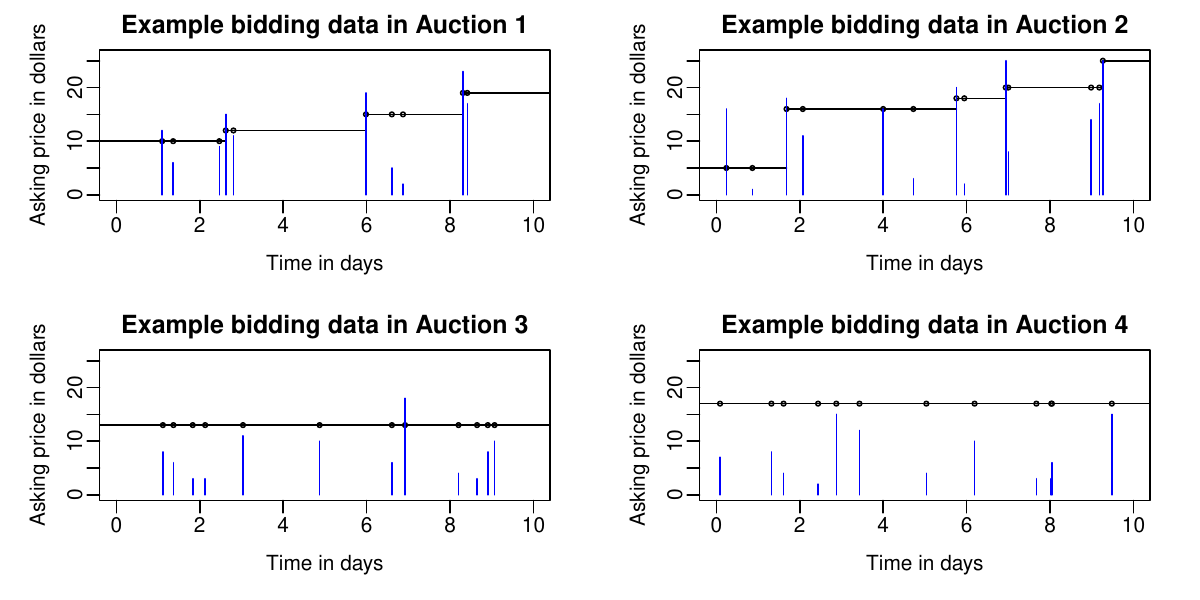}
    \caption{\textit{An illustration of 4 separate second price auctions for a given product. For each auction, true bid values are generated from a discrete Uniform distribution and starting price is respectively set to $\$10,\$5,\$13,\$17$. Blue vertical lines are the bid values, black horizontal lines are the current selling prices, and black dots are the time points when the bids are made.}}
    \label{fig:example.plot}
\end{figure}

{\it An illustration of  multiple second price auctions:}
Figure \ref{fig:example.plot} provides a concrete illustration of a setting with multiple (four) second price auctions for the same product. The data for these auctions has been simulated from a setting where the bids follow a discrete uniform distribution with lower bound 1 and upper bound 25. The waiting times between bids are generated from an exponential distribution with rate parameter $1$, and the time period for each auction is 10 days. The starting prices for the four auctions, respectively, are $\$10,\$5,\$13,\$17$. Just like in Figure~\ref{fig:example.single.plot}, the bid values and the current selling prices during the course of each auction are represented by the blue vertical lines and the black horizontal lines, respectively. The black dots on the black horizontal lines represent the time points (in days) of the bids. We can see that for auctions in the upper left and upper right, the corresponding item is sold above the starting price (at final selling price of $\$19$ and $ \$25$ respectively). On the other hand, for the auction in the lower left, the item is sold at the starting price, since there is only one bid placed above/at the starting price. Finally, for the auction in the lower right, the item is not sold since no bid is placed above or at the starting price. We will keep coming back to this integrated example to clarify and illustrate various concepts and notions that are introduced in the paper. 

{\it Quantity of interest}: Each bidder in the consumer  
population is assumed to have an independent private valuation 
(IPV) of the product. The IPV assumption in particular makes 
sense for products that are used for personal use/consumption 
(such as watches, jewelry, gaming equipment etc.) and
is commonly used in the modeling of internet auctions (see 
\cite{Song04nonparametricestimation, Hou:Rego:2007, 10.1214/11-AOAS503} and the
references therein). Economic theory suggests that the dominant strategy for a 
bidder in a second price auction is to bid one's true valuation 
(\cite{Vickrey:1961}). The quantity that we want to estimate 
from the above data is the distribution of the valuations of the 
product under consideration for the consumer population. We refer
to this as the consumer valuation distribution, and denote it by 
$F$. As noted in \cite{10.1214/11-AOAS503}, knowledge of $F$ 
provides the consumer demand curve for the product, and hence can
be used by the seller to identify the profit-maximizing price (see the discussion in Section 4.1 from \cite{10.1214/11-AOAS503}).

The problem of demand-curve/valuation distribution estimation 
using auction data has been tackled in the last two decades for 
a variety of auction frameworks, see \cite{Song04nonparametricestimation, Park:Bradlow:2005, Bradlow:Park:2007, Yao:Mela:2008, 10.1214/11-AOAS503, Backus:Lewis:2020} and the 
references therein. Some of these papers assume a parametric form 
for $F$, but as \cite{10.1214/11-AOAS503} argue, available 
auction data may often not be rich enough to verify the validity 
of the underlying parametric forms. \cite{10.1214/11-AOAS503} 
consider the second-price ascending bid auction framework for 
a single (homogeneous) product described above, and develop a 
Bayesian non-parametric approach to estimate $F$ using only the 
final selling prices in multiple auctions for a jewelry item. Note 
that the final selling price is not only the second highest placed 
bid in the auction period, it is also the second maximum order 
statistic of the (potential or placed) bids of all consumers who 
access the auction. Using only the final selling prices leads to 
identifiability problems, and \cite{10.1214/11-AOAS503} address 
this by assuming that the total number of consumers accessing the 
auction is also known. This information is available for the 
particular jewelry dataset from a third-party vendor, but is 
generally not available for most such datasets. When the total 
number of consumers accessing the auction is not known, the 
identifiability issue can be solved by using another order 
statistic (\cite{Song04nonparametricestimation}) such as the 
largest placed bid value (if available) along with the final 
selling price. \cite[Section 5]{Backus:Lewis:2020} use such an 
approach for analyzing a dataset containing compact camera auctions
on eBay.  
%\begin{center}
\begin{table}
\begin{center}
\rotatebox{90}{

\begin{tabular}{p{3cm}p{0.6cm}p{3.1cm}p{2cm}p{2.3cm}p{2.4cm}p{1.5cm}}

Paper & Auc. type & Statistical Paradigm & Utilizes \textbf{entire} standing price sequence & Needs information on unobserved bidders or bid values   
& Allows multiple bidding and sniping in same auction & Allows buy-it-now option\\
    \midrule

Bradlow \& Park (2007) & $1^{st}$ & Parametric (Bayesian) & Yes & No  & Yes & No\\
Chan et al. (2007) & $1^{st}$ & Parametric & Yes & No & Yes & Yes\\
    \midrule

Song (2004) & $2^{nd}$ & Semi and nonparametric & No (only uses $2^{nd}$ and $3^{rd}$ highest price) & No &  Yes & No\\
Yao \& Mela (2008) & $2^{nd}$ & Parametric (Bayesian) & Yes & Yes (needs highest bid value in the auction)  & Yes & No\\
George \& Hui (2012) & $2^{nd}$ & Non-parametric (Bayesian) & No (only uses final selling price)  & Yes (needs total number of participating bidders) & Yes & No\\

Backus \& Lewis (2024) & $2^{nd}$ & Non-parametric  & No (only uses $1^{st}$ and $2^{nd}$ highest price) & Yes (needs highest bid value in the auction)  & Yes & No \\

This paper & $2^{nd}$ & Semi-parametric & Yes & No & No & No\\

\end{tabular}
}
\caption{ A tabular comparison of the main features of some key related papers and the proposed methodology. }
\label{tab:reference}
\end{center}
\end{table}

Table \ref{tab:reference} provides a comparison of the main relevant 
features of key related papers and the proposed methodology in the 
paper. As the table demonstrates (and to the best of our knowledge), 
none of the existing  non-parametric/semi-parametric methods for second price auctions 
use the entire sequence of current selling price values to estimate 
the consumer valuation distribution $F$. This was the primary 
motivation for our work, as only using the final selling price 
(and possibly the maximum placed bid, if available) leaves out a 
lot of available information. Including the current selling price 
information throughout the course of the auction however, involves significant 
conceptual and methodological challenges. As observed previously, the final 
selling price (second largest placed bid) is also the second largest (potential
or placed) bid of all the consumers who accessed the auction. Hence, under 
relevant assumptions (see beginning of 
Section \ref{sec:Methodology}), the final selling price 
can be interpreted as the second largest order statistic of i.i.d. samples from
$F$. This interpretation is central to the methodology developed in 
\cite{10.1214/11-AOAS503} and \cite{Song04nonparametricestimation}. However, as the authors in \cite{10.1214/11-AOAS503}
point out, the second largest current selling price throughout the course of 
the auction is not necessarily the third largest order statistic among all 
(placed or potential) bids, unless some severely restrictive and unrealistic 
assumptions are imposed on the order in which bidders arrive in the auction. 
Hence, extending the methodology in \cite{10.1214/11-AOAS503} to include the 
entire sequence of current selling prices is not feasible; a completely new 
method of attack is needed. 

In this paper, we fill this gap in the literature and develop novel methodology
for semi-parametric estimation of the consumer valuation distribution in 
second-price ascending bid auctions which uses the entire sequence of current 
selling prices. Additionally, the total number of consumers accessing the 
auction is not assumed to be known (unlike \cite{10.1214/11-AOAS503}), and the 
highest placed bid in the auction is also not assumed to be known (unlike 
\cite{Backus:Lewis:2020}). Incorporating the above novel features is very 
challenging, and the methodological development (provided in 
Section \ref{sec:Methodology}) is quite involved. The extensive simulation 
results in Section \ref{sec:Simulation_study} demonstrate the significant 
accuracy gains in the estimation of $F$ that can be obtained by using the 
entire sequence of current selling prices as opposed to just the final selling 
price. We note that incorporating the current selling prices during the entire 
auction does come with a cost. In particular, two additional assumptions 
(compared to those in \cite{10.1214/11-AOAS503}) that need to be made about the
rate of bidder arrival and number of bids made by a single consumer. We find 
some deviations from these two assumptions in the eBay $7$-day Xbox data that 
we analyze in Section \ref{sec:Empirical_application}, but these deviations are
minor (see discussion at beginning of Section \ref{sec:Methodology}). In such 
settings, it is reasonable to expect that the advantage of incorporating 
substantial additional  data/information outweighs the cost of these 
approximations/deviations. We conclude the paper with a discussion of future 
directions in Section \ref{sec:discussion}.

\section{Methodology for learning the consumer valuation distribution $F$}\label{sec:Methodology}

We start by discussing the main assumptions needed for the 
subsequent methodological development. We consider a setting where 
we have data from several independent, non-overlapping auctions for
a single (homogeneous) product. As mentioned previously, we 
work within the IPV framework which is quite reasonable for 
items/products that are used for personal consumption. Similar to 
\cite{10.1214/11-AOAS503}, we will assume that the collection of 
bidders who access the auction is an i.i.d. sample from the 
consumer population for the corresponding product, and that the 
collection of private product valuations of these bidders is an 
i.i.d. sample from the consumer valuation distribution $F$. As 
stated in the introduction, it can be shown that the dominant 
strategy for a bidder in a second price auction is to bid his/her 
true valuation. Under this strategy, any consumer who accesses the 
auction would bid his/her valuation with no need for multiple 
bidding, and we assume this behavior. We do note that in practice, 
some consumers do not follow this strategy. For example, 
\cite{WinNT} provides an option called proxy bidding or automatic 
bidding which allows the computer to automatically place multiple 
incremental bids below a cutoff price on behalf of the consumer 
(see also 
\cite{OCKENFELS2006297,Bajari:Hortacsu:2003}). Since 
\cite{10.1214/11-AOAS503} only use the final selling price, they 
use a weaker assumption which allows for multiple bidding and 
stipulates that a consumer bids his/her true valuation sometime 
before the end of the auction. 

Our final assumption is regarding the arrival mechanism of bidders 
in the auction. We assume that consumers arrive at/access the 
auction according to a Poisson process with constant rate 
$\lambda$. Again, a rational bidder (based on economic theory) in a
second price auction should be indifferent to the timing of his/her
bid (see \cite{Milgrom:2004, BARBARO2021100553}), and this 
assumption makes sense in such settings. Again, we note that 
late-bidding (sniping) has been observed in some eBay auctions 
(see \cite{BOSE2017507, BARBARO2021100553}). 

To summarize, our assumptions are identical to those in
\cite{10.1214/11-AOAS503} with the exception of the 
single bidding-assumption and the constant rate of 
arrival assumption. While these assumptions are 
supported in general by economic theory, deviations 
from these two assumptions have been observed in
some online auctions. However, for datasets such as the
Xbox dataset analyzed in Section 
\ref{sec:Empirical_application}, these deviations are 
minor/anomalies. For example, in the Xbox dataset, only
around $10 \%$ of the bidders with placed bids show a 
multiple bidding behavior. In such settings, there is 
certainly value in using the subsequent methodology 
which uses the entire current selling price/standing 
price profile and does not assume knowledge of the 
total number of consumers accessing the auctions. If 
there is strong evidence/suspicion that the assumptions
are being extensively violated, of course the results from this 
methodology should be treated with due skepticism and 
caution. 

The subsequent methodological development in this section is quite involved, and we have tried to make
it accessible to the reader by dividing it into subsections based on the major steps, and then highlighting the key milestones within each subsection, 
wherever necessary. We start by finding the joint density of the observed data 
obtained from a single second price auction.

\subsection{Joint density of the observed data in a single second
price auction}\label{subsec:Jt_density_XTM_Single_Lambda_case}

\noindent
Consider a given (single) second price auction with starting price $r$. Hence, the 
initial selling price, denoted by $X_0$, is equal to $r$. The first time a consumer 
with bid value greater than $r$ arrives at the auction, that bid is placed but the current 
selling price remains $r$. Subsequently, the current selling price (standing price) 
changes whenever a bid greater than the existing selling price is placed. Let $M$ 
denote the number of times the selling price changes throughout the course of the 
auction. When $M > 0$, let $\{X_i\}_{i=1}^{M}$ denote the sequence of current selling
prices observed throughout the course of the auction, with $X_i$ denoting the new 
selling/standing price after the $i^{th}$ change. When $M > 0$, let $T_i$ denote the intermediate 
time between the $i^{th}$ and $(i+1)^{th}$ changes in the selling/standing price for $0 \leq i \leq M-1$. 
In particular, it follows that when $M > 0$, $T_0$ denotes the waiting time from the start of the auction 
until the moment when for the first time, the selling price changes to a higher value than the starting 
price $r$. When $M = 0$, we define $T_0 = \tau$. Finally, let $O$ be a binary random variable indicating 
whether the item is sold before the end of the auction, i.e., $O=1$ indicates the item is sold and $O=0$ 
indicates that the item is not sold. Our observed data comprises of $M$, $O$, 
$\{X_i\}_{i=1}^{M}$, and $\{T_i\}_{i=0}^{M-1}$. 

We define $T_M =\tau - \sum_{i=0}^{M-1} T_i$ as the time after the last selling price change and until
the auction closes. As discussed earlier, the number of consumers/bidders accessing a given (single) 
second price auction, denoted by $N$, remains unobserved in our setup. Based on our 
assumption regarding the arrival mechanism (Poisson process with constant rate of arrival $\lambda$) of 
bidders in the auction, we note that $N \sim \text{Poisson}(\lambda\tau)$. 

Note that there are three scenarios at the end of the auction: (a) the item is sold above the starting price
($M > 0, O = 1$), (b) the item is sold at the starting price ($M = 0, O = 1$), and (c) the item is not sold 
($M = 0, O = 0$). The following lemma provides a unified formula for the joint density of the observed data
encompassing all these three scenarios. 
\begin{lemma}\label{lemma:Result_Jt_density_XTM}
For the second price auction described above, the joint density of $M$, $O$, 
$\{X_i\}_{i=1}^{M}$, $\{T_i\}_{i=0}^{M-1}$ at values $m$, $o$, 
$\{x_i\}_{i=1}^{m}$, $\{t_i\}_{i=0}^{m-1}$ is given by
\begin{align*}
    \begin{split}
   {}& g \Big( m, o, \{x_i\}_{i=1}^{m}, \{t_i\}_{i=0}^{m-1} \Big)\\
={}& \exp(-\lambda\tau) \ 2^m \ \Big( \lambda^{m+1} \ t_0 \ \big(1 - F(x_m) \big) 
        \Big)^{1_{\{o=1\}}} \exp\bigg(\lambda\sum_{i=0}^{m}F(x_i)t_i\bigg)   \bigg(C_1 \prod_{i=1}^{m}f(x_i)\bigg)^{1_{\{m>0\}}},
        \end{split}
    \end{align*}
where $C_1$ does not depend on $F$, $x_0$ represents the starting price ($r$), $\lambda$ denotes the constant rate of arrival of the bidders throughout the course of the auction, and $f$ represents the density function corresponding to $F$.
\end{lemma}

\noindent
The proof of Lemma \ref{lemma:Result_Jt_density_XTM} is quite involved and is provided in the supplementary document. Next, we generalize our analysis to consider data from several independent, non-overlapping auctions of identical copies of a single (homogeneous) product. 

\subsection{Likelihood based on the observed data from multiple identical, non-overlapping second price auctions}\label{subsec:likelihood_2ndPrice_Auction_Single_Lambda_case}
Suppose we consider $K$ independent second price auctions of identical copies of an item (with possibly different starting prices $r_1,r_2,\ldots,r_K$). The observed data is $\{(M_{k}, O_{k}) \ , \ \{(X_{i,k} \ , \ T_{i-1,k})\}_{i=1}^{M_k}\}_{k=1}^{K}$, where $M_k$ denotes the 
number of selling price changes for the $k^{th}$ auction, $O_k$ denotes the indicator of the item being sold at the end of the $k^{th}$ auction, $\{X_{i,k}\}_{i=1}^{M_k}$ denote the selling/standing price sequence for the $k^{th}$ auction, and $\{T_{i-1,k}\}_{i=1}^{M_k}$ denote the sequence of intermediate waiting times between successive changes in the standing prices. Finally, let $T_{M_k,k} = \tau - \sum_{i=0}^{M_k - 1}T_{i,k}$, for all $k = 1, 2, \ldots, K$. Since the auctions are independent, it follows by 
Lemma \ref{lemma:Result_Jt_density_XTM} that the likelihood function of the unknown parameters $\lambda$ and $F$ for observed data values $\{(m_k, o_k) \ , \ \{(x_{i,k} \ , \ t_{i-1,k})\}_{i=1}^{m_k}\}_{k=1}^{K}$ is given by 
$Lik(\lambda, F) = \prod_{k=1}^K g ( m_k, o_k, \{x_{i,k}\}_{i=1}^{m_k}, \{t_{i,k}\}_{i=0}^{m_k-1} )$. 

Ideally, one would like to obtain estimates of $\lambda$ and $F$ by maximizing the function $Lik(\lambda, F)$. However, this likelihood function is intractable for direct maximization. A natural direction to 
proceed is to use the alternative maximization approach, which produces a sequence of iterates by alternatively maximizing $Lik$ with respect to $F$ given the current value of $\lambda$ and then maximizing $Lik$ with respect to $\lambda$ given the current value of $F$. Especially the maximization with respect to $F$ (given $\lambda$) requires intricate analysis and careful reparametrization, and we describe the details in Sections \ref{subsec:new:notation} and \ref{subsec:constraint.free.reparametrization} below.
% However, we found that such an approach can
% suffer from instability issues, which is not very surprising given the highly complicated and non-convex 
% setting. Hence, we pursue and develop a slightly different approach which consists of two major steps: 
% \begin{itemize}
%     \item Directly obtain an estimator $\hat{\lambda}$ of $\lambda$ using generalized method 
%     of moments. 
%     \item Obtain an estimate of $F$ by maximizing the profile log-likelihood $Lik(\hat{\lambda}, F)$. 
% \end{itemize}

% \noindent
% Both of the steps above,  
% This approach is computationally 
% much less expensive than alternative maximization of $\lambda$ and $F$, and as our extensive simulations 
% in Section \ref{sec:Simulation_study} show, also provides stable and accurate estimates.

\subsection{Estimation of $F$ given $\lambda$: Some new notation based on pooled standing prices across all auctions} \label{subsec:new:notation}

\noindent
Note that the function $F$ is constrained to be non-decreasing. A key 
transformation to a constraint-free parametrization (described in Section~\ref{subsec:constraint.free.reparametrization} below) is needed to facilitate the conditional maximization of $Lik$ with respect to $F$. A crucial 
precursor to this re-parametrization is introduction of some new notation 
obtained by merging the standing prices from all the $K$ different auctions 
together. Let $L = \sum_{k=1}^{K}M_k$ denote the total number of 
standing/selling price changes in all the $K$ auctions. Recall that 
$\{(m_k, o_k) \ , \ 
\{(x_{i,k} \ , \ t_{i-1,k})\}_{i=1}^{m_k}\}_{k=1}^{K}$ 
denote the observed data values, and $t_{m_k,k} = \tau - \sum_{i=0}^{m_k - 
1}t_{i,k}$ for $1 \leq k \leq K$. Let $\ell$ denote the observed value of $L$. 
We will denote by  ${\bf \bar{x}} = (\bar{x}_1, \bar{x}_2, \cdots, \bar{x}_\ell)$ the 
arrangement/ordering of the pooled collection \ 
$\{\{x_{i,k}\}_{i=1}^{m_k}\}_{k=1}^{K}$ \ 
such that $\bar{x}_1 < \bar{x}_2 < \ldots < \bar{x}_\ell$; under 
the assumption that $F$ is a continuous cdf, there should be no ties in the 
entries of ${\bf \bar{x}}$ with probability one. In other words,
we pool the standing prices from all the auctions (excluding the starting prices)
and then arrange them in ascending order as $(\bar{x}_1, \bar{x}_2, \cdots, 
\bar{x}_\ell)$. Also, for $1 \leq i \leq L$, we define $\bar{t}_i = 
t_{\bar{i},\bar{k}}$ where $\bar{i}$ and $\bar{k}$ are such that 
$\bar{x}_i = x_{\bar{i},\bar{k}}$. Further, let 
\begin{align*}
    S :={}& \text{Set of ranks/positions of} \ x_{m_k,k} \ (k=1,2,\ldots,K) \ \text{in} \ {\bf \bar{x}} \ \text{for all the auctions where the}\\
    {}& \text{item is sold above the starting price,}\\
    K_s :={}& \text{Set of auction indices for which the item is sold}.
\end{align*}

\noindent {\it Example (continued): }
Consider the example in Figure \ref{fig:example.plot}, where we have a setting with $K = 4$ independent second price auctions, 
with starting prices $(r_1, r_2, r_3, r_4) = (10, 5, 13, 17)$. The 
first auction has $m_1 = 3$ standing price changes with $(x_{1,1}, \ x_{2,1},
\ x_{3,1}) = (12, 15, 19)$, the second auction has $m_2 = 4$ standing price 
changes with $(x_{1,2}, \ x_{2,2}, \ x_{3,2}, \ x_{4,2}) = (16, 18, 20, 25)$, the item is sold
at the starting price $r_3 = 13$ in the third auction ($m_3 = 0, o_3 = 1$), and 
the item is unsold in the fourth auction ($m_4 = 0, o_4 = 0$). Pooling and 
rearranging the standing prices (excluding starting prices) from all the 
auctions, we see that $\ell = 3+4+0+0 = 7$, and 
$$
{\bf \bar{x}} = (\bar{x}_1, \bar{x}_2, \bar{x}_3, \bar{x}_4, 
\bar{x}_5, \bar{x}_6, \bar{x}_7) = (12, 15, 16, 18, 19, 20, 25). 
$$

\noindent
Note that the auction item is sold above the starting price in the first two 
auctions, and the final selling prices are $x_{m_1,1} = 19$ and $x_{m_2,2} = 
25$ respectively. Examining the positions of these two prices in ${\bf 
\bar{x}}$ gives us $S = \{5, 7\}$. Finally, $K_s = \{1,2,3\}$ is the 
collection of auction indices where the item is sold. 

Using the newly introduced notation above and Lemma \ref{lemma:Result_Jt_density_XTM}, it follows that the likelihood function is given by 
\begin{align}
Lik \left(\lambda, F \right) ={}& C^{\star}\exp\big(-K\lambda\tau\big) \ \lambda^{\ell+\left|K_s\right|} \ 2^\ell \prod_{k\in K_s}t_{0,k} \ \prod_{i\in S}\big(1 - F(\bar{x}_i)\big) \bigg[\prod_{j=1}^{\ell}f(\bar{x}_j) \bigg]^{1_{\{\ell>0\}}}\nonumber\\
{}& {\color{black}\times \prod_{k=1}^K (1 - F(r_k))^{1_\{m_k=0,o_k=1\}} \exp\bigg(\lambda\bigg(\sum_{k=1}^{K}F(r_k)t_{0,k} + \sum_{j=1}^{\ell}F(\bar{x}_j)\bar{t}_j \bigg) \bigg)},  
\label{eq:likelihood_lambda_X_F_cont}
\end{align}
where $C^{\star}$ doesn't depend on $(\lambda, F)$. Maximization of the above likelihood (given $\lambda$) over absolutely continuous CDFs leads to one of the standard difficulties in non-parametric estimation. As $F$ moves closer and closer to a CDF with a jump discontinuity at any $\bar{x}_j$, the function $Lik (\lambda, F)$ converges to 
infinity for every fixed $\lambda$. Hence, any absolutely continuous CDF with a density function cannot be a maximizer of the above profile likelihood function. Following widely used convention in the literature (see \cite{10.2307/2242287}, \cite{10.1214/aos/1176345802}), we will extend the parameter space to allow for the MLE of $F$ to be a discrete distribution function. To allow for discrete CDFs, we replace $f(\bar{x}_j)$ by $F(\bar{x}_j)- 
F(\bar{x}_j-)$. Thus the adapted likelihood can be written as
\begin{eqnarray}
& & Lik_{A} \left(\lambda, F \right) \nonumber\\
& =& C^{\star}\exp\big(-K\lambda\tau\big) \ \lambda^{\ell+\left|K_s\right|} \ 2^\ell \ \big( \prod_{k\in K_s}t_{0,k} \big) \ \prod_{i\in S}^{}\big(1 - F(\bar{x}_i)\big) \prod_{k=1}^K \big( 1 - F(r_k) \big)^{1_\{m_k=0,o_k=1\}} \nonumber\\
& & \times \exp\bigg(\lambda\bigg(\sum_{k=1}^{K}{\color{black} F(r_k)}t_{0,k} + \sum_{j=1}^{\ell}F(\bar{x}_j)\bar{t}_j \bigg)\bigg) \ \bigg[\prod_{j=1}^{\ell}\Big(F(\bar{x}_j) - F(\bar{x}_{j}-)\Big) \bigg]^{1_{\{\ell>0\}}} ,
\label{eq:likelihood_lambda_X_F}
\end{eqnarray}

\noindent
where $\bar{x}_0 = 0$. We now establish a final bit of notation 
necessary to introduce the constraint-free reparametrization of $F$. We now 
pool the $\ell + K$ standing prices from all the auctions ({\it including} the 
starting prices), i.e., 
$\{\{x_{i,k}\}_{i=0}^{m_k}\}_{k=1}^{K}$, and 
arrange them in ascending order as $z_1 < z_2 < \cdots < z_{\ell+K}$, and denote
${\bf z} := (z_1, z_2, \cdots, z_{\ell+K})$. Under the assumption that $F$ is a 
continuous cdf, there should be no ties in the entries of ${\bf \bar{x}}$ with 
probability one. The only other entries in ${\bf z}$ are the $K$ starting prices.
In practice, it is possible that there are ties in the starting prices, in which 
case we add a very small noise to the starting prices to ensure that there are no
ties in the entries of ${\bf z}$. Similar to $\bar{t}_i$, we define $\tilde{t}_i
= t_{\tilde{i},\tilde{k}}$ where $\tilde{i}$ and $\tilde{k}$ are such that 
$z_i = x_{\tilde{i},\tilde{k}}$. Further, let 
\begin{align*}
    \bar{S} :={}& \text{Set of ranks/positions of} \ x_{m_k,k} \ (k=1,2,\ldots,K) \ \text{in} \ {\bf z} \ \text{for all the auctions where the}\\
    {}& \text{item is sold {\em at or above} the starting price,}\\
    \bf{u} :={}& \{u_1,u_2,\ldots,u_\ell \}, \ \text{where } u_i = \ \text{position of} \ \bar{x}_i \ \text{in} \ {\bf z}, \ \text{i.e., } \bar{x}_i = z_{u_i}.
\end{align*}
Since the entries of ${\bf \bar{x}}$ and ${\bf z}$ both are arranged in 
ascending order, it follows that $u_1<u_2<\ldots<u_\ell$.

\noindent
{\it Example (continued):} In the example 
considered earlier in this subsection with $K=4$ auctions, by pooling the 
$4$ starting price values with entries of ${\bf \bar{x}}$ and rearranging 
them in ascending order, we obtain 
$$
{\bf z} = (5, 10, 12, 13, 15, 16, 17, 18, 19, 20, 25). 
$$

\noindent
Recall that the item is sold above the starting price only in the first two 
auctions. By identifying the positions/ranks of $x_{m_1,1}$,  $x_{m_2,2}$ and $x_{m_3,3}$ (with $m_3=0$) in 
${\bf z}$, we obtain $\bar{S} = (4, 9, 11)$. Similarly, by identifying the 
positions/ranks of the entries of ${\bf \bar{x}}$ in ${\bf z}$, we obtain 
${\bf u} = (3,5,6,8,9,10,11)$. 

It is clear from (\ref{eq:likelihood_lambda_X_F}) that for maximizing $Lik_{A}$ it is enough to search 
over the class of CDFs with jump discontinuities at elements of ${\bf \bar{x}}$, since all other CDFs will have a $Lik_A$ value of zero. The next lemma (proof 
provided in the Supplement) shows that the search for a maximizer can be further restricted to a certain  
class of CDFs with possible jump discontinuities at elements of ${\bf z}$. 
\begin{lemma} \label{restrict:search:space}
Let $\mathcal{F}_{\bf z}$ denote the class of CDFs which are piece-wise constant in $[0, z_{\ell+K}]$, such that the set of points of jump discontinuity (in $[0, z_{\ell+K}]$) is a superset of elements of ${\bf \bar{x}}$ and a subset of elements of ${\bf z}$. Then, given any $\lambda > 0$ and cdf $F$ with jump discontinuities at elements of ${\bf \bar{x}}$, there exists $\tilde{F} \in \mathcal{F}_{\bf z}$ such that $Lik_{A}(\lambda, F) \leq 
Lik_{A}(\lambda, \tilde{F})$. 
\end{lemma}

\noindent
For any $F \in \mathcal{F}_{\bf z}$, note that $Lik_{A} \left(\lambda, F \right)$ depends on $F$ only through 
$$
\{F(z_1), F(z_2)-F(z_2-), \ldots, F(z_{L+K})-F(z_{\ell+K}-)\}
$$

\noindent
or equivalently through $$
F({\bf z}) = (F(z_1), F(z_2), \cdots, F(z_{\ell+K}))
$$ 

\noindent
(since $F$ only has jump discontinuities at elements 
of ${\bf z}$ and is otherwise piece-wise constant).  
This is typical in a non-parametric setting, and we 
can hope/expect to only obtain estimates of the 
valuation distribution $F$ at the observed standing 
prices (including the starting prices). 

\subsection{Estimation of $F$ given $\lambda$: A constraint-free reparametrization}\label{subsec:constraint.free.reparametrization} 

Note that the entries of the vector $F({\bf z})$ are 
non-decreasing, and this constraint complicates the maximization of $F\mapsto Lik_{A} \left(\lambda, F \right)$. So, we transform $F\left(\bf{z}\right)$ to another $\ell+K$ dimensional parameter vector $\boldsymbol{\theta} := \left(\theta_1,\theta_2,\ldots,\theta_{\ell+K}\right)^T$ as follows: 
\begin{equation}\label{eq:defining_theta}
\theta_i = \dfrac{G(z_i)}{G(z_{i-1})} \ , \ \forall \ 1\leq i \leq (\ell+K) ,
\end{equation}

\noindent
where 
\begin{equation}\label{eq:defining_G}
    G(z_i) = 1-F(z_i) \ , \ \forall \ 1\leq i\leq (\ell+K), \ \text{and} \ G(z_0) = 1 \ \text{with} \ z_0 = 0.
\end{equation}

\noindent
Since $F$ is non-decreasing, and takes values in $[0,1]$, it follows that $\theta_i \in [0,1]$ (with the convention $0/0 := 0$). Focusing our search on the class of CDFs in $\mathcal{F}_{\bf z}$ leads to additional constraints. Since any cdf $F$ in this class has a jump discontinuity at each $\bar{x}_l = z_{u_l}$, it follows that $G(z_i) = 1 - 
F(z_i) > 0$ for $i < u_{\ell}$, and $G(z_i) < G(z_{i-1})$ 
for every $i \in {\bf u}$. In other words, we have 
$\theta_i < 1$ for $i \in {\bf u}$, and $\theta_i > 0$ for $i < u_{\ell}$. There are no other constraints on the elements of $\boldsymbol{\theta}$. Also, we can retrieve $F\left({\bf z}\right)$ given 
$\boldsymbol{\theta}$ using the following equality. 
\begin{equation}\label{eq:defining_G_in_terms_of_theta}
F\left(z_i\right) = 1 - G(z_i) = 1 - \prod_{j=1}^{i}\theta_j \ , \ \forall \ 1\leq i\leq (\ell+K).
\end{equation}
Now, using \eqref{eq:likelihood_lambda_X_F}, \eqref{eq:defining_G} and \eqref{eq:defining_theta}, we can 
rewrite the `adapted' likelihood $Lik_{A}$ in terms of $\boldsymbol{\theta}$ as follows: 
\begin{align*}
{}& Lik_{A} (\lambda, \boldsymbol{\theta} )\\
={}& C^{\star}\exp\big(-K\lambda\tau\big) \ \lambda^{\ell+\left|K_s\right|} \ 2^\ell \prod_{k\in K_s}t_{0,k}\prod_{i\in S}G(\bar{x}_i) \prod_{k=1}^K G(r_k)^{1_\{m_k=0,o_k=1\}}\\
{}& \times \exp\bigg(\lambda\bigg(\sum_{k=1}^{K}\big(1 - G(r_k)\big)t_{0,k} + \sum_{l=1}^{\ell}\big(1 - G(\bar{x}_l)\big)\bar{t}_l \bigg)\bigg) \ \bigg[\prod_{l=1}^{\ell}\Big(G(\bar{x}_{l}-) - G(\bar{x}_l)\Big) \bigg]^{1_{\{\ell>0\}}}\\
={}& C^{\star}\exp\big(-K\lambda\tau\big) \ \lambda^{\ell+\left|K_s\right|} \ 2^\ell \prod_{k\in K_s}t_{0,k}\prod_{i\in S}G(\bar{x}_i) \prod_{k=1}^K G(r_k)^{1_\{m_k=0,o_k=1\}}\\
{}& \times \exp\bigg(\lambda\bigg(K\tau - \sum_{k=1}^{K}G(r_k)t_{0,k} - \sum_{l=1}^{\ell}G(\bar{x}_l)\bar{t}_l \bigg)\bigg) \ \bigg[\prod_{l=1}^{\ell}\Big(G(\bar{x}_{l}-) - G(\bar{x}_l)\Big) \bigg]^{1_{\{\ell>0\}}}\\
={}& C^{\star} \ \lambda^{\ell+\left|K_s\right|} \ 2^\ell \prod_{k\in K_s}t_{0,k} \prod_{i\in S}G(\bar{x}_i) \prod_{k=1}^K G(r_k)^{1_\{m_k=0,o_k=1\}}\\
{}& \times \exp\bigg(-\lambda\bigg(\sum_{k=1}^{K}G(r_k)t_{0,k} + \sum_{l=1}^{\ell}G(\bar{x}_l)\bar{t}_l \bigg)\bigg) \ \bigg[\prod_{l=1}^{\ell}\Big(G(\bar{x}_{l}-) - G(\bar{x}_l)\Big) \bigg]^{1_{\{\ell>0\}}}\\
={}& C^{\star} \ \lambda^{\ell+\left|K_s\right|} \ 2^\ell \prod_{k\in K_s}t_{0,k} \prod_{i\in \bar{S}}G(z_i) \ \exp\bigg(-\lambda\sum_{i=1}^{\ell+K}G(z_i)\tilde{t}_i \bigg) \ \bigg[\prod_{l=1}^{\ell}\Big(G(z_{u_l}-) - G(z_{u_l})\Big) \bigg]^{1_{\{\ell>0\}}},
\end{align*}
where $u_0 = 0$. Using \eqref{eq:defining_G_in_terms_of_theta}, we get 
\begin{align}
Lik_{A} ( \lambda, \boldsymbol{\theta} )
={}& C^{\star} \ \lambda^{\ell+\left|K_s\right|} \ 2^\ell \bigg(\prod_{k\in K_s}t_{0,k}\bigg) \bigg(\prod_{i\in \bar{S}}\prod_{j=1}^{i}\theta_j\bigg) \ \exp\bigg(-\lambda\sum_{i=1}^{\ell+K}\tilde{t}_i\bigg(\prod_{j=1}^{i}\theta_j\bigg)\bigg) \nonumber\\
{}& \times \bigg[\prod_{l=1}^{\ell}\bigg((1 - \theta_{u_l})\prod_{j=1}^{u_{l}-1}\theta_j\bigg) \bigg]^{1_{\{\ell>0\}}}.  \label{eq:likelihood_lambda_theta}
\end{align}

\noindent
A straightforward argument shows that 
$$
Lik_{A} ( \lambda, (\theta_1, \theta_2, \ldots, \theta_{\ell+K})) = Lik_{A} ( c \lambda, (c^{-1} \theta_1, \theta_2, \ldots, \theta_{\ell+K}))
$$

\noindent
for any constant $c > 0$. To address this identifiability issue, we note 
that no data is observed below $z_1$, which is the smallest starting price 
among all auctions in the data. Hence, it is quite reasonable to assume 
that $F(z_1) = 0$, which is equivalent to setting $\theta_1 = 1$. 

With the identifiability issue resolved, the goal now is to maximize $Lik_{A}$ with respect to $\boldsymbol{\theta}$, where each entry of $\boldsymbol{\theta}$ is in $[0,1]$, $\theta_i < 1$ for $i \in {\bf u}$, and $\theta_i > 0$ for $i < u_{\ell}$. We achieve this using the coordinate-wise ascent algorithm. The details 
of this algorithm are derived in Section \ref{sec:coordinate_max_2ndPrice_Auction_Single_Lambda_case}. 

\section{Maximizing $Lik_{A} ( \lambda, \boldsymbol{\theta} )$: Coordinate ascent algorithm}\label{sec:coordinate_max_2ndPrice_Auction_Single_Lambda_case}

\noindent
Applying natural logarithm on both sides of the equation in \eqref{eq:likelihood_lambda_theta}, we get 
\begin{align}
\ln (Lik_{A} ( \lambda, \boldsymbol{\theta} ))
={}& \ln(C^{\star}) + \big(\ell+\left|K_s\right|\big)\ln(\lambda) + \ell\ln (2) + \sum_{k\in K_s}\ln(t_{0,k}) + \sum_{i\in\bar{S}}\sum_{j=1}^{i}\ln(\theta_j) \nonumber\\
{}& - \lambda\sum_{i=1}^{\ell+K}\tilde{t}_i\bigg(\prod_{j=1}^{i}\theta_j\bigg) + 1_{\{\ell>0\}} \bigg[ \sum_{l=1}^{\ell}\ln (1 - \theta_{u_l}) + \sum_{l=1}^{\ell}\sum_{j=1}^{u_{l}-1}\ln\theta_j \bigg],
\label{eq:log_likelihood}
\end{align}

\noindent
where $u_0 = 0$. We now introduce notation which allows for a more compact and accessible representation of $\ln(Lik_{A})$.  
Recall that ${\bf z}$ is obtained by pooling all the $K$ starting prices and the $\ell = \sum_{k=1}^K m_k$ 
`non-starting' standing prices (elements of ${\bf \bar{x}}$), and $u_j$ represents the position of 
the $\bar{x}_i$ in ${\bf z}$ for $1 \leq j \leq \ell+K$. In particular, $u_\ell$ is the position of 
$\bar{x}_\ell$, the largest `non-starting' standing price across all the $K$ auctions in ${\bf z}$. In 
other words, $\bar{x}_\ell = z_{u_\ell}$. It is possible that $u_\ell < \ell + K$. For example, in 
settings where the starting price in one of the auctions where the item is unsold is larger than 
$\bar{x}_\ell$, it follows that $\bar{x}_\ell$ is not the largest entry in ${\bf z}$ and $u_\ell < 
\ell+K$. With this background, we define 
\begin{align}
l_i &= 0 \quad \text{if} \quad 1\leq i \leq u_1 - 1, \nonumber\\
l_i &= j \quad \text{if} \quad u_{j} \leq i \leq u_{j+1} - 1 \ , \ \text{for} \ 
i=u_1, u_1 +1,\ldots,u_{\ell} - 1, \nonumber\\
\text{and} \ l_i &= \ell \quad \text{if} \quad u_{\ell} \leq i \leq \ell+K.
\label{eq:notations_in_log_likelihood_expression}
\end{align}

\noindent
In other words, note that $u_1 < u_2 < \cdots < u_\ell$ induce an ordered 
partition of the set $\{1,2, \cdots, u_\ell -1\}$ into $\ell$ disjoint subsets 
$$
\{1, \cdots, u_1 - 1\}, \{u_1, \cdots, u_2 - 1\}, \cdots, \{u_{\ell-1}, \cdots, 
u_{\ell} - 1\}. 
$$

\noindent
Hence, any $1 \leq i \leq u_\ell$ belongs to one of the subsets in the above 
partition, and $l_i$ is defined to be one less than the position of that subset in the 
partition. For $u_\ell \leq i \leq \ell+K$ we define 
$l_i = \ell$. 

\noindent
{\it Example (continued):} 
In the example with $K=4$ auctions from 
Figure \ref{fig:example.plot}
% Section~\ref{subsec:new:notation}
, $u_\ell = \ell+K = 11$, and 
$$
(l_1, l_2, l_3, l_4, l_5, l_6, l_7, l_8, l_9, l_{10}, l_{11}) = 
(0,0,1,1,2,3,3,4,5,6,7). 
$$

\noindent
Using the above notation, it follows from \eqref{eq:log_likelihood} that 
\begin{align}
\ln(Lik_{A} ( \lambda, \boldsymbol{\theta} ))
={}& \ln(C^{\star}) + \big(\ell+\left|K_s\right| \big) \ln(\lambda) + \ell\ln (2) + \sum_{k\in K_s}\ln(t_{0,k}) + \sum_{i=1}^{\ell+K}\left|Q_i\right|\ln(\theta_i) \nonumber\\
{}& -\lambda\sum_{i=1}^{\ell+K}\tilde{t}_i \bigg(\prod_{j=1}^{i}\theta_j \bigg) + 1_{\{\ell>0\}}\bigg[ \sum_{l=1}^{\ell}\ln (1 - \theta_{u_l}) + \sum_{i=1}^{\ell+K}(\ell - l_i)\ln \theta_i \bigg],
\label{eq:log_likelihood_lambda_theta}
\end{align}

\noindent
where
\begin{align*}
Q_i := \left\{j\in \bar{S} : j \geq i\right\} = \text{Set of $j\in \bar{S}$ which are greater than or equal to $i$}. 
\end{align*}

\noindent
Note that 
$$
\left|Q_1\right| = \left|\bar{S}\right| = \text{Number of auctions where the item is sold {\em at or above} the starting price.}
$$

\noindent
To maximize $\ln(Lik_{A} ( \lambda, \boldsymbol{\theta} ))$, we pursue
the coordinate-wise ascent approach where each iteration of the algorithm cycles
through maximization of $\ln(Lik_{A} ( \lambda, \boldsymbol{\theta} ))$ with respect to the co-ordinate $\theta_i$ (with other entries 
of $\boldsymbol{\theta}$ fixed at their current values) for every $1 \leq i \leq
\ell+K$. We now show that each of these $\ell+K$ coordinate-wise maximizers are 
available in closed form.

\subsection{Coordinate-wise maximizers for $Lik_{A} ( \lambda, \boldsymbol{\theta} )$} \label{subsec:coordinate_max_theta}

\noindent
We start with the maximization with respect to $\lambda$. Simple calculation shows that for given $F$ (or $\boldsymbol{\theta}$), $Lik_{A} \left(\lambda, \boldsymbol{\theta} \right)$ is maximized at 

$$\lambda=\frac{\ell+\left|K_s\right| }{\sum_{i=1}^{\ell+K}\tilde{t}_i \bigg(\prod_{j=1}^{i}\theta_j \bigg)}.$$

Based on the algebraic structure of $Lik_{A} ( \lambda, \boldsymbol{\theta} )$,  
we divide the coordinate-wise maximization steps into three groups: One with 
$\theta_k$ when $k \in \bf{u}$, where $\bf{u}$ is defined to be the set $\{u_1,u_2, \ldots, u_{\ell} \}$, the second with $\theta_k$ when $k \notin \bf{u}$ and $k \leq\max(\bar S)$, and the third with 
$\theta_k$ when $k \notin u$ and $k > \max(\bar S)$. We discuss each case in detail separately below. 

\medskip

\textbf{Case I: Maximization w.r.t. $\theta_i$ for $i \in \bf{u}$}. \ \ 
If ${\bf u}$ is non-empty, then $\ell > 0$. For any $i \in \bf{u}$, taking derivative of the 
expression for $\ln(Lik_{A} ( \lambda, \boldsymbol{\theta} ))$ in 
\eqref{eq:log_likelihood_lambda_theta} w.r.t. $\theta_i$ and equating it to zero
gives us the following
\begin{align}
{}& \dfrac{\partial\big[\ln(Lik_{A} ( \lambda, \boldsymbol{\theta} )) \big]}{\partial\theta_i} =  0 \nonumber\\
\Leftrightarrow{}& - \lambda\sum_{\tilde{i}=i}^{\ell+K}\tilde{t}_{\tilde{i}} \Bigg(\prod_{\underset{j\neq i}{j=1}}^{\tilde{i}}\theta_j \Bigg) + \dfrac{\big(\left|Q_i\right| +  (\ell - l_i) \big)}{\theta_i} - \dfrac{1}{1 - \theta_i} = 0 \nonumber\\
%\implies & \ - \ \lambda\sum_{i=k}^{\ell+K}\tilde{t}_i\left(\prod_{\underset{j\neq k}{j=1}}^{i}\theta_j\right) \ + \ \dfrac{\left(\left|Q_k\right| +  1_{\{\ell>0\}}\left(\ell - l\right)\right)}{\theta_k} \ - \  \dfrac{1_{\{\ell>0\}}\left(\prod_{\underset{j\neq k}{j=u_{l-1}+1}}^{u_l}\theta_j\right)}{\left(1 - \prod_{j=u_{l-1}+1}^{u_l}\theta_j\right)} \ = \ 0 \nonumber\\
\Leftrightarrow{}& - A_i + \dfrac{B_i}{\theta_i} - \dfrac{1}{1 - \theta_i} = 0, 
\label{eq:coordinate_max_theta_equation}
\end{align}
where
\begin{align}
A_i ={}& \lambda\sum_{\tilde{i}=i}^{\ell+K}\tilde{t}_{\tilde{i}}\Bigg(\prod_{\underset{j\neq i}{j=1}}^{\tilde{i}}\theta_j \Bigg) > 0 \nonumber\\
B_i ={}& \left|Q_i\right| +  (\ell - l_i) > 0.
\label{eq:coordinate_max_theta_quadratic_equation_notations}
\end{align}

\noindent
Since $\theta_i \leq 1$ and $\theta_i > 0$, it follows that 
\begin{align}
    {}& \dfrac{\partial\big[\ln(Lik_{A} ( \lambda, \boldsymbol{\theta} )) \big]}{\partial\theta_i} = 0 \nonumber\\
    \implies{}& A_i\theta_i^2 - \big(A_i + B_i + 1 \big)\theta_i + B_i = 0. 
    \label{eq:coordinate_max_theta_quadratic_equation}
\end{align}

\noindent
Since $B_i > 0$, it follows that the discriminant of the quadratic 
equation \eqref{eq:coordinate_max_theta_quadratic_equation}, denoted by $D_i$,  
satisfies 
\begin{align}
D_i ={}& \big(A_i + B_i + 1 \big)^2 - 4A_iB_i \nonumber\\
={}& \big(A_i - B_i + 1 \big)^2 + 4 B_i \ > \ 0. 
\label{eq:coordinate_max_theta_quadratic_equation_discriminant}
\end{align}

\noindent
Hence, the quadratic equation \eqref{eq:coordinate_max_theta_quadratic_equation}
has two real roots, namely,
\begin{equation}\label{eq:coordinate_max_theta_quadratic_equation_roots}
\theta_i = \dfrac{\big(A_i + B_i + 1 \big) \pm \sqrt{D_i}}{2A_i}.
\end{equation}

\noindent
Since $\sqrt{D_i} > A_i - B_i + 1$ by \eqref{eq:coordinate_max_theta_quadratic_equation_discriminant}, it follows that 
\begin{eqnarray*}
\dfrac{\big(A_i + B_i + 1 \big) + \sqrt{D_i}}{2A_i} 
\geq \dfrac{2\big(A_i + 1 \big)}{2A_i} > 1, 
\end{eqnarray*}

\noindent
since $A_i >0$. Hence the larger root with the positive sign for the 
square-root discriminant always lies outside the set of allowable values for 
$\theta_i$. The smaller root with the negative sign can be shown to be strictly positive since 
$(A_i + B_i + 1)^2 - D_i = 4 A_i B_i > 0$. Also, if $A_i \geq B_i + 1$, then 
$$
A_i + B_i + 1 - \sqrt{D_i} < A_i + B_i + 1 \leq 2 A_i. 
$$

\noindent
If $A_i < B_i + 1$, then using $A_i > 0$ we get 
\begin{eqnarray*}
& & (B_i + 1 - A_i)^2 = (B_i + 1 + A_i)^2 - 4A_i B_i - 4A_i < D_i\\
&\Rightarrow& (B_i + 1 + A_i) - \sqrt{D_i} < 2A_i. 
\end{eqnarray*}

\noindent
It follows that the smaller root lies in $(0,1)$. Since 
$$
\dfrac{\partial^2 \big[\ln(Lik_{A} ( \lambda, \boldsymbol{\theta} ) ) \big]}{\partial\theta_i^2} = - \dfrac{B_i}{\theta_i^2} - \dfrac{1}{(1 - \theta_i )^2} \ < \ 0, 
$$

\noindent
it follows that the smaller root is the unique maximizer of $\ln(Lik_{A} ( \lambda, \boldsymbol{\theta} ))$ with 
respect to $\theta_i$. To conclude, the unique maximizer of $\ln(Lik_{A} ( \lambda, \boldsymbol{\theta} ))$ with 
respect to $\theta_i$ is given by 
\begin{equation}\label{eq:coordinate_max_theta_expression_1}
    \hat{\theta}_i = \dfrac{\big(A_i + B_i + 1 \big) - \sqrt{D_i}}{2A_i},
\end{equation}
where $A_i$ and $B_i$ are as defined in \eqref{eq:coordinate_max_theta_quadratic_equation_notations}.\\

\textbf{Case II: Maximization w.r.t. $\theta_i$ for $i \notin \bf{u}$ and $i \leq \max(\bar S)$}.  For any $\theta_i$ with $i \notin \bf{u}$ and $i < u_\ell$ (other than $\theta_1$, which is set to $1$), the coefficient of $\ln(\theta_i)$, given by $\left|Q_i\right| +  1_{\{\ell>0\}}(\ell - l_i)$, is strictly positive, while $i <u_\ell \leq  \max(\bar S)$. Even when $i \geq u_\ell$, $(\ell - l_i)=0$, since $i \leq \max(\bar S)$, $|Q_i|>0$, still the coefficient is strictly positive. Again taking derivative of the log-likelihood expression in \eqref{eq:log_likelihood_lambda_theta} w.r.t. $\theta_i$ and equating it to zero gives us
\begin{align*}
{}& \dfrac{\partial\big[\ln(Lik_{A} ( \lambda, \boldsymbol{\theta} )) \big]}{\partial\theta_i} = 0\\
\Leftrightarrow{}& - \lambda\sum_{\tilde{i}=i}^{\ell+K}\tilde{t}_{\tilde{i}} \Bigg(\prod_{\underset{j\neq i}{j=1}}^{\tilde{i}}\theta_j \Bigg) + \dfrac{\big( \left|Q_i\right| +  1_{\{\ell>0\}}(\ell - l_i) \big)}{\theta_i} = 0\\
\Leftrightarrow{}& \theta_i = \dfrac{\big( \left|Q_i\right| +  1_{\{\ell>0\}}(\ell - l_i) \big)}{A_i},
\label{eq:coordinate_max_theta_1_quadratic_equation}
\end{align*}

\noindent
where $A_i$ is as defined in 
\eqref{eq:coordinate_max_theta_quadratic_equation_notations}. Note that 
${( \left|Q_i\right| +  1_{\{\ell>0\}}(\ell - l_i) )}/{A_i}$ is positive but not guaranteed to be less than or equal to $1$. However, since 
\begin{equation*}
\dfrac{\partial^2 \big[\ln(Lik_{A} ( \lambda, \boldsymbol{\theta} )) \big]}{\partial\theta_i^2} = - \dfrac{\big( \left|Q_i\right| +  1_{\{\ell>0\}}(\ell - l_i) \big)}{\theta_i^2} \ < \ 0,
\end{equation*}

\noindent
it follows that ${\partial [\ln(Lik_{A} ( \lambda, \boldsymbol{\theta} )
)]}/{\partial\theta_i} > 0$, i.e., 
$\ln(Lik_{A} ( \lambda, \boldsymbol{\theta} ))$ is an increasing 
function of $\theta_i$ for $\theta_i < {(\left|Q_i\right| +  1_{\{\ell>0\}}(\ell - l_i))}/{A_i}$. Hence, the unique maximizer of $\ln(Lik_{A} ( \lambda, \boldsymbol{\theta} ))$ with respect to 
$\theta_i$ is given by 
\begin{equation}
    \hat{\theta}_i = \min \bigg\{1, \dfrac{\big( \left|Q_i\right| +  1_{\{\ell>0\}}(\ell - l_i) \big)}{A_i} \bigg\}. 
    \label{eq:coordinate_max_theta_expression_2}
\end{equation}

\textbf{Case III: Maximization w.r.t. $\theta_i$ for $i \notin \bf{u}$ and $i > \max(\bar S)$}. In this case 
$|Q_i| = 0$ and $l_i = \ell$. It follows from \eqref{eq:log_likelihood_lambda_theta} that 
$\ln(Lik_{A} ( \lambda, \boldsymbol{\theta} ))$ is maximized with respect to $\theta_i$ at $0$. Hence, we set 
\begin{equation}
    \hat{\theta}_i = 0. 
    \label{eq:coordinate_max_theta_expression_3}
\end{equation}

\noindent
This amounts to estimating $F(z_i)$ for $i > \max(\bar S)$ by $1$. {\color{black} Note that any such $z_i$ corresponds to a 
starting price which is greater than the largest final selling price for all auctions in the data (including auctions where the item is sold at the starting price)}. Since the data 
offers no evidence that the support of the true valuation distribution $F$ extends up to $z_i$, setting 
the estimate of $F(z_i)$ to $1$ indeed seems a sensible choice in this non-parametric setting. 

A crucial aspect of coordinate-wise maximization of non-convex functions is effective initialization of parameter values. We first derive a generalized method of moments based initial estimator for $\lambda$, and then use it to obtain principled initial estimators for the components of 
$\boldsymbol{\theta}$.

\subsection{Initial estimator of $\lambda$: Generalized method of moments} 
\label{subsec:consistent_estimator_lambda}

\noindent
Consider first a single second price auction with starting price $r$, and recall that $M$
denotes the number of times the selling price changes throughout the course of the 
auction. Our goal is to find a function $h$ such that $E[h(M)] = \lambda$. To this end, 
we consider the process of consumers accessing the auction whose bid value is greater 
than or equal to $r$. Since we are assuming that the consumers bid their true private 
value, it follows that the proportion of such consumers in the population of all 
customers is $1 - F(r)$, and this ``thinned'' process of arriving consumers with bid 
values greater than $r$ is a Poisson process with rate $\lambda (1 - F(r))$. Let $N_r$ 
represent the total number of consumers who access the auction in the period $[0, \tau]$
and have bid values greater than the starting price $r$. Then $N_r \sim 
\mbox{Poisson}(\lambda \tau (1 - F(r)))$. Moreover, given $N_r=n$, let $A_i$ ($i=1,2,\ldots,n$) represent the event that the current selling price changes after the $i^{\text{th}}$ consumer (with bid greater than $r$) accesses the auction. Let, $\textbf{1}_{A_i}$ be the indicator function of the occurrence of the event $A_i$.

Note that $E [M \mid N_r=0 ] = 0 = E [M \mid N_r=1 ]$, and for $n \geq 2$, we have 
\begin{align}
    E \big[M \mid N_r=n \big] = E \big[\text{Number of selling price changes} \mid N_r=n \big] 
    ={}& E\bigg[\sum_{i=1}^{n} \textbf{1}_{A_i} \Big| N_r=n \bigg] \nonumber\\
    \stackrel{(a)}{=}& \sum_{i=2}^{n} P\big( A_i \mid N_r=n\big) \nonumber\\
    \stackrel{(b)}{=}& \sum_{i=2}^{n}\frac{2(i-1)}{i(i-1)} \nonumber\\ ={}& 2\sum_{i=2}^{n}i^{-1}. \label{eq:conditional.expectation.M.given.N}
\end{align}

\noindent
Here $(a)$ follows from the fact that two bids above $r$ are needed for the first change
in the standing/selling price. For $(b)$, note that the arrival of $i^{\text{th}}$ 
consumer with bid greater than $r$ changes the selling price if and only if the 
corresponding bid is the highest or second highest among the $i$ starting price exceeding
bids. Note that these bid values are i.i.d. with distribution $F$ truncated above $r$. 
There are $i(i-1)$ possible choices for the joint positions of the highest and second 
highest bids. The $i^{th}$ bid is the highest bid for $(i-1)$ of these choices, and the 
second highest bid for another $(i-1)$ choices, leading us to $(b)$. It follows from 
(\ref{eq:conditional.expectation.M.given.N}) that 
$$
E[M] = 2 E \bigg[ 1_{\{N_r > 1\}} \sum_{i=2}^{N_r}i^{-1} \bigg] =: g \big(\lambda \tau \big(1 - F(r) \big) \big) \mbox{ (say). }
$$

\noindent
Note that $1_{\{N_r > 1\}} \sum_{i=2}^{N_r}i^{-1}$ is increasing in $N_r$, and 
$N_r$ is stochastically increasing in terms of its mean parameter $\lambda \tau 
(1-F(r))$. Hence $g$ is a strictly increasing function, and 
\begin{equation} \label{lambda_generalized_method_of_moments}
\lambda=g^{-1} (E[M])/\tau (1 - F(r)). 
\end{equation}

\noindent
If the starting price $r$ is negligible, for example compared to the smallest final selling price seen in the data set, then it is reasonable to assume that $F(r) \approx 0$ (According to \cite{Starting:Price:2006} this is the optimal strategy for seller). Suppose now, we consider the data from $K$ independent second price auctions of 
identical copies of an item, with $M_k$ denoting the number of standing/selling price 
changes throughout the course of the $k^{th}$ auction, and with $r_k$ the starting price 
for the $k^{th}$ auction for $1\leq k\leq K$. {Let $K_r$ be the collection of all auction indices with negligible starting prices.} Then, it follows that from (\ref{lambda_generalized_method_of_moments}) that 
$$
\hat{\lambda} := \tau^{-1} \ g^{-1}\bigg(|K_r|^{-1} \sum_{k\in K_r}M_k\bigg),
$$
should be a reasonable generalized method of moments estimator for $\lambda$.

\begin{figure}
    \centering
    \includegraphics[width = 0.5\linewidth]{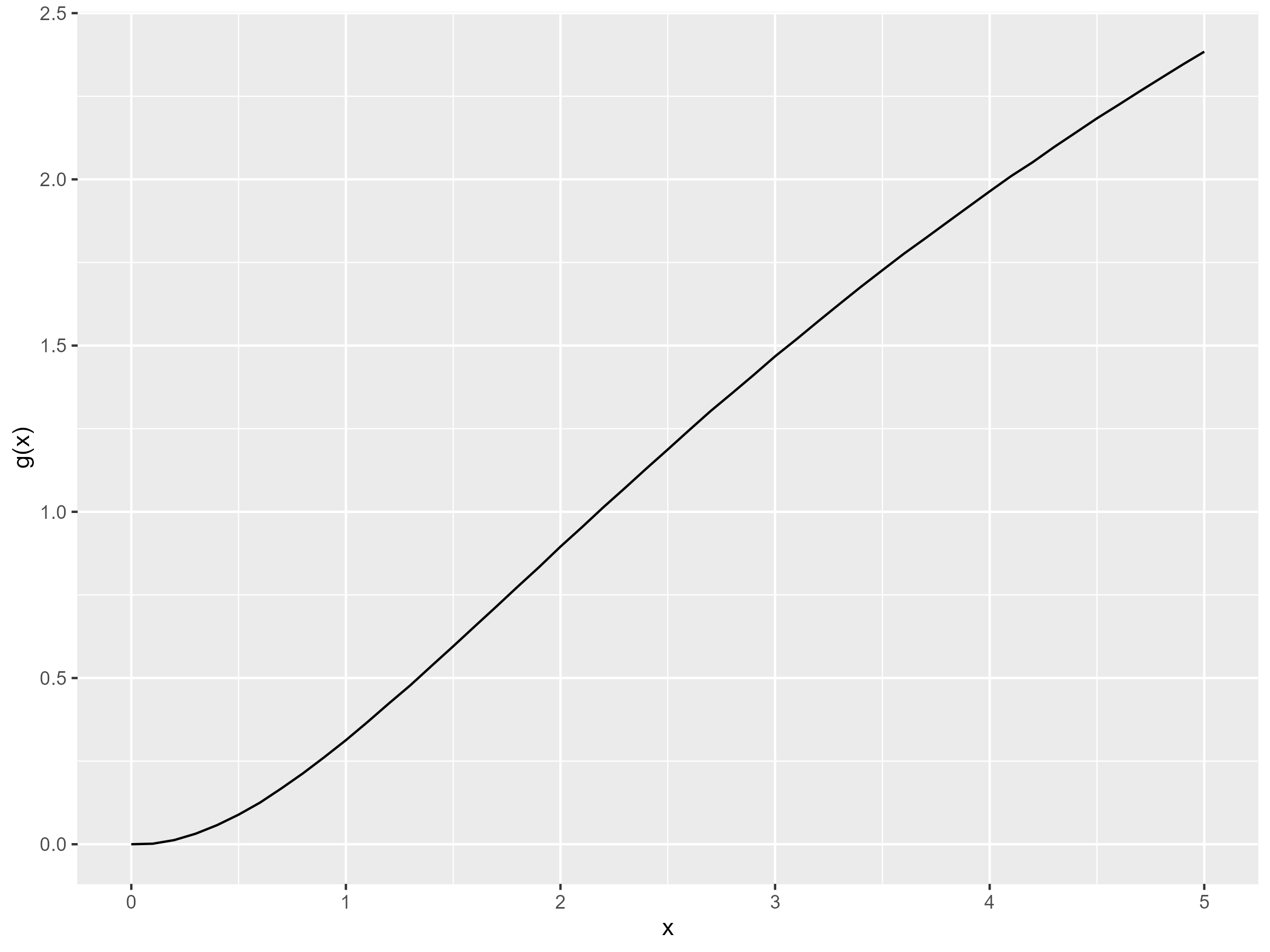}
    \caption{Plot of $g$ over the interval $\left[0,5\right]$.}
    \label{fig:plot.of.g}
\end{figure}

Of course, the function $g$ is not available in closed form and needs to be computed 
using numerical methods. A natural approach, given the definition of $g$ as a Poisson 
expectation, is Monte Carlo. Indeed, we computed $g(x)$ for every $x$ on a fine grid 
(with spacing $0.1$) ranging from $0$ to $5$. This Monte Carlo computation of $g$ is a {\it one-time process} that required minimal computational effort. The resulting plot of $g$ is provided in Figure~\ref{fig:plot.of.g}.

\subsection{Constructing an initial estimate $\hat{F}_{init}$ of $F$ (and 
$\boldsymbol{\theta}^{(0)}$ of $\boldsymbol{\theta}$) based exclusively on final selling prices and 
first observed bids}
\label{subsec:Initialization_theta_F_2ndPrice_Auction_Single_Lambda_case}

\noindent
The details of all the steps of the coordinate ascent maximization algorithm for $Lik_{A}$ are explicitly 
derived above in Section \ref{subsec:coordinate_max_theta}. However, as mentioned previously, a crucial detail which needs to be 
worked out is a `good' initial starting point for the algorithm. Especially for highly non-convex 
maximization such as in the current setting, the choice of the initial/starting value can play a critical
role in the quality of the final estimate produced by the coordinate ascent algorithm. 
In this section, we construct an initial estimate of $F$ based on the empirical distribution functions of both the final selling prices and the first observed bids (i.e., the price when for the first time the standing price jumps to a higher value from its respective starting price), respectively. Note that the 
methodology developed in \cite{10.1214/11-AOAS503} also relies exclusively on the final selling prices. However, that approach requires  the knowledge of the total number of consumers accessing each of the $K$ auctions. We do not assume this 
knowledge in the current setting, and need to overcome this additional challenge. Also, as stated above, we will use the first non-starting standing prices (first observed bids) to improve the quality of our initial estimator of 
$F$. 

Once the initial estimate $\hat{F}_{init}$ of $F$ is constructed, we can easily construct the initial estimate $\boldsymbol{\theta}^{(0)}:= (\theta_1^{(0)}, \theta_2^{(0)}, \ldots, \theta_{\ell+K}^{(0)})^T$ of $\boldsymbol{\theta}$ using \eqref{eq:defining_theta} as follows:
\begin{equation}\label{eq:initial.estimate.theta}
    \theta_i^{(0)} = \frac{1 - \hat{F}_{init} (z_i)}{1 - \hat{F}_{init} (z_{i-1})}, \ \forall \ 1\leq i \leq (\ell+K).
\end{equation}

\noindent
We now describe in detail the various steps involved in construction of 
$\hat{F}_{init}$. 

\medskip

\noindent
\textbf{Step I: Construct an estimate of $F$ based on the empirical distribution function of \textit{only} the final selling prices of auctions with relatively small starting prices.} \ \ First, consider a single second price auction with starting price $X_0 = r$. As in 
Section \ref{subsec:consistent_estimator_lambda} consider the process of consumers accessing the auction whose bid value is greater 
than or equal to $r$, and let $N_r$ represents the number of such consumers that access the auction in the period $[0, \tau]$. As observed in 
Section \ref{subsec:consistent_estimator_lambda}, this thinned process of arriving consumers is a Poisson process with rate $\lambda (1 - F(r))$, and 
$N_r \sim Poisson(\lambda \tau (1 - F(r)))$. We derive the conditional cdf of the
final selling price $X_M$ given $X_0 = r, N_r \geq 2$ as a function of $F(x)$. 
For this purpose, note that 
\begin{align}
    P (X_M \leq x \mid X_0 = r, N_r \geq 2)
    ={}& \sum_{n=2}^{\infty} P(X_M \leq x \mid N_r=n, X_0 = r) P(N_r=n \mid N_r\geq 2) \nonumber\\
    ={}& \sum_{n=2}^{\infty} P(X_M \leq x, R > x \mid X_0 = r, N_r=n) P(N_r=n \mid N_r\geq 2) \nonumber\\
    {}& + \sum_{n=2}^{\infty} P(X_M \leq x, R \leq x \mid X_0 = r, N_r=n) P(N_r=n \mid N_r\geq 2) . 
    \label{eq:conditional.cdf.of.SellingPrice.given.N}
\end{align}

\noindent
Recall that $R$ denotes the maximum placed bid during the course of the auction 
(we do not observe it), and that the valuation distribution of customers arriving
in the thinned Poisson process discussed above is the truncated version of $F$ at
$r$, denoted by 
$$
F_r (x) := \frac{F(x) - F(r)}{1 - F(r)} \ 1_{\{x > r\}}. 
$$

\noindent
Now, note that the event $\{X_M \leq x, R > x\}$ is equivalent to the constraint 
that the largest order statistic of the valuations of all the customers arriving 
via the thinned Poisson process is greater than $x$, but second largest order 
statistic is less than or equal to $x$. Similarly, the event $\{X_M \leq x, R 
\leq x\}$ is equivalent to the constraint that the largest order statistic of the
valuations of all the customers arriving via the thinned Poisson process is less than or equal to $x$. With $\lambda_r = \lambda (1 - F(r))$, it follows from \eqref{eq:conditional.cdf.of.SellingPrice.given.N} that 
\begin{align}
    {}& P( X_M \leq x \mid X_0 = r, N_r \geq 2 ) \nonumber\\
    ={}& \frac{1}{P(N\geq 2)}\bigg[\sum_{n=2}^{\infty} \frac{n(F_r(x))^{n-1}(1 - F_r(x))\exp(-{\lambda_r}\tau)({\lambda_r}\tau)^n}{n!} + \sum_{n=2}^{\infty} \frac{(F_r(x))^{n}\exp(-{\lambda_r}\tau)({\lambda_r}\tau)^n}{n!} \bigg] \nonumber\\
    ={}& \frac{1}{P(N\geq 2)}\bigg[{\lambda_r}\tau\exp(-{\lambda_r}\tau)(1 - F_r(x))\sum_{n=2}^{\infty} \frac{({\lambda_r}\tau F_r(x))^{n-1}}{(n-1)!} + \exp(-{\lambda_r}\tau)\sum_{n=2}^{\infty} \frac{({\lambda_r}\tau F_r(x))^n}{n!}\bigg] \nonumber\\
    ={}& \frac{{\lambda_r}\tau\exp(-{\lambda_r}\tau)(1 - F_r(x)) \big(\exp({\lambda_r}\tau F_r(x)) - 1 \big) + \exp(-{\lambda_r}\tau) \big(\exp({\lambda_r}\tau F_r(x)) - {\lambda_r}\tau F_r(x) - 1 \big)}{P(N\geq 2)} \nonumber\\
    ={}& \frac{\exp(-{\lambda_r}\tau) \ \Big({\lambda_r}\tau(1 - \eta)\big(\exp({\lambda_r}\tau \eta) - 1 \big) + \exp({\lambda_r}\tau \eta) - {\lambda_r}\tau \eta - 1 \Big)}{1 - \exp(-{\lambda_r}\tau) - {\lambda_r}\tau \exp(-{\lambda_r}\tau)}, \nonumber
\end{align}

\noindent
where $\eta = F_r(x)$. Let,
\begin{equation}\label{eq:cdf.of.SellingPrice}
    G_{{\lambda_r}}(\eta) := \frac{\exp(-{\lambda_r}\tau) \ \Big({\lambda_r}\tau(1 - \eta)\big(\exp({\lambda_r}\tau \eta) - 1 \big) + \exp({\lambda_r}\tau \eta) - {\lambda_r}\tau \eta - 1 \Big)}{1 - \exp(-{\lambda_r}\tau) - {\lambda_r}\tau \exp(-{\lambda_r}\tau)}.
\end{equation}
Note that 
\begin{align}
    \frac{d}{d\eta}G_{{\lambda_r}}(\eta) ={}& \frac{({\lambda_r}\tau)^2 (1 - \eta)\exp({\lambda_r}\tau \eta) - {\lambda_r}\tau \big(\exp({\lambda_r}\tau \eta) - 1\big) + {\lambda_r}\tau\exp({\lambda_r}\tau \eta) - {\lambda_r}\tau}{\exp({\lambda_r}\tau) - (1 + {\lambda_r}\tau )} \nonumber\\
    ={}& \frac{({\lambda_r}\tau)^2(1 - \eta)\exp({\lambda_r}\tau \eta)}{\exp({\lambda_r}\tau) - (1 + {\lambda_r}\tau )} \ > \ 0 \quad \text{for} \ \eta \in (0,1).
   \label{eq:derivative.G.lambda.wrt.eta}
\end{align}

\noindent
It follows that $G_{\lambda_r}$ is a strictly increasing function for $\eta \in 
[0,1]$. 

Now, coming back to our setting with $K$ independent auctions, suppose that  
we have multiple auctions with a given starting price $r$ (or close to $r$) where the 
item is sold above the starting price. Then based on the Glivenko-Cantelli lemma, 
\eqref{eq:cdf.of.SellingPrice} and \eqref{eq:derivative.G.lambda.wrt.eta}, we can 
use the function $G_{\lambda_r}^{-1}$ (with an appropriate estimate of $\lambda_r$)
applied to empirical cdf of the final selling prices of these auctions to estimate 
$F_r (x)$ for $x > r$. Setting the estimate of $F(r)$ to be zero, we can then 
obtain an estimate of $F(x)$ for $x > r$. Clearly, we would like to choose $r$ to 
be as small as possible. 

With this background, let $r_q$ denote the $q^{th}$ quantile of the 
starting prices among $\{r_1, r_2, \cdots, r_K\}$. Here $q \in 
(0,1)$ is a user-specified constant, and we denote by   $V(q)$ the set of 
indices of starting prices which lie within $[0, r_q]$. Ideally, one would like to have a reasonable number of auctions with very small/negligible starting prices. For example, in the Xbox data analyzed in Section~\ref{sec:Empirical_application}, roughly $25\%$ of the 
auctions have a starting price less than $\$1$ (the smallest final selling price 
is $\$28$). Let 
$$
G_{SP} (x) := \frac{1}{|V(q)|}\sum_{k \in V(q)} 
1_{\{ X_{m_k,k} \ \leq \ x \}},
$$

\noindent
be the empirical distribution function of the final selling prices for auctions in 
$V(q)$. Based on the above discussion we construct the estimator 
$\hat{F}_{SP}$
of $F$ as 
\begin{equation} \label{selling:price:estimator}
    \hat{F}_{SP} (x) = G_{\hat{\lambda}}^{-1}(G_{SP} (x)), \ \forall \ x\in \R.
\end{equation}

\noindent
 Here $\hat{\lambda}$ is our initial estimate of $\lambda$. 
%  Since $G_{\hat{\lambda}}^{-1} (0) = 0$, {\color{red}it follows that $\hat{F}_{SP} (x) = 0$ 
% for $x \leq r_{q}$}.
In fact, $\hat{F}_{SP} (x) = 0$ for all values below the 
smallest final selling price for auctions corresponding to $V(q)$. There
are likely many observed standing prices in the $K$ auctions which are below this 
smallest final selling price, and these values can/should be used to improve the 
estimator $\hat{F}_{SP}$. This process is described in the next step. 

\medskip

\noindent
\textbf{Step II: Incorporate the first non-starting standing prices into the 
construction of the initial estimate of $F$.} \ \ Consider again, to begin with, a 
single second price auction with starting price $r$, and the associated thinned Poisson process of arriving consumers with valuation greater than $r$. Letting $Y_1, Y_2 $ represent valuations of the first two arriving consumers in the thinned process, we have 
\begin{equation} \label{eq:cdf.of.First.Observed.Bid}
 P(X_1 \leq x \mid X_0 = r, N_r \geq 2) = P(\min \{Y_1, Y_2\} \leq x) =  1 - (1 - F_r(x))^2 .
\end{equation}

\noindent
Similar to Step I, let 
$$
G_{FP} (x) := \frac{1}{|V(q)|}\sum_{k \in V(q)} 
1_{\{X_{1,k} \ \leq \ x \}},
$$

\noindent
be the empirical distribution function of the first non-starting standing prices for
auctions in $V(q)$. Based on \eqref{eq:cdf.of.First.Observed.Bid}, we 
construct the estimator $\hat{F}_{FP}$ of $F$ as 
\begin{equation} \label{eq:initial.estimate.F.First.Observed.Bids}
    \hat{F}_{FP} (x) = 1 - \sqrt{1 - G_{FP} (x)}, \ \forall \ x\in \R. 
\end{equation}

\noindent
{Note that $G_{FP} (z_1) = 0$, which implies that $\hat{F}_{FP} (z_1) = 
0$, where $z_1$ is the smallest starting price.} 
However, 
$\hat{F}_{FP} (x) > 0$ when $x$ is larger than the smallest first non-starting 
standing price among auctions in $V(q)$. This smallest non-starting 
standing price is often considerably smaller than the smallest final selling price, 
and hence $\hat{F}_{FP}$ can be combined with $\hat{F}_{SP}$ of Step I, to get a 
better initial estimate of $F$ as follows. 

\medskip

\noindent
\textbf{Step III: Combining the two different initial estimates, namely, 
$\hat{F}_{FP}$ and $\hat{F}_{SP}$.} \ \ Let $p_1$ and $p_2$ respectively represent 
the largest non-starting standing price and the smallest final selling price for 
auctions in $V(q)$. As discussed previously, $\hat{F}_{SP}$ 
underestimates $F$ below $p_2$ and $\hat{F}_{FP}$ overestimates $F$ above $p_1$. 
Let $c$ be the largest real number $\leq \min\{p_1, p_2\}$ such that 
$\hat{F}_{FP} (c) \leq \hat{F}_{SP}(p_1)$. Then, we 
define a function $\hat{F}_{(0)}$ based on $\hat{F}_{FP}$ and $\hat{F}_{SP}$ as 
follows:
\begin{equation}\label{eq:initial.estimate.F}
    \hat{F}_{(0)}(x) = 
        \begin{cases}
        \hat{F}_{FP} (x) & \text{if} \ \ x\leq c\\
        \hat{F}_{SP} (x) & \text{if} \ \ x>p_2\\
        \hat{F}_{FP} (c) + \Big( \frac{\hat{F}_{SP} (p_1) - \hat{F}_{FP} (c)}{p_1 - c} \Big)(x-c) & \text{if} \ \ c<x\leq p_2.
        \end{cases}
\end{equation}

\noindent
This function $\hat{F}_{(0)}$ in \eqref{eq:initial.estimate.F} combines the strengths of
the two estimators $\hat{F}_{FP}$ and $\hat{F}_{SP}$, and gives a more balanced 
estimator of $F$ over all regions. Finally, since $G_{FP}, G_{SP}$ are step 
functions, so are $\hat{F}_{FP}, \hat{F}_{SP}$. It follows based on 
\eqref{eq:initial.estimate.F} that $\hat{F}_{(0)}$ is a step function as well, 
and has jumps only at the first non-starting standing prices and final selling prices for auctions in $V(q)$. A continuous version of this 
estimator, denoted by $\hat{F}_{init}$ can be obtained by linear interpolation 
of the values between the jump points. 
{Since $z_1 \leq c$, it follows that $\hat{F}_{(0)}(z_1)=\hat{F}_{FP}(z_1)=0$.}

\iffalse

get the values of $\hat{F}_{MLE}$ at any other real numbers in $(0,z_1)$, $(z_i,z_{i+1})$ for $1\leq i\leq (\ell+K-1)$, and $(z_{\ell+K},\infty)$. 

{\color{red} For the sake of 
computational stability (to avoid too many initial values of $1$ in 
${\boldsymbol \theta}^{(0)}$) in the coordinate ascent algorithm, we transform 
$\hat{F}_{(0)}$ to a  continuous function by linear interpolation and obtain the 
desired initial estimate, denoted by $\hat{F}_{init}$, of $F$. After getting the initial estimate $\hat{F}_{init}$ of $F$, we can easily get the initial estimate 
$\boldsymbol{\theta}^{(0)}$ of $\boldsymbol{\theta}$ using 
\eqref{eq:initial.estimate.theta}.} 

\fi

\subsection{The Coordinate ascent algorithm for maximizing $Lik_{A}$} \label{subsec:coordinate.ascent.algorithm}

\noindent
All the developments and derivations in the earlier subsections can now be compiled and
summarized via the following coordinate ascent algorithm to maximize 
$Lik_{A} ( \lambda, \boldsymbol{\theta} )$. 
\begin{algorithm}\label{algo:coordinate.ascent.algorithm}
Coordinate ascent algorithm:
\begin{enumerate}
    \item[Step~1.] Start with initial value $\lambda^{(0)} = \hat{\lambda}$ (Section \ref{subsec:consistent_estimator_lambda}) and $\boldsymbol{\theta}^{(0)} = (\theta_1^{(0)}, \theta_2^{(0)}, \ldots, \theta_{\ell+K}^{(0)})^T$ (Section \ref{subsec:Initialization_theta_F_2ndPrice_Auction_Single_Lambda_case}), and a user defined tolerance level $\epsilon > 0$.  
    {Note $\theta_1^{(0)}=1$ since $\hat{F}_{(0)}(z_1)=0$.}
    %and calculate $\ln(Lik_{A}(\boldsymbol{\theta}^{(0)} ))$ using \eqref{eq:log_likelihood_lambda_theta}.
    \vspace{0.1in}
    \item[Step~2.] Set $m = 0$. 
    %For any integer $m\geq 0$, suppose we have $\boldsymbol{\theta}^{(m)}$ such that its first $i_0$ many indices are same as that of $\boldsymbol{\theta}^{(0)}$, i.e., $\theta^{(m)}_i = \theta^{(0)}_i$ for $1\leq i \leq i_0$. We also calculate the corresponding function value $\ln(Lik_{A} (\boldsymbol{\theta}^{(m)}))$ using \eqref{eq:log_likelihood_lambda_theta}.
    \vspace{0.1in}
    \item[Step~3.] Set $\theta_1^{(m+1)}=1$ (for identifiability). For any $2 \leq i \leq (\ell+K)$, sequentially obtain $\theta_i^{(m+1)}$ from \eqref{eq:coordinate_max_theta_expression_1}, \eqref{eq:coordinate_max_theta_expression_2} and 
        \eqref{eq:coordinate_max_theta_expression_3} by using the coordinate values in \\$(\theta_1^{(m+1)},\ldots, \theta_{i-1}^{(m+1)}, \theta_{i+1}^{(m)}, \ldots, \theta_{\ell+K}^{(m)})^T$ to compute $A_i, B_i, C_i$.
    \vspace{0.1in}
    \item[Step~4.] Update $\lambda^{(m+1)}=\big(\ell+\left|K_s\right| \big)/\left(\sum_{i=1}^{\ell+K}\tilde{t}_i \bigg(\prod_{j=1}^{i}\theta_j^{(m+1)} \bigg)\right)$
    
    \item[Step~5.]If 
    \begin{equation*}
        \ln\big(Lik_{A} \big(\lambda^{(m+1)},\boldsymbol{\theta}^{(m+1)} \big)\big) - \ln\big(Lik_{A} \big(\lambda^{(m)}, \boldsymbol{\theta}^{(m)} \big)\big) > \epsilon,
    \end{equation*}
    set $m \leftarrow m+1$. Otherwise, go to Step$~6$.
    \vspace{0.1in}
    \item[Step~6.] Set $\hat{\boldsymbol{\theta}}_{MLE} = \boldsymbol{\theta}^{(m+1)}$ and $\hat{\lambda}_{MLE} = \lambda^{(m+1)}$. 
    \end{enumerate}
\end{algorithm}
Once we get $\hat{\boldsymbol{\theta}}_{MLE}$, we can easily get the corresponding maximum likelihood estimator of $F$ as follows 
\begin{equation}\label{eq:F.MLE.at.Z}
\hat{F}_{MLE} (z_i) = 1 - \prod_{j=1}^{i}\hat{\theta}_{MLE, j} \ , \ \forall \ 1\leq i\leq (\ell+K).
\end{equation}

\noindent
As explained above, the adapted objective function $Lik_{A}$ and the search space $\mathcal{F}_{\bf z}$ 
of CDFs with relevant jump discontinuities are artifacts of the semi-parametric approach that we pursue. 
However, once the estimates of the valuation distribution at elements of ${\bf z}$ are obtained using 
Algorithm \ref{algo:coordinate.ascent.algorithm}, a continuous estimate on the entire valuation 
distribution can be constructed via interpolation. In particular, we use the values of $\hat{F}_{(MLE)}$ 
at $z_i$'s, $\hat{F}_{(MLE)} (0) = 0$, and linear interpolation to construct a continuous estimator of the
population valuation distribution over the entire real line.

\section{Simulation study}\label{sec:Simulation_study}

\noindent
In this section we consider various choices of the true underlying valuation 
distribution $F$, e.g., uniform, piecewise uniform, pareto, gamma, and beta 
distributions, which are commonly used in marketing research. 
{
We then 
illustrate and compare the performance of the 
MLE  $\hat{F}_{MLE}$, the initial estimate $\hat{F}_{init}$ and the Pólya Tree estimate $\hat F_{PT}$ estimate from \cite{10.1214/11-AOAS503} with the 
corresponding true valuation distribution $F$. Note that the Bayesian 
methodology in \cite{10.1214/11-AOAS503} (which uses only the final selling 
prices in each auction) requires the knowledge of the total number of 
consumers accessing the auction. This does not hold in our motivating 
application, but in order to compare these methods we will also  assume such a knowledge in our 
synthetic data evaluations for $\hat F_{PT}$ below.
% The estimator $\hat{F}_{SP}$ in 
% (\ref{selling:price:estimator}) is based only 
% on final selling prices, and does not need the knowledge of total number of 
% customers accessing the auction. The estimator $\hat{F}_{init}$, which 
% improves $\hat{F}_{SP}$ by combining it appropriately with the first 
% non-starting standing price based estimator $\hat{F}_{FP}$ in 
% (\ref{eq:initial.estimate.F}).   DO WE STILL KEEP THE LAST TWO SENTENCE?
% Hence, $\hat{F}_{init}$ will be used as a representative/adaptation of the final selling price based approach of  \cite{10.1214/11-AOAS503} for the setting considered in this paper. 
}

\subsection{Data generation}\label{subsec:Data_simulation}

\begin{figure}
    \centering
    \includegraphics[width = 0.5\linewidth]{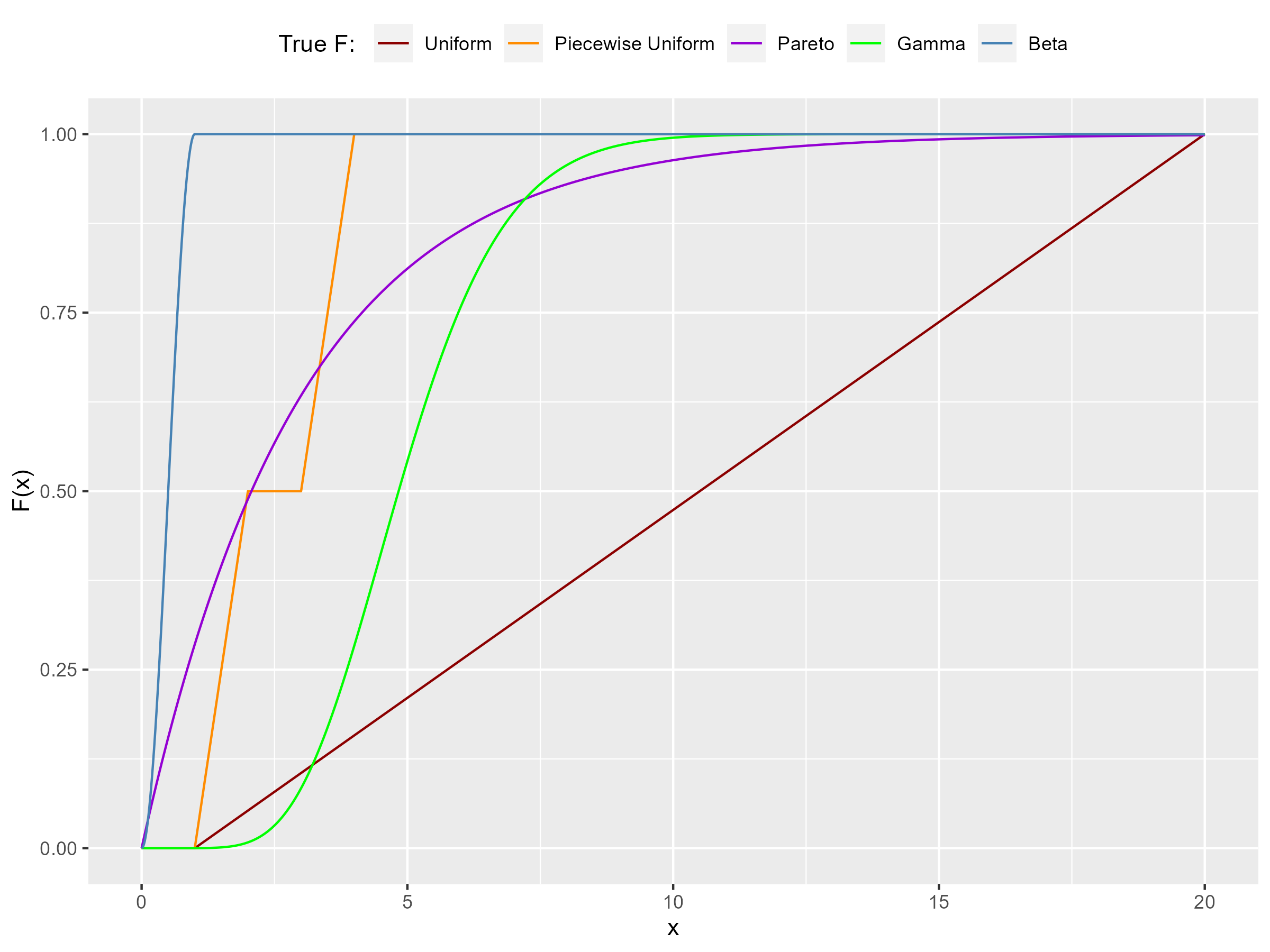}
    \caption{\textit{``True'' underlying valuation distribution functions used in the simulation studies.}}
    \label{fig:true.F.plots}
\end{figure}

We conducted five sets of simulation experiments, each using data simulated 
from a different choice of the underlying valuation distribution $F$. The 
cumulative distribution functions corresponding to the five choices of true 
underlying valuation distributions are shown in 
Figure~\ref{fig:true.F.plots}. 

For the first set of simulations, the underlying $F$ is a Uniform$\left(1, 20\right)$ distribution. For the second set of simulations, the underlying $F$ is an equally weighted mixture of the Uniform$\left(1, 2\right)$ and Uniform$\left(3, 4\right)$ distributions. From a managerial/marketing perspective, this corresponds to a market with two distinct consumer segments with different average valuations. For the third set of simulations, the true underlying $F$ is a Pareto distribution with location parameter $3$ and dispersion parameter $100$. For the fourth set of simulations, the true  underlying $F$ is a Gamma distribution with shape parameter $10$ and rate parameter $2$. For the last and fifth set of simulations, the underlying $F$ is a Beta distribution with its two positive shape parameters being equal to $2$, i.e., Beta$(2,2)$ distribution.

From each of the five true underlying $F$'s, we consider two settings, with  $K = 100$ and $K = 1000$ independent auctions of identical copies of an item. Varying the sample size here sheds light on the relationship between the sample size and the precision of the  MLE $\hat{F}_{MLE}$, the initial estimate $\hat{F}_{init}$ and the Pólya Tree estimate $\hat{F}_{PT}$ in estimating the true valuation distribution $F$. For each auction, we took the auction window ($\tau$) to be 100 units, and the constant rate ($\lambda$) of arrival of bidders to be equal to $1$. We then simulated the inter-arrival times between bidders from an exponential distribution with rate parameter $\lambda = 1$, and drew the bidders' valuations from $F$, keeping track of the entire sequence of standing prices and the intermediate times between jumps in the standing price for all independent auctions involved. This sequence of standing prices throughout the course of the auction, and the intermediate times between standing price changes are then treated as the {\it observed dataset} that is subsequently used to compute the initial estimator and the  MLE. 
For each choice of true $F$ and number of auctions $K$, $100$ replicated 
datasets are generated. 

Since the data is generated to be consistent with the modeling assumptions, one expects $\hat{F}_{MLE}$, which utilizes all available information, to have a superior performance than the initial estimator $\hat{F}_{init}$, which only uses the final selling price and first non-starting standing price for each auction, and the Pólya Tree estimator $\hat{F}_{PT}$ of \cite{10.1214/11-AOAS503}, which only uses final selling prices. The goal of these simulations is to examine extensively how much improvement can be obtained from our proposed method by incorporating the additional information in a variety of settings.  

\subsection{Simulation Results} \label{subsec:simulation.with.various.true.F}

\noindent
For each replicated dataset generated (as described in the previous 
subsection), we apply our semi-parametric methodology to obtain the initial 
estimate $\hat{F}_{init}$, the MLE $\hat{F}_{MLE}$, and the Pólya Tree estimator $\hat{F}_{PT}$ in \cite{10.1214/11-AOAS503}. The goal 
is to compare the accuracy of each of these estimators with respect to the respective true valuation distribution $F$. 

\begin{figure}
    \centering
    \includegraphics[width = \linewidth]{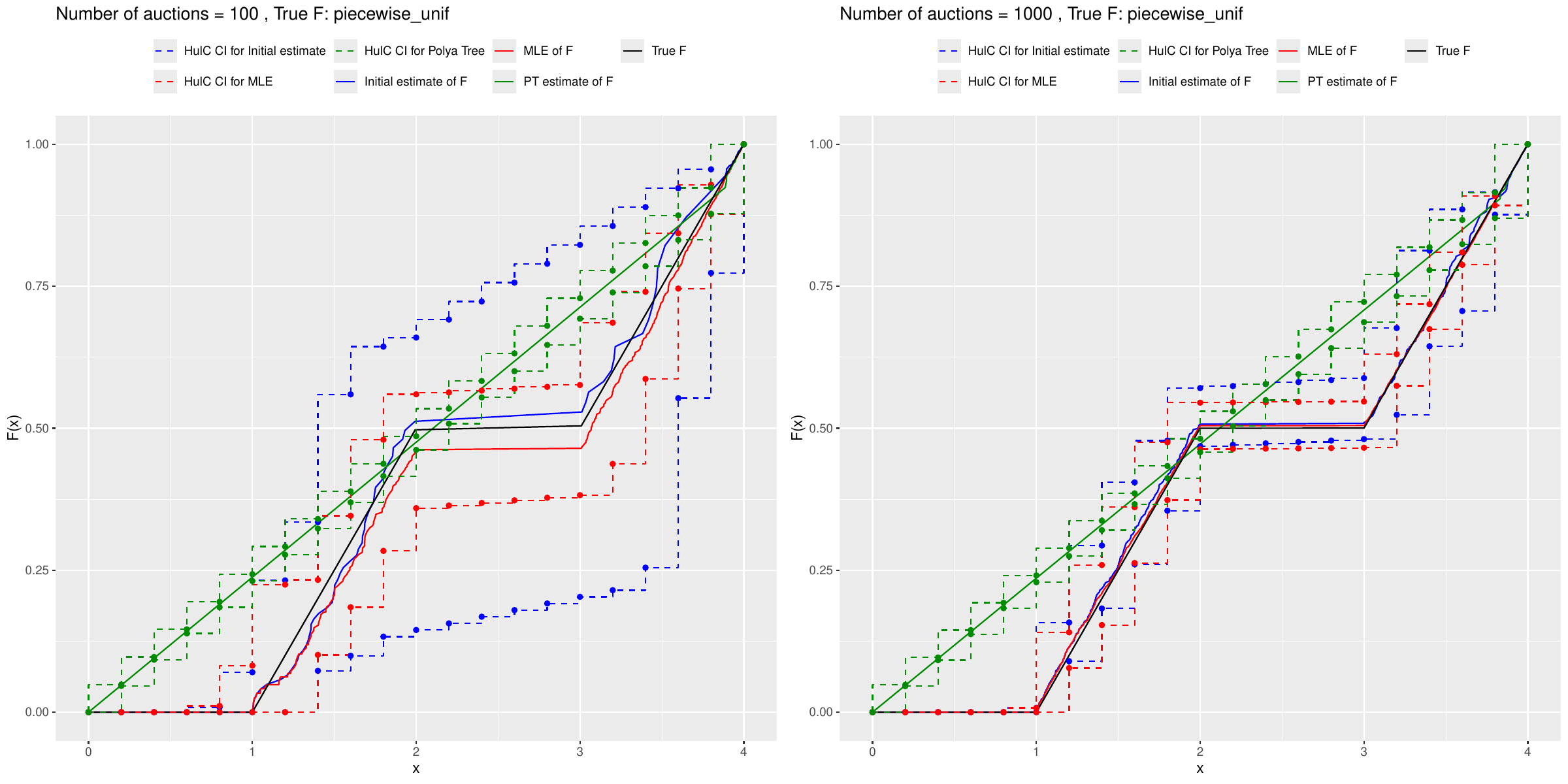}
    \caption{\textit{Plot of the  MLE $\hat{F}_{MLE}$ (red), initial estimator $\hat{F}_{init}$ (blue), Pólya Tree estimator (green) and the true valuation distribution $F$ (taken to be piece-wise Uniform) for a random chosen replicate with $K=100$ (left) and $K=1000$ independent auctions (right).  $90\%$-HulC confidence intervals or credible interval (for Pólya Tree) are also provided for both estimators (dotted lines, matching colors).}}
    \label{fig:piecewiseunif}
\end{figure}

\begin{figure}
    \centering
    \includegraphics[width = \linewidth]{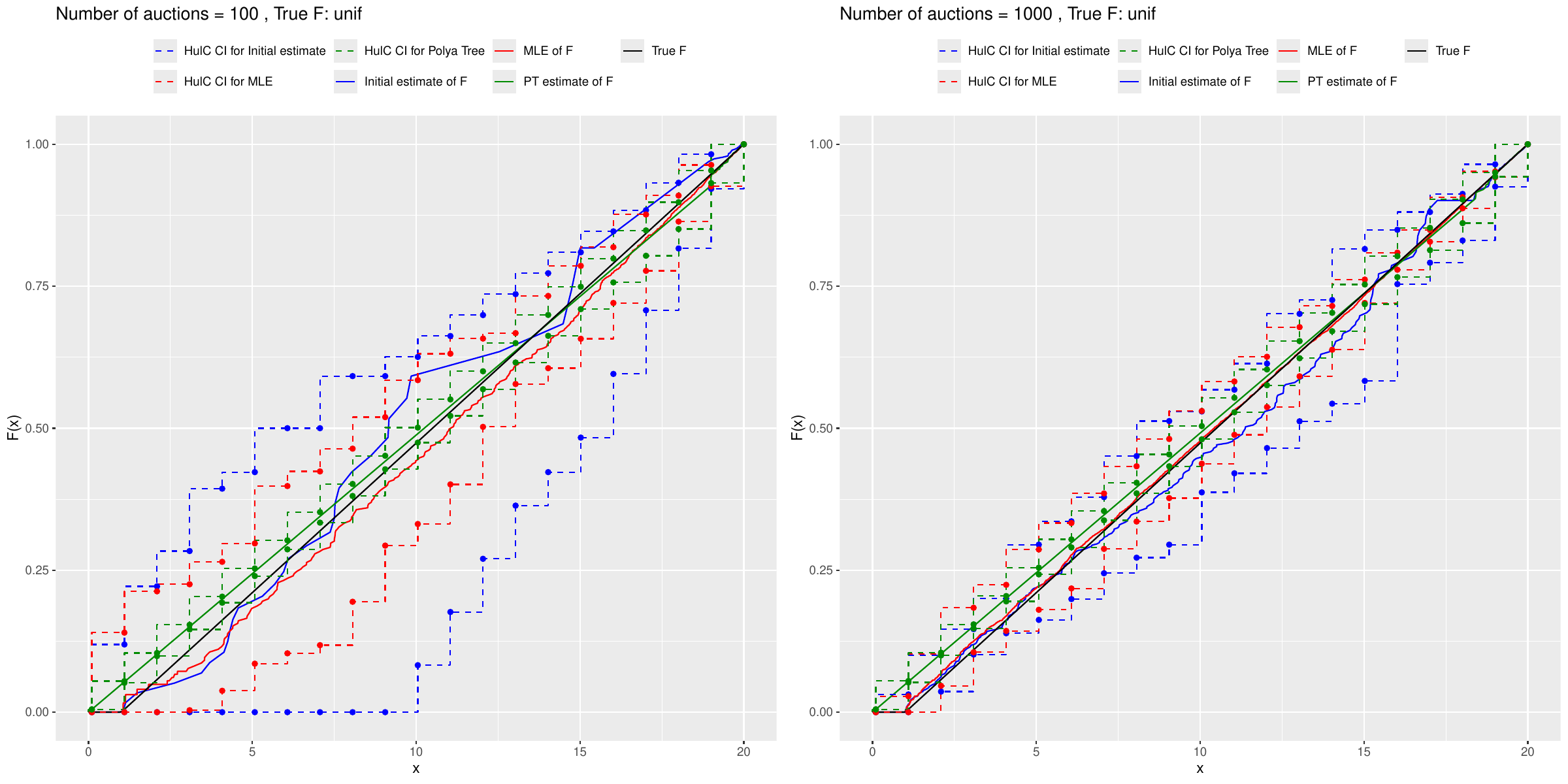}
    \caption{\textit{Plot of the  MLE $\hat{F}_{MLE}$ (red), initial estimator $\hat{F}_{init}$ (blue), Pólya Tree estimator (green) and the true valuation distribution $F$ (taken to be Uniform$(1,20)$) for a random chosen replicate with $K=100$ (left) and $K=1000$ independent auctions (right). $90\%$-HulC confidence intervals or credible interval (for Pólya Tree) are also provided for both estimators (dotted lines, matching colors).}}
    \label{fig:unif}
\end{figure}

\begin{figure}
    \centering
    \includegraphics[width = \linewidth]{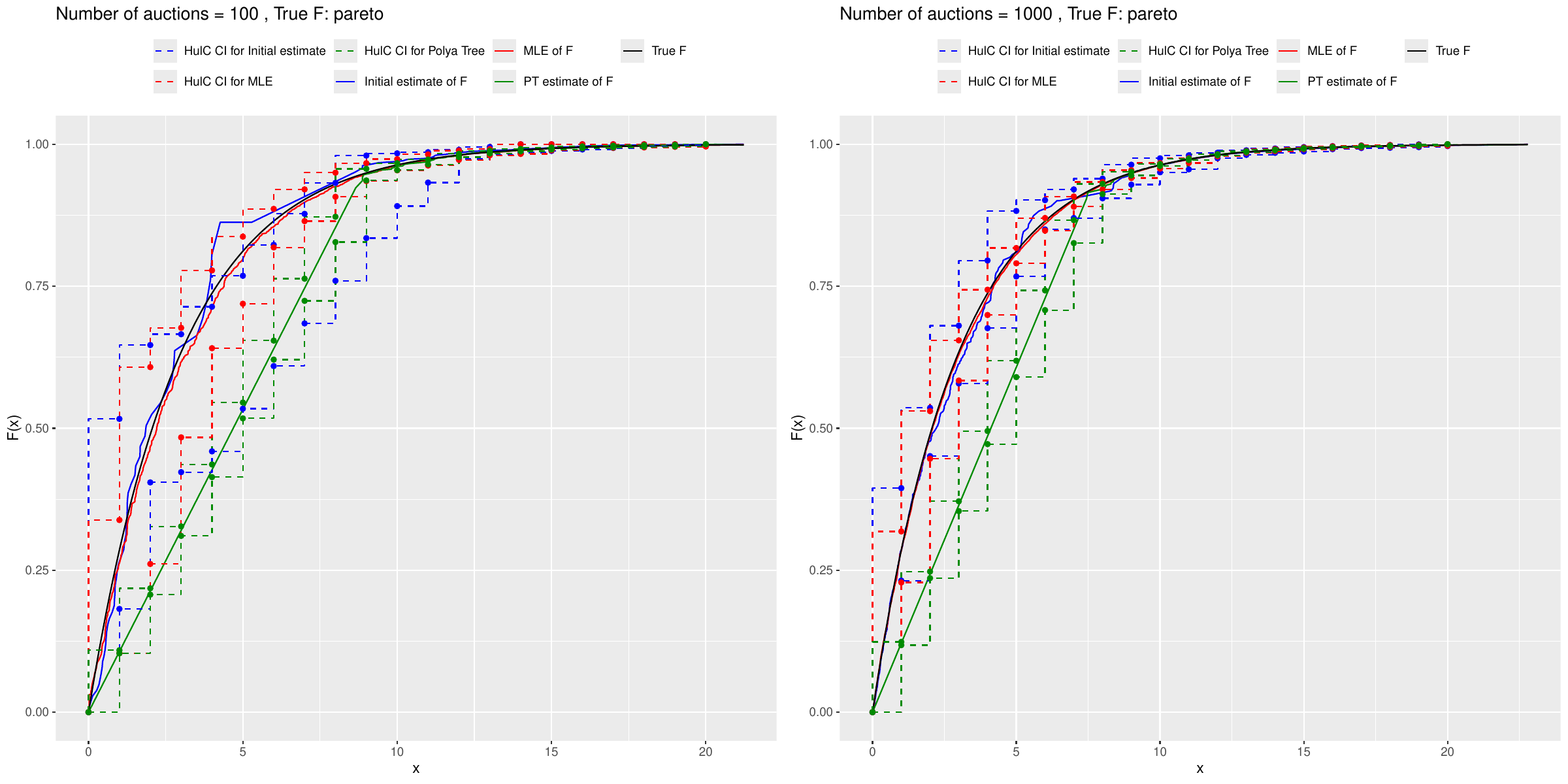}
    \caption{\textit{Plot of the  MLE $\hat{F}_{MLE}$ (red), initial estimator $\hat{F}_{init}$ (blue), Pólya Tree estimator (green) and the true valuation distribution $F$ (taken to be Pareto(location = 3,dispersion = 100)) for a random chosen replicate with $K=100$ (left) and $K=1000$ independent auctions (right).  $90\%$-HulC confidence intervals or credible interval (for Pólya Tree) are also provided for both estimators (dotted lines, matching colors).}}
    \label{fig:pareto}
\end{figure}

\begin{figure}
    \centering
    \includegraphics[width = \linewidth]{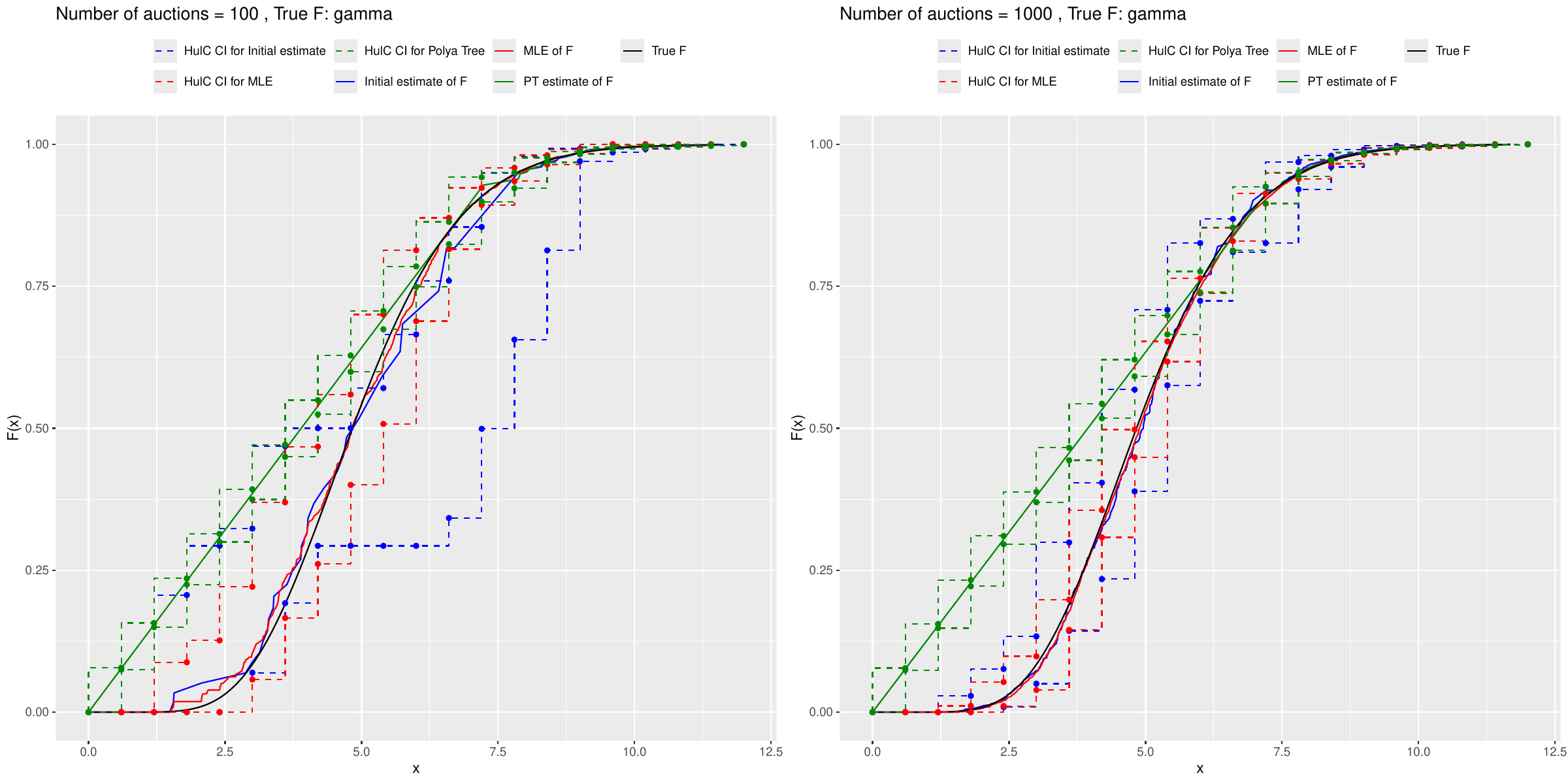}
    \caption{\textit{Plot of the  MLE $\hat{F}_{MLE}$ (red), initial estimator $\hat{F}_{init}$ (blue), Pólya Tree estimator (green) and the true valuation distribution $F$ (taken to be Gamma(shape = 10, rate = 2)) for a random chosen replicate with $K=100$ (left) and $K=1000$ independent auctions (right).  $90\%$-HulC confidence intervals or credible interval (for Pólya Tree) are also provided for both estimators (dotted lines, matching colors).}}
    \label{fig:gamma}
\end{figure}

\begin{figure}
    \centering
    \includegraphics[width = \linewidth]{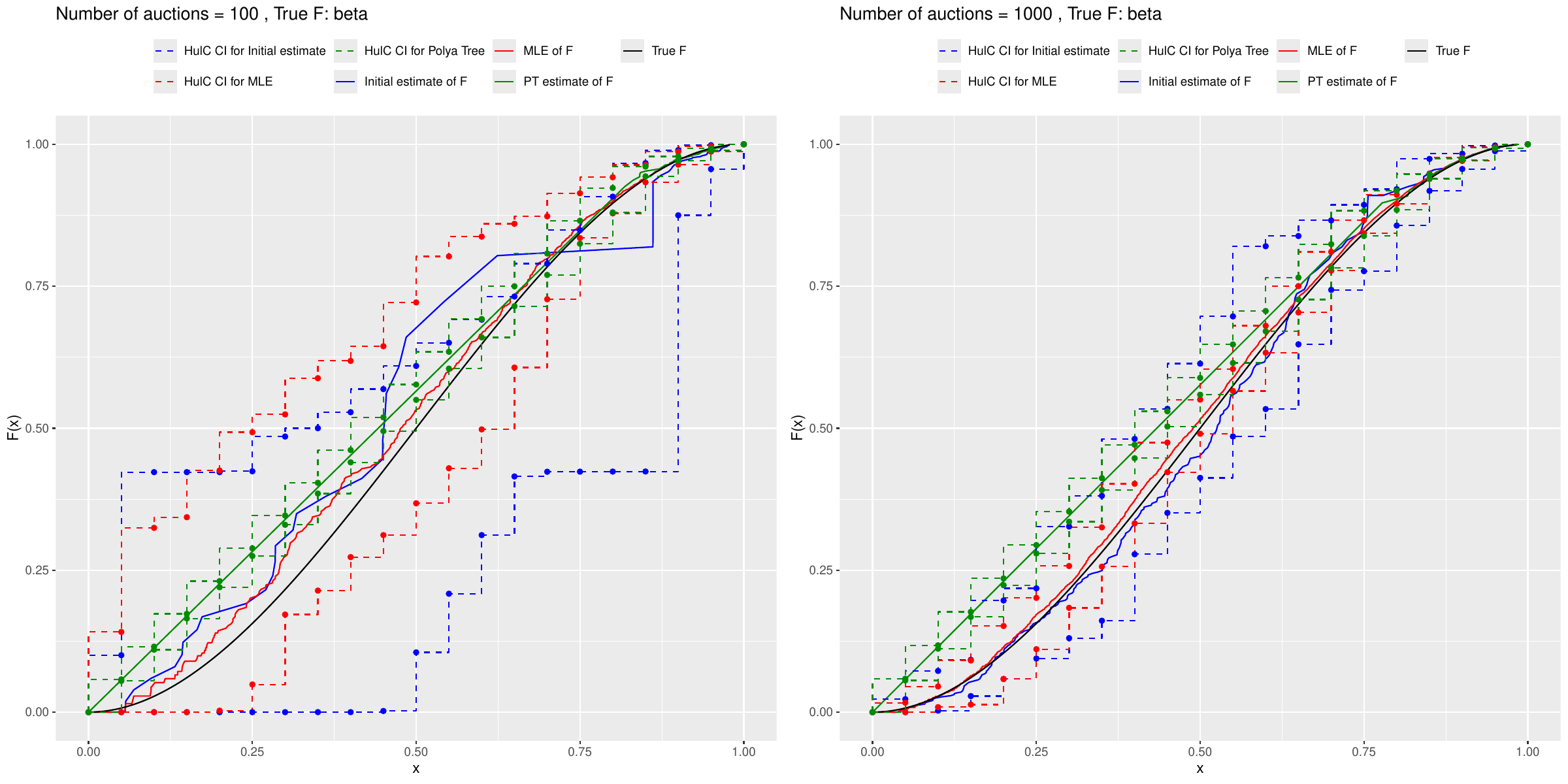}
    \caption{\textit{Plot of the  MLE $\hat{F}_{MLE}$ (red), initial estimator $\hat{F}_{init}$ (blue), Pólya Tree estimator (green) and the true valuation distribution $F$ (taken to be Beta(2,2)) for a random chosen replicate with $K=100$ (left) and $K=1000$ independent auctions (right).  $90\%$-HulC confidence intervals or credible interval (for Pólya Tree) are also provided for both estimators (dotted lines, matching colors).}}
    \label{fig:beta}
\end{figure}

We first provide a visual illustration of the results by choosing a random replicate out of $100$ for each of the $10$ simulation settings ($5$ true valuation distributions, and $2$ settings for the total number of auctions $K$). For the left part in Figure~\ref{fig:piecewiseunif}, we consider a randomly chosen replicate from the setting where the true valuation distribution $F$ is piece-wise Uniform and $K = 100$. The estimates $\hat{F}_{MLE}$, $\hat{F}_{init}$, $\hat{F}_{PT}$ and the true valuation distribution $F$ are plotted. We provide the $90\%$ confidence intervals for $\hat{F}_{MLE}$ and $\hat{F}_{init}$ based on the HulC approach developed in \cite{kuchibhotla2021hulc}.  We also provide the 90\% Bayesian credible intervals for 
$\hat{F}_{PT}$. It can be seen that $\hat{F}_{MLE}$ is much closer to the true valuation distribution $F$ compared to $\hat{F}_{init}$ and $\hat{F}_{PT}$ at almost all values in the interval $(1,20)$. The right part of Figure~\ref{fig:piecewiseunif} provides a similar plot for a randomly chosen replicate generated from the piece-wise Uniform and $K=1000$ setting. As expected, we see that the bias of both $\hat{F}_{MLE}$ and $\hat{F}_{init}$ reduces drastically when we increase the number of independent auctions $K$ from $100$ to $1000$, and $\hat{F}_{MLE}$ still overall provides a much more accurate estimate of the true valuation distribution $F$. We provide similar plots for a randomly chosen replicate from the eight other settings (with true $F$ being Uniform$\left(1,20\right)$, Pareto, Gamma and Beta, and with $K = 100, 1000$) in Figures~\ref{fig:unif}, \ref{fig:pareto}, \ref{fig:gamma}, and \ref{fig:beta}, and see that a similar phenomenon holds for all these settings. 

\noindent
The above plots based on single chosen replicates are illustrative, but need 
to be complemented with performance evaluation averaged over all the $100$ 
replicates in each of the $10$ simulation settings. In Table \ref{auction:performance:evaluation}, for each simulation setting, we provide both the Kolmogorov-Smirnov distance (KS-distance) and the Total variation distance (TV-distance) between the true valuation distribution $F$ and various estimators. In this comparison, we also include two parametric methods, the Gamma-F method and the Truncated Normal-F (TN-F) method, used in \cite{10.1214/11-AOAS503}. Both methods use final selling prices, and similar to Pólya Tree, require the knowledge of the total number of potential bidders who access the auction. The distances for all five methods (MLE, Initial, Pólya Tree, Gamma-F and TN-F) are averaged over the $100$ respective replications. The results show that the  MLE (based on the entire collection of standing prices) uniformly 
outperforms the initial estimator (based only on final selling prices and 
first non-starting standing prices) and the other three final selling price based estimators (Pólya Tree, Gamma-F and TN-F) in all the simulation settings. This strongly suggests that if the additional assumptions of single bidding and constant arrival rate seem to largely hold, it is worth using the proposed methodology which incorporates the extra information available in the form of all standing prices within the auction period.

\begin{table}[!h]
  \centering
  \small
  
  \begin{tabular}{@{} ll ccccc @{}}
    \toprule
    & & \multicolumn{5}{c}{\textbf{KS distance }} \\
      % & \multicolumn{3}{c}{\textbf{Total Variation }} \\
    \cmidrule(lr){3-7} 
    Distribution & $K$ 
      & $\ \quad $MLE$\ \ $   & Initial & Pólya Tree  
      & Gamma-F & TN-F\\ 
    \midrule
 Uniform & 100  &  0.048 &  0.138 &  0.050 &  0.266 &  0.732 \\ 
   & 1000  &  0.015 &  0.051 &  0.050 &  0.302 &  0.727 \\ 
 [\smallskipamount] Piecewise-Uniform & 100  &  0.048 &  0.104 &  0.239 &  0.385 &  0.708 \\ 
   & 1000  &  0.015 &  0.035 &  0.237 &  0.397 &  0.716 \\ 
 [\smallskipamount] Pareto & 100  &  0.048 &  0.114 &  0.280 &  0.061 &  0.131 \\ 
   & 1000  &  0.017 &  0.041 &  0.251 &  0.021 &  0.124 \\ 
 [\smallskipamount] Gamma & 100  &  0.045 &  0.134 &  0.290 &  0.314 &  0.239 \\ 
   & 1000  &  0.014 &  0.051 &  0.296 &  0.348 &  0.229 \\ 
 [\smallskipamount] Beta & 100  &  0.054 &  0.146 &  0.121 &  0.281 &  0.488 \\ 
   & 1000  &  0.018 &  0.057 &  0.123 &  0.382 &  0.478 \\ 
  \cmidrule(lr){3-7}    & & \multicolumn{5}{c}{\textbf{Total Variation}} \\    \cmidrule(lr){3-7} 
 Uniform & 100  &  0.075 &  0.211 &  0.058 &  0.296 &  0.738 \\ 
   & 1000  &  0.033 &  0.089 &  0.060 &  0.313 &  0.734 \\ 
 [\smallskipamount] Piecewise-Uniform & 100  &  0.091 &  0.151 &  0.486 &  0.445 &  0.674 \\ 
   & 1000  &  0.065 &  0.082 &  0.479 &  0.433 &  0.691 \\ 
 [\smallskipamount] Pareto & 100  &  0.057 &  0.174 &  0.281 &  0.057 &  0.13 \\ 
   & 1000  &  0.019 &  0.069 &  0.249 &  0.021 &  0.123 \\ 
 [\smallskipamount] Gamma & 100  &  0.067 &  0.228 &  0.317 &  0.316 &  0.241 \\ 
   & 1000  &  0.022 &  0.083 &  0.309 &  0.351 &  0.229 \\ 
 [\smallskipamount] Beta & 100  &  0.074 &  0.237 &  0.133 &  0.288 &  0.483 \\ 
   & 1000  &  0.025 &  0.090 &  0.129 &  0.385 &  0.472 \\ 
    \bottomrule
  \end{tabular}
  \caption{Kolmogorov-Smirnov (KS) distance and Total variation  distance between various estimators ($\hat{F}_{MLE}$, $\hat{F}_{init}$, Pólya Tree estimator(\emph{PT}), the Gamma method based on final selling prices - \emph{Gamma-F}, the Truncated-Normal method based on final selling prices - \emph{TN-F}) and the true valuation distribution $F$, averaged over 100 replications within each of the 10 simulation settings.}
  \label{auction:performance:evaluation}
\end{table}

For uncertainty quantification, we provide the coverage rate for 90\% HulC-based intervals for the proposed {\it MLE} and {\it Init} approaches, and also the coverage rate for 90\% credible intervals for the Pólya Tree ({\it PT}) approach of George and Hui (2012), under varied simulation settings. Note that for every approach, we compute the coverage rate as the proportion of replications for which the {\it entire} true distribution function is contained within the respective upper and lower confidence bands (with Bonferroni correction used to adjust for multiple testing). The results show that the proposed {\it MLE} approach provides significantly better coverage than the other two approaches. We also computed the band area (area between the lower and upper confidence bands) for each of the approaches, and the respective values (averaged over all replications) are provided in Table \ref{auction:coverage:evaluation} as well. These values reveal that the {\it Init} approach yields significantly wider confidence bands than the {\it MLE} approach, yet suffers from lower coverage. On the other hand, the {\it PT} approach yields significantly narrower bands compared to the {\it MLE} approach,  but {has zero empirical coverage}. So overall, the proposed {\it MLE} approach strikes a much better balance overall than the other two estimators. 

The HulC approach assumes median unbiasedness of the underlying estimators, but $\hat{F}_{init}$ and $\hat{F}_{MLE}$ are not median unbiased - adjusting for median bias in HulC can be computationally expensive and leads to wider confidence bands. Note that near the boundaries, where the value of the underlying true CDF is very close to zero or one, median unbiasedness is a tall ask for these non-parametric/semi-parametric estimators. Initial studies did show that the estimated median bias for $\hat{F}_{MLE}$ was very small across the simulation settings for points away from the boundary (points within 5th to 95th percentile range). The last three columns of Table \ref{auction:coverage:evaluation} provide `truncated coverage' for the three methods  across all settings, i.e., coverage calculation is restricted only to points within the 5th to 95th percentile range of the true distribution $F$. The results show that the observed coverage for $\hat{F}_{init}$ and $\hat{F}_{MLE}$ moves closer to the nominal coverage of 90\% in many settings when we restrict away from the boundary. The comparative performance of the three methods remains the same. These results suggest that is not much advantage, in terms of coverage, by adjusting for median bias of the proposed MLE estimator in HulC. 
\begin{table}[!h]
  \centering
  \small
  
  \begin{tabular}{@{} ll rrl rrl rrl @{}}
    \toprule
    & & \multicolumn{3}{c}{\textbf{Coverage }} 
      & \multicolumn{3}{c}{\textbf{Band Area }} 
      & \multicolumn{3}{c}{\textbf{Truncated Coverage }}\\
    \cmidrule(lr){3-5} \cmidrule(l){6-8} \cmidrule(l){9-11}
    Distribution & $K$ 
      & MLE   & Initial & PT  
      & MLE   & Initial & PT  
      & MLE    & Initial  & PT\\ 
    \midrule
Uniform & 100 & 0.65 & 0.63 & 0.00 & 3.61 & 6.66 & 0.45 & 0.92 & 0.90 & 0.00 \\  & 1000 & 0.94 & 0.84 & 0.00 & 1.09 & 3.14 & 0.42 & 0.94 & 0.84 & 0.00 \\[\smallskipamount]Piecewise-& 100 & 0.97 & 0.83 & 0.00 & 0.61 & 0.98 & 0.09 & 0.90 & 0.77 & 0.00 \\ Uniform & 1000 & 0.98 & 0.96 & 0.00 & 0.18 & 0.38 & 0.09 & 0.98 & 0.95 & 0.00 \\[\smallskipamount]Pareto & 100 & 0.56 & 0.01 & 0.00 & 1.25 & 2.90 & 0.27 & 0.89 & 0.01 & 0.00 \\  & 1000 & 1.00 & 0.93 & 0.00 & 0.45 & 1.18 & 0.20 & 0.97 & 0.88 & 0.00 \\[\smallskipamount]Gamma & 100 & 0.42 & 0.05 & 0.00 & 0.90 & 2.09 & 0.20 & 0.89 & 0.07 & 0.00 \\  & 1000 & 0.97 & 0.81 & 0.00 & 0.30 & 0.92 & 0.18 & 0.97 & 0.86 & 0.00 \\[\smallskipamount]Beta & 100 & 0.36 & 0.18 & 0.00 & 0.15 & 0.32 & 0.02 & 0.96 & 0.65 & 0.00 \\  & 1000 & 0.91 & 0.74 & 0.00 & 0.05 & 0.14 & 0.02 & 0.97 & 0.81 & 0.00 \\[\smallskipamount]
    \bottomrule
  \end{tabular}
  \caption{
  Empirical coverage,  average band area, and empirical truncated coverage for the HulC confidence intervals of estimators $\hat{F}_{MLE}$, $\hat{F}_{init}$, and for the credible interval of estimator $\hat{F}_{PT}$ (denoted as PT) , averaged over 100 replications within each of the 10 simulation settings. Truncated coverage refers to coverage calculated only on points within the 5th to 95th percentile range of the true distribution $F$.
  % Empirical coverage, average band area, and Empirical truncated coverage (explanat) for the estimators across 10 simulation settings with each has 100 replications {\color{red} gpt, PT(only here with explanation), 2 digits, in the paper also}.
  }
  \label{auction:coverage:evaluation}
\end{table}

\subsection{Evaluation based on profit-maximizing price}

\noindent
As noted in \cite{10.1214/11-AOAS503}, a key advantage of knowing the valuation distribution $F$ is that it enables managers to determine the profit-maximizing price for a product in a given market. In particular, if $c$ is the production cost, then the expected (per person) profit at a price $x$ is given by 
$$
\pi(x)=(1-F(x))(x-c)
$$

\noindent
In particular, $\pi(x)$ is the total expected profit divided by the total number of potential customers in the underlying population. As \cite{10.1214/11-AOAS503} note, $\pi(x)$ is the product of the proportion of potential customers with valuation above price $x$ (i.e., $1 - F(x)$), and the profit for each sale (i.e., $x-c$). Hence, the profit maximizing price $x_F$, is defined as 
$$
x_F =\mbox{argmax}_x (1-F(x))(x-c). 
$$

\noindent
Based on an estimate $\hat{F}$ of $F$, the underlying profit maximizing price $x_F$ can be estimated by 
$$
x_{\hat{F}}=\mbox{argmax}_x (1-\hat F(x))(x-c). 
$$

\noindent
Based on these considerations, for each of the ten simulation settings—corresponding to five choices of the true valuation distribution $F$ and two choices of the total number of auctions $K$—we now additionally evaluate each method along two dimensions: (a) the proximity of the estimated profit‑maximizing price $x_{\hat F}$ to the true profit‑maximizing price $x_F$, and (b) the proximity of the estimated (per person) maximum profit
\[
\hat{\pi}(x_{\hat F}) = \bigl(1-\hat F(x_{\hat F})\bigr)\bigl(x_{\hat F}-c\bigr)
\]
to the true profit ($\pi(x_{\hat F})$) achieved when pricing at $x_{\hat F}$.

The second criterion in (b) is particularly relevant in practice. When $\hat F$ is used as the estimated valuation distribution, $\hat{\pi}(x_{\hat F})$ is the natural estimate of the maximum attainable (per person) profit. However, the actual (and unobserved) profit realized by the practitioner when setting the price $x_{\hat F}$ is $\pi(x_{\hat F})$. From a managerial perspective, it is therefore important that $\hat{\pi}(x_{\hat F})$ be close to $\pi(x_{\hat F})$ (see, e.g., \cite{walker1978financial}; \cite{burns2001sales}). 

We use five methods in our comparison. These include the proposed {\em MLE} approach, the {\it Initial } approach, the Pólya Tree approach of \cite{10.1214/11-AOAS503} and two parametric approaches based on final selling prices used in \cite{10.1214/11-AOAS503} - the Gamma-F method, and the Truncated Normal-F (TN-F) method.

\begin{figure}[htbp]
    \centering
    \includegraphics[width=\textwidth]{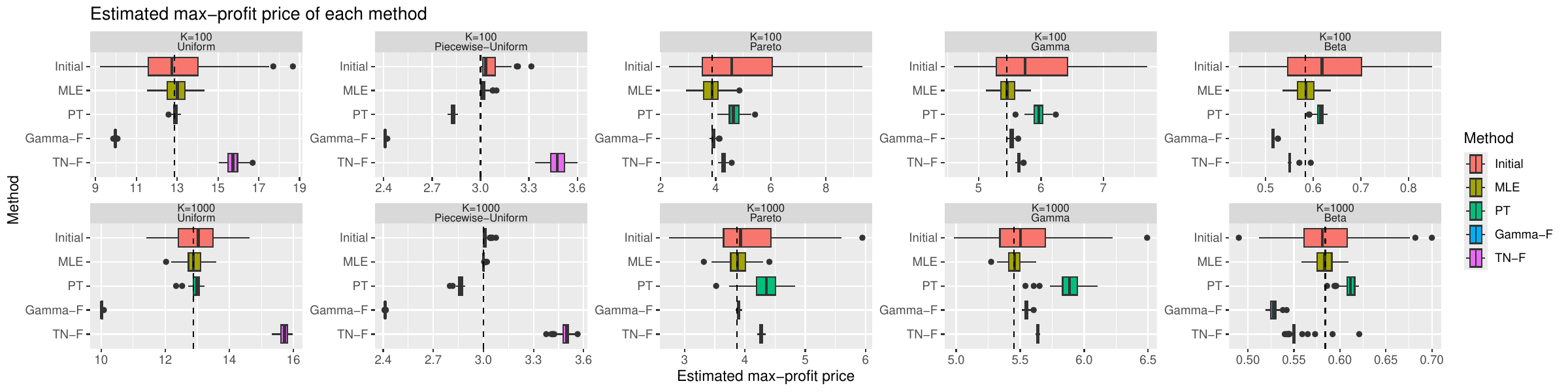} 
    \caption{Box plots for the estimated profit maximizing price $x_{\hat F}$ for each of the five estimators, $\hat{F}_{MLE}$, $\hat{F}_{init}$, Pólya tree estimator ({PT}), the Gamma method based on final selling prices ({Gamma-F}), and the Truncated-Normal method based on final selling prices ({TN-F}). Results are based on 100 replications within each of the 10 simulation settings. The dashed lines represent the true profit maximizing price $x_F$ in each scenario.}
    \label{fig:est_price}
\end{figure}

Figure \ref{fig:est_price} provides box plots for the estimated (per person) profit maximizing price $x_{\hat F}$ of each method in each of the 10 simulation settings. The production cost $c$ was set to the 25\% quantile of the underlying true distribution, and the results described below exhibit similar patterns even when $c$ varies from the 10\% to the 50\% quantile (see Supplemental Section S5 for more details). 

It can be seen that the estimated profit‑maximizing price produced by the MLE method is generally very close to the true optimal price, while exhibiting lower variability than the Initial method. In contrast, the Pólya tree method tends to overestimate the optimal price across nearly all settings (Piecewise Uniform being the sole exception), whereas the Gamma method typically underestimates it (Gamma being the sole exception). The Truncated-Normal method tends to have a much larger bias compared to the other methods.

% As for the actual profit, MLE always reach the best profit, while Initial is always slight worse, Pólya Tree can be significantly worse in Gamma setting, and Gamma method can be significantly worse in most settings. 
Table \ref{tab:est_profit} provides the difference between $\hat{\pi}(x_{\hat{F}})$, the estimated (per person) profit 
at the price $x_{\hat{F}}$, and $\pi(x_{\hat{F}})$, the corresponding actual 
profit, averaged over 100 replications. We can see that the MLE consistently delivers the most accurate estimates across all 
settings. Its mean deviance exceeds $0.05$ only in the Uniform setting with 
$K=100$; even in that case, it remains the best-performing method among all 
estimators considered. The initial estimator generally performs comparably 
but exhibits noticeably larger deviations.

In contrast, the Gamma method systematically underestimates the maximum profit 
in nearly all settings, with the Pareto case being the sole exception. The 
Pólya tree estimator displays more variable behavior: it tends to overestimate 
profits most severely in the Pareto setting, while underestimating profits in the Uniform and the Piecewise Uniform settings. The Truncated-Normal method performs the worst 
overall, showing poor accuracy relative to the other estimators in most 
simulation settings. 

\begin{table}[!h]
  \centering
  \small
  
  \begin{tabular}{@{} l l ccccc @{}}
    \toprule
    & &   \multicolumn{5}{c}{\textbf{Deviance of Estimated Maximum Profit  $\hat\pi(x_{\hat F})-\pi(x_{\hat F})$}} \\
    \cmidrule(lr){3-7}
    Distribution & $K$ 
      & $\ \quad $MLE$\ \ $   & Initial & Pólya Tree &Gamma-F &TN-F\\ 
    \midrule
    Uniform & 100  &  {\color{red} +0.064}  &  {\color{red} +0.235}  &  {\color{blue} -0.119}  &  {\color{blue} -1.094}  &  {\color{red} +7.250}  \\ 
   & 1000  &  {+0.022}  &  {\color{red} +0.115}  &  {\color{blue} -0.096}  &  {\color{blue} -1.286}  &  {\color{red} +7.164}  \\ 
 [\smallskipamount] Piecewise-Uniform & 100  &  {+0.006}  &  {+0.007}  &  {\color{blue} -0.240}  &  {\color{blue} -0.157}  &  {\color{red} +1.399}  \\ 
   & 1000  &  {+0.001}  &  {+0.008}  &  {\color{blue} -0.242}  &  {\color{blue} -0.192}  &  {\color{red} +1.426}  \\ 
 [\smallskipamount] Pareto & 100  &  {+0.027}  &  {\color{red} +0.191}  &  {\color{red} +0.915}  &  {+0.033}  &  {\color{red} +0.370}  \\ 
   & 1000  &  {+0.009}  &  {\color{red} +0.055}  &  {\color{red} +0.739}  &  {-0.006}  &  {\color{red} +0.347}  \\ 
 [\smallskipamount] Gamma & 100  &  {+0.013}  &  {\color{red} +0.118}  &  {+0.022}  &  {\color{blue} -0.256}  &  {\color{blue} -0.117}  \\ 
   & 1000  &  {+0.003}  &  {+0.026}  &  {-0.014}  &  {\color{blue} -0.277}  &  {\color{blue} -0.117}  \\ 
 [\smallskipamount] Beta & 100  &  {+0.002}  &  {+0.017}  &  {-0.003}  &  {\color{blue} -0.050}  &  {\color{blue} -0.071}  \\ 
   & 1000  &  {+0.000}  &  {+0.004}  &  {-0.005}  &  {\color{blue} -0.063}  &  {\color{blue} -0.071}  \\ 
% Uniform & 100  &  {\color{red} +0.11}  &  {\color{red} +0.36}  &  {\color{blue} - 0.20}  &  {\color{blue} - 0.89}  &  {\color{red} +9.90}  \\ 
%  [\smallskipamount]  & 1000  &  {+0.03}  &  {\color{red} +0.17}  &  {\color{blue} - 0.17}  &  {\color{blue} - 1.25}  &  {\color{red} +9.78}  \\ 
%  [\smallskipamount] Piecewise-Uniform & 100  &  {+0.00}  &  {+0.01}  &  {\color{blue} - 0.22}  &  {\color{blue} - 0.17}  &  {\color{red} +1.03}  \\ 
%  [\smallskipamount] & 1000  &  {+0.00}  &  {+0.00}  &  {\color{blue} - 0.20}  &  {\color{blue} - 0.19}  &  {\color{red} +1.08}  \\ 
%  [\smallskipamount] Pareto & 100  &  {+0.04}  &  {\color{red} +0.20}  &  {\color{red} +1.07}  &  {+0.02}  &  {\color{red} +0.43}  \\ 
%  [\smallskipamount]  & 1000  &  {+0.01}  &  {\color{red} +0.06}  &  {\color{red} +0.88}  &  {- 0.01}  &  {\color{red} +0.40}  \\ 
%  [\smallskipamount] Gamma & 100  &  {+0.04}  &  {\color{red} +0.08}  &  {\color{blue} - 0.23}  &  {\color{blue} - 0.62}  &  {\color{blue} - 0.41}  \\ 
%  [\smallskipamount]  & 1000  &  {+0.00}  &  {+0.03}  &  {\color{blue} - 0.28}  &  {\color{blue} - 0.67}  &  {\color{blue} - 0.41}  \\ 
%  [\smallskipamount] Beta & 100  &  {+0.00}  &  {+0.01}  &  {- 0.02}  &  {\color{blue} - 0.06}  &  {\color{blue} - 0.11}  \\ 
%  [\smallskipamount]      & 1000  &  {+0.00}  &  {+0.00}  &  {- 0.02}  &  {\color{blue} - 0.09}  &  {\color{blue} - 0.11}  \\
    \bottomrule
  \end{tabular}
  \caption{Deviance of the estimated maximum profit—defined as the difference between 
the estimated (per person) profit $\hat{\pi}(x_{\hat{F}})$ and the actual (per person) profit 
$\pi(x_{\hat{F}})$ evaluated at the estimated optimal price $x_{\hat{F}}$—for 
each of the five estimators: $\hat{F}_{\mathrm{MLE}}$, $\hat{F}_{\mathrm{init}}$, 
the Pólya tree estimator, the Gamma method based on final selling prices (\emph{Gamma-F}), and the Truncated-Normal method based on final selling prices (\emph{TN-F}). 
Results are averaged over 100 replications for each of the 10 simulation 
settings. Positive values indicate overestimation of the maximum profit, 
while negative values indicate underestimation. Estimates are highlighted 
in red when the mean deviance exceeds $0.05$, and in blue when the mean 
deviance is below $-0.05$.
  }
  \label{tab:est_profit}
\end{table}

\section{Empirical application}\label{sec:Empirical_application}

\noindent
In this section we apply our method to estimate the true valuation 
distribution of an Xbox based on actual data obtained from second-price 
auctions on eBay. In Section \ref{subsec:xbox_auction_data_overview}, we 
provide an overview of the data, and discuss features and adjustments to 
ensure its suitability for the methodology developed in the paper. In 
Section \ref{subsec:xbox_auction_data_analysis}, we apply our semi-parametric 
methodology on the data set and present the findings, and perform additional 
performance analysis. 

\subsection{Data overview}\label{subsec:xbox_auction_data_overview}

\noindent
The data set on eBay on online auctions of Xbox game consoles was obtained 
from the \href{https://www.modelingonlineauctions.com/datasets}{Modeling Online Auctions} data repository. More specifically, we focus on a data set which provides information for $93$ online auctions of identical Xbox game consoles where each auction lasts for $7$ days. For each auction, a user's bid is recorded {\it only if it changes the standing price in the auction}. For each such bid, the following information is provided: \textit{auctionid} (unique auction identifier), \textit{bid} (dollar value of the bid), \textit{bidtime} (the time, in days, that the bid was placed), \textit{bidder} (bidder eBay username), \textit{bidderrate} (internal eBay rating of the bidder),  \textit{openbid} (the starting price for the auction, set by the seller), and \textit{price} (the final selling price for the auction). While the standing 
price values throughout the course of the auction were not directly provided, 
they can be easily inferred from the successful bid values from the 
\textit{bid} column and the starting price. Also, the \textit{bidtime} column 
directly provides the sequence of times at which there is a change in the 
standing price. 

As mentioned in the introduction, we found that a minor fraction of bidders
(less than $10\%$ of the total) placed multiple bids. Many of these bids are 
consecutive bids by the same bidder to ensure that they become the leader in 
the auction. Note that we observe only 'successful' bids, i.e., bids which 
change the standing price of the auction. If a successful bidder (post 
bidding) observes that the standing price of the auction has changed to their 
bid (plus a small increment), it can be inferred that this bid is currently 
the second highest. Hence, through a proxy bidding system offered by eBay, the
bidder could choose to incrementally push up their bid until they become the 
leader in the option (the standing price becomes less than their latest bid). 
The proxy system also needs to be provided with a ceiling value, above which 
no bids are to be submitted. This value is very likely the bidder's true 
valuation of the product. With this in mind, and to adapt the data as much as 
possible to our single bidding assumption, we remove all the previous bids of 
such multiple bidders from the data, and keep only the final bid. Finally, 
there are a couple of auctions where the first successful bid values are same 
as the starting prices (\textit{openbid}) of the corresponding auctions. To 
ensure compliance with our requirement of no ties, and for uniformity, we added small random noise from Uniform$\left(0,0.01\right)$ to all the bids across all the auctions. Since the 
total number of bidders accessing the auctions is not available, the final selling price based 
methodology in \cite{10.1214/11-AOAS503} is not applicable. As in the simulations, we will use the 
initial estimator $\hat{F}_{init}$, which is computed using only the final selling prices 
and first observed bids in all auctions, as a representative of this methodology in the 
current setting.

\subsection{Analysis of Xbox data}\label{subsec:xbox_auction_data_analysis}

Using the Xbox 7-day auctions dataset with slight modifications as mentioned 
in Section~\ref{subsec:xbox_auction_data_overview}, we now compute the initial
estimate $\hat{F}_{init}$ and the  MLE $\hat{F}_{MLE}$. For the 
estimation of $\lambda$ (see Section \ref{subsec:consistent_estimator_lambda})
we need to choose a subset of auctions whose starting prices are negligible in 
the given context. We found that the smallest final selling price in all the 
auctions is \$25 and the median final selling price in all the auctions is 
\$125. Given this, we chose all auctions with starting price less than 
\$1 ($16$ out of $93$) for obtaining the generalized method of moments based 
estimator of $\lambda$, and also for the computing the final selling price 
based estimator $\hat{F}_{SP}$ (see Step I in 
Section \ref{subsec:Initialization_theta_F_2ndPrice_Auction_Single_Lambda_case}). Recall that 
$\hat{F}_{SP}$ is one of the components used to compute the initial estimator 
$\hat{F}_{init}$. 

\begin{figure}
    \centering
    \includegraphics[width = \linewidth]{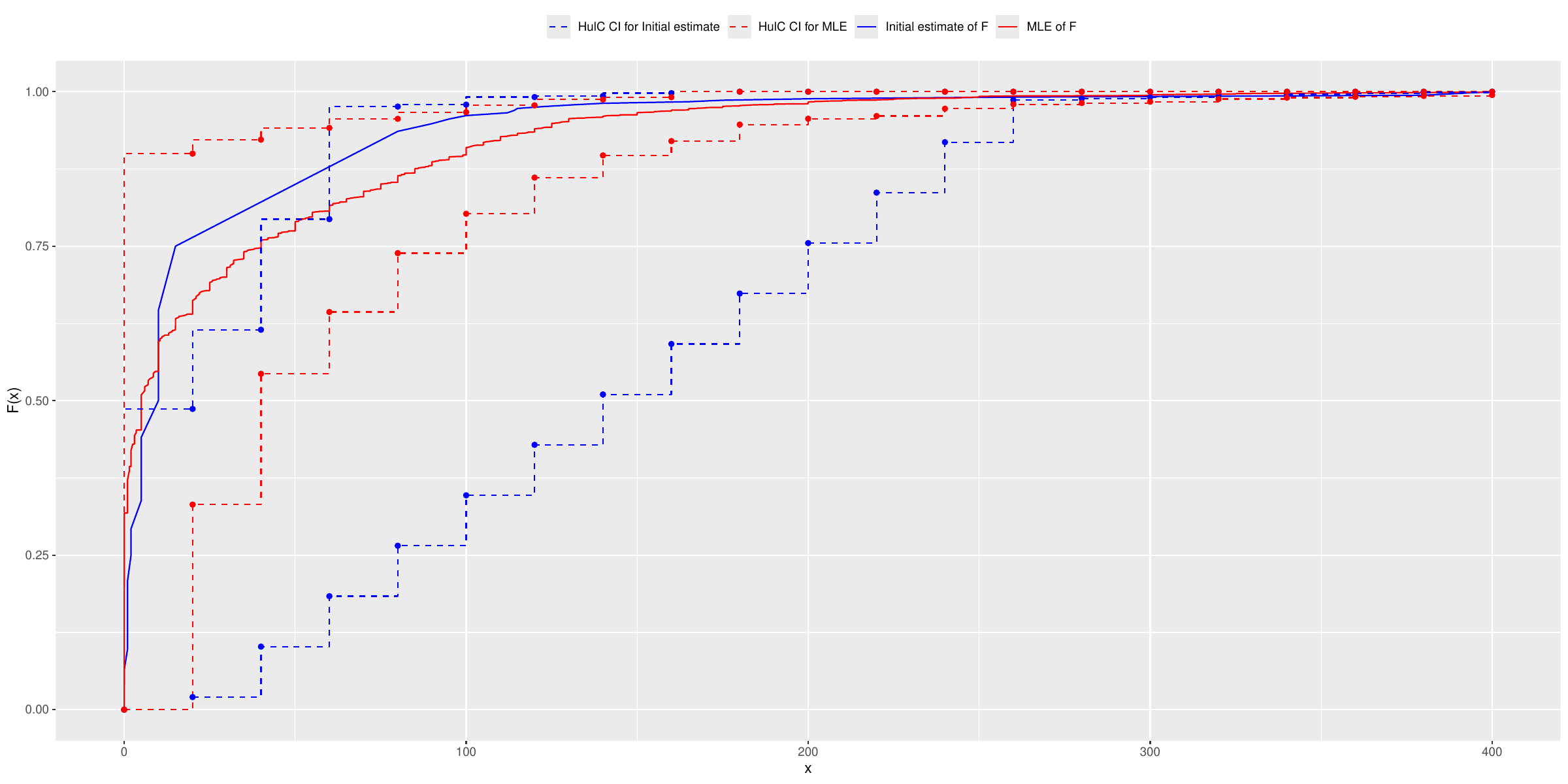}
    \caption{\textit{Plot of Initial estimate (solid blue line, based only on final selling prices and first observed bids) vs.  MLE (solid red line, based on entire sequence of standing prices) of F and their corresponding $90\%$ HulC confidence regions (dotted blue and red lines) for the Xbox 7-day auctions dataset.}}
    \label{fig:Xbox_whole_data_Init_vs_MLE}
\end{figure}

The plots of the initial estimate $\hat{F}_{init}$ and the  MLE 
$\hat{F}_{MLE}$ of the (unknown) true valuation distribution along with 
the corresponding $90\%$ HulC confidence regions are provided in 
Figure~\ref{fig:Xbox_whole_data_Init_vs_MLE}. Similar to the phenomenon 
observed in the simulations in 
Section~\ref{subsec:simulation.with.various.true.F}, we notice that the HulC 
confidence region of $\hat{F}_{MLE}$ is narrower than that of 
$\hat{F}_{init}$, indicating comparatively smaller variance of $\hat{F}_{MLE}$. 
% {\color{red} We 
% see that the two estimates are reasonably different: the total variation 
% distance between them is $0.4140$ and the KS distance 
% between them is $0.4686$}.
Another interesting 
observation is that the curves for these two estimates cross exactly once, 
with $\hat{F}_{ MLE} (x)$ dominated by $\hat{F}_{init} (x)$ after the 
crossing point, and vice-versa before the crossing point. This implies that $\hat{F}_{init}$ stochastically dominates $\hat{F}_{MLE}$. In other words,
the final selling price/first non-starting standing price based initial estimator 
signifies higher Xbox valuations than the MLE estimator based on the entire 
collection of standing prices throughout the course of the auctions. 

%Moreover, we calculate the Wasserstein distance between $F^{(0)}$ and $\hat{F}_{MLE}$ to get an idea of the distance between them. The value of the Wasserstein distance in this case is approximately $12.68$.

Unlike the simulation setting, the true valuation distribution is obviously 
not known here. However, we still undertake a limited performance evaluation 
and comparison exercise for the two approaches. As discussed previously in 
Section \ref{subsec:Data_simulation}, if the modeling assumptions are largely 
unviolated (which seems to be the case) one would expect the MLE to do better 
than the initial estimator. The goal of this limited evaluation is again to 
understand the amount of improvement, and also to examine the stability of 
both estimators. For this purpose, we split the entire Xbox dataset into 
training and test sets. In particular, we consider two choices of splits 
namely, $1:1$ and $2:1$, for the ratio of auctions in training vs. test data. For each splitting proportion, 100 random splits are performed. For each split, estimates $\hat{F}_{MLE}$ and $\hat{F}_{init}$ are obtained from the training data, along with the estimate $\hat{\lambda}_{MLE}$ of the Poisson arrival rate. For each test auction and each of two estimates of $F$ ({\it init} and {\it MLE}), 100 pseudo auctions with the same starting price are generated. The mean squared error of the corresponding 100 final selling prices is computed around the observed final selling price in the test auction. These relative mean squared errors are then averaged over all the test auctions. The mean squared errors are finally averaged over these 100 replications/splits to obtain an overall MSE for each of the two methods. The MSEs for both $\hat{F}_{MLE}$ and $\hat{F}_{init}$ are normalized by the MSE of the preliminary estimator $\hat{F}_{FP}$ (based solely on the first non-starting standing prices). 
% The whole process above is repeated with a 2:1 ratio partition between the training and test set. 
The resulting normalized MSE values are provided in Table~\ref{tab:MSE}. { In both settings, {\it MLE}  yields a better estimate of the final selling price compared to the {\it init}.}

% \begin{figure}
%     \centering
%     \includegraphics[width = \linewidth]{plots/Xbox_train_test.pdf}
%     \caption{\textit{Plot of $\hat{F}_{init, \ training}$ (solid blue line) vs. $\hat{F}_{ MLE, \ training}$ (solid red line) vs. $\hat{F}_{MLE, \ test}$ (dotted red line) for two different (1:1 and 2:1) splits into training and test sets of the Xbox dataset.}}
%     \label{fig:Xbox_train_test}
% \end{figure}

\begin{table}
    \centering
    \begin{tabular}{@{} c c c @{}}
    \toprule
    train : test &$1:1$  &$2:1$ \\
    \midrule
    $MSE(\hat F_{init})/MSE(\hat F_{FP})$ & $0.1535$ & $0.1565$\\
    $MSE(\hat F_{MLE})/MSE(\hat F_{FP})$ & $0.1265$ & $0.1218$\\
    %$MSE(\hat F_{PT})/MSE(\hat F_{FP})$ &$0.1369$&$0.1307$\\
    \bottomrule
    \end{tabular}
    \caption{\textit{Mean squared error (MSE) between data generated based on starting price in the test set and estimated $\hat \lambda_{MLE}$, $\hat \lambda_{init}$, $\hat{F}_{ MLE, \ test}$ and each of $\hat{F}_{init, \ training}$,  $\hat{F}_{MLE, \ training}$ and  $\hat{F}_{PT, \ training}$ (with number of bidders generated from MLE model) respectively, compared with data generated from $\hat{F}_{FP, \ training}$,  averaged over $100$ replications of the random split with same proportion, and in each replicate the data is generated 100 times based on the same set of starting price.}}
    \label{tab:MSE}
\end{table}

\section{Discussion and future research}\label{sec:discussion}

\noindent
In this paper we have developed a semi-parametric methodology for estimating 
the consumer valuation distribution using second price auction data. Unlike 
the approach in \cite{10.1214/11-AOAS503}, our methodology uses the collection
of current selling price values throughout the course of the auctions, and 
does not require knowledge of the total number of bidders accessing the 
auction. Extensive simulations demonstrate that, when the 
modeling assumptions are true, using our approach can lead to significantly 
better performance than estimators based on just final
selling prices and first observed bids. Two additional assumptions (compared 
to \cite{10.1214/11-AOAS503}) which preclude multiple bidding 
and postulate constant rate of arrival of the consumers to the auction are 
needed. Many real-life second price auctions see only minor departures from 
these assumptions, which are supported by economic theory. However, if there 
is evidence of major violation, results from the proposed methodology should 
be used cautiously. Generalizing our methodology by relaxing one or both of 
these assumptions is a topic of future research. One possible direction which 
we plan on exploring is allow two different rates for the bidder arrival 
process, with a transition between these two rates happening sometime 
during the auction period. 

Another pertinent direction for future research is the incorporation of covariates in the model; see, for example, \cite{Bajari:Hortacsu:2003}. The population valuation distribution for the same product might depend on factors such as condition of the item and seller statistics. One possible path for incorporating this in our model is through {a semi-parametric approach where a linear combination of the relevant covariates is added as a location parameter for the cdf $F$}. In particular, the population valuation cdf, evaluated at $t$, with $K$ relevant covariate values $x_1, x_2, \ldots, x_K$ is given by $F \left( t - \sum_{i=1}^K \beta_i x_i \right)$, see e.g., \cite{groeneboom2018current}. The corresponding likelihood maximization will now involve the unknown regression parameters $\{\beta_i\}_{i=1}^K$ in addition to $\lambda$ and $F$. While the conditional maximization of $\lambda$ and $F$ will involve similar calculations as those developed in this paper, conditional maximization of the regression coefficients will need more careful thought and analysis.

%%%%%%%%%%%%%%%%%%%%%%%%%%%%%%%%%%%%%%%%%%%%%%
%% Single Appendix:                         %%
%%%%%%%%%%%%%%%%%%%%%%%%%%%%%%%%%%%%%%%%%%%%%%
%\begin{appendix}
%\section*{???}%% if no title is needed, leave empty \section*{}.
%\end{appendix}
%%%%%%%%%%%%%%%%%%%%%%%%%%%%%%%%%%%%%%%%%%%%%%
%% Multiple Appendixes:                     %%
%%%%%%%%%%%%%%%%%%%%%%%%%%%%%%%%%%%%%%%%%%%%%%
%\begin{appendix}
%\section{???}
%
%\section{???}
%
%\end{appendix}

%%%%%%%%%%%%%%%%%%%%%%%%%%%%%%%%%%%%%%%%%%%%%%
%% Support information, if any,             %%
%% should be provided in the                %%
%% Acknowledgements section.                %%
%%%%%%%%%%%%%%%%%%%%%%%%%%%%%%%%%%%%%%%%%%%%%%
%\begin{acks}[Acknowledgments]
% The authors would like to thank ...
%\end{acks}
%%%%%%%%%%%%%%%%%%%%%%%%%%%%%%%%%%%%%%%%%%%%%%
%% Funding information, if any,             %%
%% should be provided in the                %%
%% funding section.                         %%
%%%%%%%%%%%%%%%%%%%%%%%%%%%%%%%%%%%%%%%%%%%%%%
\begin{funding}
Rohit Patra's work was partially supported by NSF grant DMS-2210662. Kshitij Khare's work was partially supported by NSF grant DMS-2410677. 
\end{funding}

%%%%%%%%%%%%%%%%%%%%%%%%%%%%%%%%%%%%%%%%%%%%%%
%% Supplementary Material, including data   %%
%% sets and code, should be provided in     %%
%% {supplement} environment with title      %%
%% and short description. It cannot be      %%
%% available exclusively as external link.  %%
%% All Supplementary Material must be       %%
%% available to the reader on Project       %%
%% Euclid with the published article.       %%
%%%%%%%%%%%%%%%%%%%%%%%%%%%%%%%%%%%%%%%%%%%%%%
\begin{supplement}
\stitle{Supplement for ``A semi-parametric approach for estimating consumer valuation distributions using second price auctions''.}
\sdescription{In this supplementary article, we provide proofs of some of the lemmas in the paper, and some additional simulation results.}
\end{supplement}

%%  imsart-nameyear.bst  will be used to                   %%
%%  MR numbers will be added by VTeX.                      %%
%%                                                         %%
%%  Use \cite{...} to cite references in text.             %%
%%                                                         %%
%%%%%%%%%%%%%%%%%%%%%%%%%%%%%%%%%%%%%%%%%%%%%%%%%%%%%%%%%%%%%

%% if your bibliography is in bibtex format, uncomment commands:
\bibliographystyle{imsart-nameyear} % Style BST file
\bibliography{References}       % Bibliography file (usually '*.bib')

%% or include bibliography directly:
% \begin{thebibliography}{}
% \bibitem[\protect\citeauthoryear{???}{???}]{b1}
% \end{thebibliography}

\end{document}

%\externaldocument{Auctions-aoas}

\begin{frontmatter}
\title{Supplement for ``A semi-parametric approach for estimating consumer valuation distributions using second price auctions''}
%\title{A sample article title with some additional note\thanksref{T1}}
\runtitle{Estimating Valuation distributions using Second Price Auctions}
%\thankstext{T1}{A sample of additional note to the title.}

\begin{aug}
%%%%%%%%%%%%%%%%%%%%%%%%%%%%%%%%%%%%%%%%%%%%%%%
%% Only one address is permitted per author. %%
%% Only division, organization and e-mail is %%
%% included in the address.                  %%
%% Additional information can be included in %%
\author[A]{\fnms{Sourav} \snm{Mukherjee}\ead[label=e2]{sm37sds@gmail.com}},
\author[A]{\fnms{Ziqian} \snm{Yang}\ead[label=e4]{zi.yang@ufl.edu}},
\author[B]{\fnms{Rohit K} \snm{Patra}\ead[label=e1]{rkumarpatra@gmail.com}},
\and
\author[A]{\fnms{Kshitij} \snm{Khare}\ead[label=e3]{kdkhare@stat.ufl.edu}}
\address[A]{Department of Statistics, University of Florida}\printead{e2,e4,e3}

\address[B]{LinkedIn Inc}\printead{e1}
%%%%%%%%%%%%%%%%%%%%%%%%%%%%%%%%%%%%%%%%%%%%%%
%% Addresses                                %%
%%%%%%%%%%%%%%%%%%%%%%%%%%%%%%%%%%%%%%%%%%%%%%

%\address[B]{???, \printead{e2,e3}}
\runauthor{Mukherjee, Yang, Patra, and Khare}
\end{aug}

%\begin{abstract}
%In this supplemental article, we provide the proof of Lemma 2.1 and Lemma 2.2 from the main paper. 
%\end{abstract}

%\begin{keyword}
%\kwd{Second Price Auction, Non-parametric estimation}
%\end{keyword}

\end{frontmatter}
%%%%%%%%%%%%%%%%%%%%%%%%%%%%%%%%%%%%%%%%%%%%%%
%% Please use \tableofcontents for articles %%
%% with 50 pages and more                   %%
%%%%%%%%%%%%%%%%%%%%%%%%%%%%%%%%%%%%%%%%%%%%%%
%\tableofcontents

%%%%%%%%%%%%%%%%%%%%%%%%%%%%%%%%%%%%%%%%%%%%%%
%%%% Main text entry area:

\section{Proof of Lemma 2.1}\label{sec:proof_of_lemma2.1}

\begin{proof}
We first introduce some additional notation. Let $\{\Delta_i\}_{i=1}^{M-1}$ represent the number of bidders accessing the auction between the $i^{th}$ and $(i+1)^{th}$ changes in the selling price, and let $\Delta_0$ represent the number of bidders accessing the auction until the first time when the selling price changes to a higher value from the starting price $r$. Also, let $\{S_i\}_{i=2}^{N}$ represent the time between the arrival of $(i-1)^{th}$ and $i^{th}$ bidders accessing the auction, and $S_1$ let represents the waiting time of the arrival of the first bidder from the start of the auction. Recall that a bidder accessing the auction is allowed to place a bid only if the bid value is greater than the current selling price. We now consider three possible scenarios at the end of the auction.\\ 
\noindent
\\\textbf{Case I: When the item is sold above the starting price $\mathbf{(M>0, O=1)}$.}\quad In this case, the number of times the selling price changes throughout the course of the auction i.e., $M$, is positive. Now, let us first derive the conditional density of the standing prices $\{X_i\}_{i=1}^{M}$ given $\{\Delta_i\}_{i=0}^{M-1}$, $M$, 
$O = 1$, and $N=n$. 
\begin{itemize}
\item Since, $\Delta_0$ is the number of bidders until the first time that the standing price changes to a higher value than the starting price $r$, it means that there are $(\Delta_0 - 2)$ many bids that are less than $r$, and only two bids are higher than $r$ with $X_1$ being the second highest bid. Also, the first bid which is higher than $r$ can occur at $(\Delta_0 -1)$ many places. 
%Thus the density of $X_1$ given $\{\Delta_i\}_{i=0}^{M-1}$, $M$, 
%$O = 1$, and $N=n$ is given by $(\Delta_0 -1)F(r)^{\Delta_0 -2}f(X_1)$. 
\item For $X_2$ to be the next standing price After $X_1$ being the current second highest bid, the next $(\Delta_1 - 1)$ bids must be less than $X_1$ and the $(\Delta_0 + \Delta_1)^{\text{th}}$ bid should be higher than $X_1$. 
%Otherwise the second highest bid would increase to $X_2$ much sooner. Hence, the probability of occurring the next $\Delta_1$ bids would be $F(X_1)^{\Delta_1 - 1}f(X_2)$. 
\item Continuing on like this, we should have the last $(n-\sum_{i=0}^{M-1}\Delta_i)$ many bids less than $X_M$ after $X_M$ becomes the standing price (and the second highest bid of the entire auction) with the (unobserved) highest bid $R$ occurring somewhere before. 
%So, the probability would be $F(X_M)^{n-\sum_{i=0}^{M-1}\Delta_i} (1-F(X_M))$.
\end{itemize}

It follows that the conditional density of 
$\{X_i\}_{i=1}^{M}$ given $\{\Delta_i\}_{i=0}^{M-1}$, $M$, 
$O = 1$, and $N=n$ is given by 
\begin{align}
={}& (\Delta_0 - 1) F(r)^{\Delta_0 - 2} F(X_1)^{\Delta_1 - 1}F(X_2)^{\Delta_2 - 1}\times \ldots \times F(X_M)^{n - \sum_{i=0}^{M-1}\Delta_i} \nonumber\\
{}& \times \big(1 - F(X_M)\big) \prod_{i=1}^{M}f(X_i) \nonumber\\
={}& \dfrac{(\Delta_0 - 1)}{F(r)} \ \big(1 - F(X_M)\big) F(X_M)^{n - \sum_{i=0}^{M-1}\Delta_i} \prod_{i=1}^{M}f(X_i) \prod_{i=0}^{M-1}F(X_i)^{\Delta_i-1},
\label{eq:conditional_jt_density_X_delta_M_single_auction}
\end{align}
where $X_0 = r$. Note that the above holds only if $M\leq (n-1)$, $\Delta_0 \geq 2$, $\Delta_1, \Delta_2, \ldots, \Delta_{M-1} \geq 1$, and $\sum_{i=0}^{M-1}\Delta_i \leq n$. 

For a collection of i.i.d. random variables $Y_1, Y_2, 
\cdots, Y_n$, the distribution of number of changes in the 
running second maximum, and location of these changes in the 
index set $\{1,2, \cdots, n\}$ is invariant under any strictly monotone transformation on the $Y_i$s. If $F$ is absolutely continuous, then $F^{-1}$ exists and is strictly increasing. Note $\{F^{-1} (Y_i)\}_{i=1}^n$ is a collection of i.i.d. Uniform$[0,1]$ random variables. Applying the above conclusions to our context with $Y_i$ being the valuation of 
the $i^{th}$ bidder accessing the auction, it follows that the distribution of $\{\Delta_i\}_{i=0}^{M-1}$, $M$, 
$O$ given $N=n$ does not depend on $F$. Using  (\ref{eq:conditional_jt_density_X_delta_M_single_auction}), 
it follows that the joint density of $M$, $\{X_i\}_{i=1}^{M}$,  $\{\Delta_i\}_{i=0}^{M-1}$, $O$ at values $m$ (with $m > 0$), $o=1$, $\{x_i\}_{i=1}^{m}$, $\{\delta_i\}_{i=0}^{m-1}$ given $N = n$ is equal to
\begin{eqnarray*}
C_1 \dfrac{(\delta_0 - 1)}{F(r)} \ \big(1 - F(x_m)\big) F(x_m)^{n - \sum_{i=0}^{m-1}\delta_i} \prod_{i=1}^{m}f(x_i) \prod_{i=0}^{m-1}F(x_i)^{\delta_i-1}, 
\end{eqnarray*}

\noindent
assuming that the arguments satisfy the constraints $m\leq 
(n-1)$, $\delta_0 \geq 2$, $\delta_1, \delta_2, \ldots, 
\delta_{m-1} \geq 1$, and $\sum_{i=0}^{m-1} \delta_i \leq 
n$ (otherwise the value of the joint density is $0$). Here 
the term $C_1$ is independent of $F$.

Since bidders are assumed to arrive at the auction via a Poisson process with rate $\lambda$, it follows that the number of potential bidders $N$ in any auction follows a Poisson$(\lambda\tau)$ distribution. Also, conditional on $N=n$, note that $\{S_i\}_{i=1}^{n}$ are i.i.d. exponential random variables with rate $\lambda$. Hence, the joint density of the partial sum $\bigg(S_1,S_1+S_2,\ldots, \sum_{i=1}^{n}S_i \bigg)$ given $N=n$ is
\begin{equation}\label{eq:conditional_jt_density_PartialSum_S}
    \dfrac{n!}{\tau^n}, \ \text{where} \ S_i \geq 0 \ \forall \ i \ \text{and} \ \sum_{i=1}^{n}S_i \leq \tau.
\end{equation}

\noindent
It follows that 
\begin{equation}\label{eq:conditional_jt_density_PartialSum_S=ordered_uniform}
    \bigg(S_1,S_1+S_2,\ldots, \sum_{i=1}^{n}S_i \bigg) \ \underset{=}{d} \ \big(U_{(1)},U_{(2)}, \ldots, U_{(n)} \big)
\end{equation}
given $N=n$, where $\{U_i\}_{i=1}^{n}$ are i.i.d.  Uniform$[0,\tau]$, and $(U_{(1)},U_{(2)}, \ldots, U_{(n)})$ are the corresponding order statistics. 

Since $T_i$ denotes the intermediate time between the 
$i^{th}$ and $(i + 1)^{th}$ changes in the standing price 
for $0 \leq i \leq M-1$, it can be easily seen that
\begin{equation*}
   T_0 = \sum_{i=1}^{\Delta_0}S_i , \ T_1 = \sum_{i=\Delta_0+1}^{\Delta_0+\Delta_1}S_i , \ T_2 = \sum_{i=\Delta_0+\Delta_1+1}^{\Delta_0+\Delta_1+\Delta_2}S_i ,  \ldots  , \ T_{M-1} = \sum_{i=\Delta_0+\Delta_1+\ldots+\Delta_{M-2}+1}^{\Delta_0+\Delta_1+\ldots+\Delta_{M-1}}S_i,
\end{equation*}
and $T_M = \tau - \sum_{i=0}^{M-1}T_i$. Since $\{S_i\}_{i=1}^n$ and $(M, \{X_i\}_{i=1}^{M}, O,  \{\Delta_i\}_{i=0}^{M-1})$ are independent given $N=n$. it follows that 
\begin{equation}\label{eq:conditional_jt_density_PartialSum_T=ordered_uniform}
    \bigg( T_0, T_0+T_1, \ldots, \sum_{i=0}^{M-1}T_i \bigg)^T \ \underset{=}{d} \ \big( U_{(J_0)},U_{(J_1)}, \ldots, U_{(J_{M-1})} \big)^T,
\end{equation}

\noindent
given $N=n$, $\{X_i\}_{i=1}^{M}$, $\{\Delta_i\}_{i=0}^{M-1}$,
$M$ and $O$. Here $J_k = \sum_{i=0}^{k}\Delta_i$ for 
$k=0,1,\ldots,M-1$. 

From \eqref{eq:conditional_jt_density_PartialSum_S} and \eqref{eq:conditional_jt_density_PartialSum_S=ordered_uniform}, joint density of $(U_{(J_0)},U_{(J_1)}, \ldots, U_{(J_{M-1})})^T$ given $N=n$, $\{X_i\}_{i=1}^{M}$, $\{\Delta_i\}_{i=0}^{M-1}$,
$M$ and $O$ is equal to 
\begin{equation}\label{eq:conditional_jt_density_T_single_auction_0}
    f_{(U_{(J_0)}, \ldots, U_{(J_{M-1})})}(u_0,\ldots,u_{M-1}) = \dfrac{(\tau - u_{M-1})^{n - \sum_{i=0}^{M-1}\Delta_i}}{B(\Delta)\tau^n} \prod_{i=0}^{M-1}(u_i - u_{i-1})^{\Delta_i-1},
\end{equation}
where $u_{-1} = 0$, and
$$B(\Delta) = \dfrac{\big(n-\sum_{i=0}^{M-1}\Delta_i \big)! \prod_{i=0}^{M-1}(\Delta_i - 1)!}{n!}.$$
From \eqref{eq:conditional_jt_density_PartialSum_T=ordered_uniform} and \eqref{eq:conditional_jt_density_T_single_auction_0}, it follows that the conditional density of $(T_0, T_0+T_1, \ldots, \sum_{i=0}^{M-1}T_i)$ given $N=n$, $\{X_i\}_{i=1}^{M}$, $\{\Delta_i\}_{i=0}^{M-1}$, $M$, and $O$ is equal to 
\begin{equation}\label{eq:conditional_jt_density_T_single_auction}
\dfrac{(T_M)^{n - \sum_{i=0}^{M-1}\Delta_i}}{B({\Delta})\tau^n} \prod_{i=0}^{M-1}T_i^{\Delta_i-1},
\end{equation}
where $B({\Delta})$ is as defined above. 

Since the Jacobian of the transformation from $(T_0, T_0+T_1, \ldots, \sum_{i=0}^{M-1}T_i)^T$ to \\$(T_0, T_1, \ldots, T_{M-1})^T$ is $1$, combining  \eqref{eq:conditional_jt_density_X_delta_M_single_auction} and \eqref{eq:conditional_jt_density_T_single_auction} it follows that the joint density of $M$, $\{T_i\}_{i=0}^{M-1}$, $\{X_i\}_{i=1}^{M}$, $\{\Delta_i\}_{i=0}^{M-1}$, $O$ at values $m$ (with $m > 0$), $\{t_i\}_{i=0}^{m-1}$, $\{x_i\}_{i=1}^{m}$, $\{\delta_i\}_{i=0}^{m-1}$, $o=1$ given $N = n$ is equal to 
\begin{align}
{}& \frac{C_1 (\delta_0 - 1)}{B(\delta)\tau^n F(r)} \big(1 - F(x_m)\big) \big(F(x_m)t_m\big)^{n - \sum_{i=0}^{m-1}\delta_i} \prod_{i=0}^{m-1}t_i^{\delta_i-1} \prod_{i=1}^{m}f(x_i) \prod_{i=0}^{m-1}F(x_i)^{\delta_i-1} \nonumber\\
={}& \frac{C_1 (\delta_0 - 1)\big(1 - F(x_m)\big)}{B({\delta})\tau^n F(r)} \big(F(x_m)t_m\big)^{n - \sum_{i=0}^{m-1}\delta_i} \prod_{i=0}^{m-1}\big(F(x_i)t_i\big)^{\delta_i-1} \prod_{i=1}^{m}f(x_i) \nonumber\\
={}& \frac{C_1 n! \big(1 - F(x_m)\big)\big(F(x_m)t_m\big)^{n - \sum_{i=0}^{m-1}\delta_i}(\delta_0 - 1)\big(F(r)t_0\big)^{\delta_0 - 1}}{\tau^nF(r)\big(n - \sum_{i=0}^{m-1}\delta_i \big)!(\delta_0 - 1)!} \prod_{i=1}^{m-1} \frac{\big(F(x_i)t_i\big)^{\delta_i-1}}{(\delta_i - 1)!}\prod_{i=1}^{m}f(x_i) \nonumber\\
={}& \frac{C_1 n!t_0 \big(1 - F(x_m)\big)}{\tau^n} \frac{\big(F(x_m)t_m\big)^{n - \sum_{i=0}^{m-1}\delta_i}}{\big(n - \sum_{i=0}^{m-1}\delta_i \big)!} \frac{\big(F(r)t_0 \big)^{\delta_0 - 2}}{(\delta_0-2)!} \prod_{i=1}^{m-1}\frac{\big(F(x_i)t_i \big)^{\delta_i-1}}{(\delta_i - 1)!} \prod_{i=1}^{m}f(x_i),
\label{eq:conditional_jt_density_T_X_delta_M_single_auction}
\end{align}
where $x_0 = r$, and the arguments satisfy the constraints assuming that the arguments satisfy the constraints $m\leq 
(n-1)$, $\delta_0 \geq 2$, $\delta_1, \delta_2, \ldots, 
\delta_{m-1} \geq 1$, $\sum_{i=0}^{m-1} \delta_i \leq 
n$, and $\sum_{i=0}^{m-1}t_i \leq \tau$ (otherwise the value of the joint density is $0$). 

Now, summing over $\delta_i$'s in \eqref{eq:conditional_jt_density_T_X_delta_M_single_auction} such that, $\delta_0 \geq 2$, $\delta_1, \delta_2, \ldots, \delta_{m-1} \geq 1$, $\sum_{i=0}^{m-1}\delta_i \leq n$; the joint density of $M$, $\{X_i\}_{i=1}^{M}$, $\{T_i\}_{i=0}^{M-1}$, and $O$ at values $m$ (with $m > 0$), $\{t_i\}_{i=0}^{m-1}$, $\{x_i\}_{i=1}^{m}$, $o=1$ given $N = n$ is equal to 
\begin{equation}\label{eq:conditional_jt_density_X_T_M_single_auction}
\dfrac{C_1 n!t_0\big(1 - F(x_m)\big)}{\tau^n (n-m-1)!} \bigg(\sum_{i=0}^{m}F(x_i)t_i \bigg)^{n-m-1} \prod_{i=1}^{m}f(x_i),
\end{equation}
where $x_0 = r$ and the arguments satisfy the constraints 
$m \leq (n-1)$ and $t_m = \tau - \sum_{i=0}^{m-1}t_i \geq 0$. Moreover, since $N \sim \text{Poisson}(\lambda\tau)$, we have
\begin{equation}\label{eq:pmf_N}
P(N=n) = \exp(-\lambda\tau)\dfrac{(\lambda\tau)^n}{n!}.
\end{equation}
Combining \eqref{eq:conditional_jt_density_X_T_M_single_auction} and \eqref{eq:pmf_N}, we get the joint density of $M$, $\{X_i\}_{i=1}^{M}$, $\{T_i\}_{i=0}^{M-1}$, $O$, and $N$ at values $m$ (with $m > 0$), $\{t_i\}_{i=0}^{m-1}$, $\{x_i\}_{i=1}^{m}$, $o=1$, and $n$ is equal to 
\begin{equation}\label{eq:jt_density_X_T_M_N}
C_1 \exp(-\lambda\tau)\dfrac{\lambda^n t_0 \big(1 - F(x_m)\big)}{(n-m-1)!} \bigg(\sum_{i=0}^{m}F(x_i)t_i \bigg)^{n-m-1} \prod_{i=1}^{m}f(x_i),
\end{equation}
where $x_0 = r$ and the arguments satisfy the constraints 
$m \leq (n-1)$ and $t_m = \tau - \sum_{i=0}^{m-1}t_i \geq 0$. 
Finally, summing over $n$ in \eqref{eq:jt_density_X_T_M_N} such that $n\geq (m+1)$, we get the joint density of $\{X_i\}_{i=1}^{M}$, $\{T_i\}_{i=0}^{M-1}$, $M$, and $O$ at values $m$ (with $m > 0$), $\{t_i\}_{i=0}^{m-1}$, $\{x_i\}_{i=1}^{m}$, and $o=1$ is equal to 
\begin{align}
{}& C_1 \exp(-\lambda\tau) \lambda^{m+1} T_0 \big(1 - F(x_m)\big) \exp\bigg(\lambda\sum_{i=0}^{m}F(x_i)t_i \bigg) \prod_{i=1}^{m}f(x_i) \nonumber\\
={}& C_1 \exp(-\lambda\tau) \Big( \lambda^{m+1} t_0 \big(1 - F(x_m) \big) 
        \Big) \exp\bigg(\lambda\sum_{i=0}^{M}F(x_i)t_i \bigg) \nonumber\\
{}& \times \bigg(\prod_{i=1}^{M}f(x_i) \bigg),
\label{eq:jt_density_X_T_M}
\end{align}
where $x_0=r$, $t_m = \tau - \sum_{i=0}^{m-1}t_i \geq 0$.\\
\noindent
\\\textbf{Case II: When the item is sold at the starting price $\mathbf{(M=0, O=1)}$.}\quad In this case, the only bid which is higher than the starting price remains unobserved and the value of $M = 0$. Moreover, we have $X_0 = r = X_M$, $T_0 = \tau = T_M$, $\Delta_0 = N$ and $N\geq 1$. Since the probability that $M = 0, O = 1$ given $N=n$ equals 
\begin{equation}\label{eq:jt.density.X.delta.given.N.item.sold.At.startingPrice}
    n\big(F(r)\big)^{n-1}\big(1 - F(r)\big).
\end{equation}

\noindent
for $n \geq 1$, it follows using \eqref{eq:pmf_N} and \eqref{eq:jt.density.X.delta.given.N.item.sold.At.startingPrice} that the joint density of $M$, $X_0$, $T_0$, $O$, and $N$ at values $0$, $r$, $\tau$, $1$ and $n$ is equal to 
\begin{align}
    {}& n\big(F(r)\big)^{n-1}\big(1 - F(r)\big)\exp(-\lambda\tau)\dfrac{(\lambda\tau)^n}{n!} \nonumber\\
    ={}& \lambda\tau\big(1 - F(r)\big) \exp(-\lambda\tau) \dfrac{\big(\lambda\tau F(r)\big)^{n-1}}{(n-1)!}.
    \label{eq:jt.density.X.T.N.item.sold.At.startingPrice}
\end{align}
Summing over $n$ in \eqref{eq:jt.density.X.T.N.item.sold.At.startingPrice} for  $n \geq 1$, we get the joint density of $M$, $X_0$, $T_0$, and $O$ at values $0$, $r$, $\tau$ and $1$ equals 
\begin{align}
    \exp(-\lambda\tau) \lambda\tau \big(1 - F(r)\big)  \exp \big(\lambda\tau F(r) \big) 
    \label{eq:jt.density.X.T.item.sold.At.startingPrice}
\end{align}
\noindent
\\\textbf{Case III: When the item is not sold $\mathbf{(M=0, O=0)}$.}\quad This situation can occur if either all the bids are less than the starting price or no bidding happened at all. In any case $M = 0$. Additionally, we have $X_0 = r = X_M$, $T_0 = \tau = T_M$, $\Delta_0 = N$ and $N\geq 0$. Since the probability that $M = 0, O = 0$ given $N=n$ equals 
\begin{equation}\label{eq:jt.density.X.delta.given.N.item.NotSold}
    \big(F(r) \big)^{n}.
\end{equation}
for $n \geq 0$, it follows using \eqref{eq:pmf_N} and \eqref{eq:jt.density.X.delta.given.N.item.NotSold} that the joint density of $M$, $X_0$, $T_0$, $O$ and $N$ at values $0$, $r$, $\tau$, $0$ and $n$ is equal to
\begin{align}
    {}& \big(F(r) \big)^{n} \exp(-\lambda\tau)\dfrac{(\lambda\tau)^n}{n!} \nonumber\\
    ={}& \exp(-\lambda\tau) \dfrac{\big(\lambda\tau F(r) \big)^{n}}{n!}.
    \label{eq:jt.density.X.T.N.item.NotSold}
\end{align}
Summing over $n$ in \eqref{eq:jt.density.X.T.N.item.NotSold} such that $n\geq 0$, we get the joint density of $M$, $X_0$, $T_0$, and $O$ at values $0$, $r$, $\tau$ and $0$ equals 
\begin{align}
    \exp(-\lambda\tau)  \exp \big(\lambda\tau F(r) \big) \label{eq:jt.density.X.T.item.NotSold}
\end{align}

\noindent
Finally, the expressions in \eqref{eq:jt_density_X_T_M}, \eqref{eq:jt.density.X.T.item.sold.At.startingPrice}, and \eqref{eq:jt.density.X.T.item.NotSold} altogether conclude the proof of Lemma 2.1.
\end{proof}

\section{Proof of Lemma 2.2}\label{sec:proof_of_lemma2.2}

\noindent
For every $1 \leq l \leq \ell$, we define $\tilde{F}(\bar{x}_l) := F(\bar{x}_l)$. In other words, $\tilde{F}(z_{u_l}) := F(z_{u_l})$ for every 
$1 \leq l \leq \ell$. Also, let $\tilde{F}(z_0) = 
F(z_0) = 0$ (with $z_0 := 0$ and $u_0 := 0$). 
Fix $1 \leq l \leq \ell$ arbitrarily. We now define 
$\tilde{F}$ on $(z_{u_{l-1}}, z_{u_l})$. Note that any
element of ${\bf z}$ in this open interval has to be 
a starting price for one of the auctions in the 
dataset. First, 
$$
\tilde{F}(x) := F(z_{u_{l-1}}) \; \mbox{ for } z_{u_{l-1}} < x < z_{{u_{l-1}}+1}. 
$$

\noindent
If $u_{l-1}+1 = u_l$, the defining task is 
accomplished. Otherwise, for every $i$ such that 
$u_{l-1} + 1 \leq i \leq u_l - 1$, we define 
$$
\tilde{F}(x) = F(z_i) \; \mbox{ for } z_{i} \leq x 
< z_{i+1}. 
$$

\noindent
Hence, $\tilde{F}$ has now been defined on $[0, z_{u_\ell}]$. 

We now consider two scenarios. If $u_\ell = \ell+K$, 
then define $\tilde{F}(x) = F(x)$ for $x > z_{u_\ell}$. It follows from the above construction that $\tilde{F} \in \mathcal{F}_{\bf z}$. For every 
$1 \leq l \leq \ell$, note that 
$$
\tilde{F}(\bar{x}_l -) = \tilde{F}(z_{u_l} -) = 
F(z_{u_l} - 1) \leq F(z_{u_l} -). 
$$

Since $\tilde{F}(z_{u_l}) = F(z_{u_l})$, it follows that 
$$
\tilde{F}(z_{u_l}) - \tilde{F}(z_{u_l} -) \geq 
F(z_{u_l}) - F(z_{u_l} -), 
$$

\noindent
or equivalently
$$
\tilde{F}(\bar{x}_l) - \tilde{F}(\bar{x}_l -) \geq 
F(\bar{x}_l) - F(\bar{x}_l -).  
$$

\noindent
Since $\tilde{F}$ and $F$ match on all elements of 
${\bf z}$ by the above construction, we also have  
$\tilde{F}(r_k) = F(r_k)$ for every $1 \leq k \leq K$.
It follows by Eq. (2.4) in the main paper that 
$Lik_{PA}(F) \leq Lik_{PA}(\tilde{F})$. 

On the other hand, if $u_\ell < \ell + K$, we define 
$$
\tilde{F}(x) = F(z_{u_\ell}) \mbox{ for } z_{u_\ell} < x < z_{u_\ell+1}
$$

and 
$$
\tilde{F}(x) = 1 \mbox{ for } z_{u_\ell+1} \leq x. 
$$

\noindent
Hence, $\tilde{F}$ and $F$ match on all elements of 
$\{z_i\}_{i=1}^{u_\ell}$, and $\tilde{F}$ dominates $F$ on all 
elements of $\{z_i\}_{i=u_\ell+1}^{\ell+K}$. 
By the exact same arguments as in the first scenario, it follows that 
$\tilde{F}(z_{u_l}) - \tilde{F}(z_{u_l} -) \geq 
F(z_{u_l}) - F(z_{u_l} -)$ for every $1 \leq l \leq \ell$. It again follows by Eq. (2.4) in the main paper 
that $Lik_{PA}(F) \leq Lik_{PA}(\tilde{F})$. 

The above analysis assumes that $\ell > 0$. If $\ell = 0$, then the vector ${\bf \bar{x}}$ is empty. It follows from Eq. (2.4) in the main paper that $Lik_{PA}(F)$ depends on $F$ only through $\{F(r_k)\}_{k=1}^K$, and is non-decreasing in each of these $K$ elements. In this case, let $\tilde{F}$ denote the CDF corresponding to the distribution which puts a point mass at zero. Then, $\tilde{F} \in 
\mathcal{F}_{\bf z}$ and $Lik_{PA}(F) \leq Lik_{PA}(\tilde{F})$. \hfill$\Box$

\section{Additional simulation experiments: Settings with high expected number of participants per auction} \label{app:sec:largelambda}

\noindent
While we do consider settings with a fairly large number of auctions ($1000$ auctions) in simulations, the expected number of participants in each auction is around $100$ (we use a Poisson process with rate of arrival $1$ over $100$ time units, i.e., $\lambda = 1$ and $\tau = 100$). {\color{black} To explore the performance of our proposed method in setting where there is a larger expected number of participants}, we conducted additional simulation studies where the arrival rate of the underlying Poisson process was set to $10$ and $50$. The corresponding expected number of participants in the auction would then be $1000$ and $5000$ respectively.  In the process of conducting these simulations, we discovered that the following minor modifications are needed to our optimization algorithm to increase efficiency and address numerical issues. 

The first modification involves the computation of 
$G_\lambda^{-1}$ (an ingredient in the computation of $\hat{F}_{SP}$, 
and hence $\hat{F}_{init}$). Recall that the function $G_\lambda$ is defined by 
$$
G_{{\lambda}}(\eta) := \frac{\exp(-{\lambda}\tau) \ \Big({\lambda}\tau(1 - \eta)\big(\exp({\lambda}\tau \eta) - 1 \big) + \exp({\lambda}\tau \eta) - {\lambda}\tau \eta - 1 \Big)}{1 - \exp(-{\lambda}\tau) - {\lambda}\tau \exp(-{\lambda}\tau)}. 
$$

\noindent
When $\lambda \tau$ becomes large, it turns out that $G_{\lambda}(\eta)$ takes values very close to $0$ when $\eta$ is not sufficiently close to 1. For example, when $\lambda\tau=1000$, we find that $G_{\lambda}(\eta) <5\times 10^{-4}$ for $\eta \leq 0.99$; however when $\eta\to 1$, $G_{\lambda}(\eta)\to 1$. So, near $\eta = 1$, the value taken by $G_\lambda (\eta)$ has a sudden spike from almost zero to almost one. This may lead to instability in the numerical inversion of $G_\lambda$. Fortunately, when $\lambda \tau$ is large, one can use the approximation 
\begin{equation} \label{G_lambda_approximation}
\begin{aligned}
    G_{{\lambda}}(\eta) &\approx \exp(-{\lambda}\tau) \ \Big({\lambda}\tau(1 - \eta)\exp({\lambda}\tau \eta)  + \exp({\lambda}\tau \eta)\Big)\\
    &=\exp\big(-    {\lambda}\tau(1-\eta)\big)(\lambda\tau(1 - \eta) + 1), 
\end{aligned}
\end{equation}

% . In particular, since $G_{SP} (x)$ is the ecdf function for 1000 final selling prices, then the smallest non-zero value taken by this function is $10^{-3}$, which is greater than $G_{\lambda}(0.99)$. 

\noindent
which in particular works very well near $\eta \approx 1$. For example, when $\eta=0.99$, the error in this approximation is of the order $10^{-18}$ when $\lambda \tau = 1000$, and is expected to be even smaller for larger $\eta$ values and for larger $\lambda \tau$ values. Further, based on the approximation in (\ref{G_lambda_approximation}), it can be shown that 
\begin{equation}\begin{aligned}
    G^{-1}_{{\lambda}}(x) &\approx 1-\frac{1}{\lambda\tau}\Big(-1-W\Big(-\frac x e\Big)\Big)
\end{aligned}
\end{equation}

\begin{table}[tb]
  \centering
  \small
  
  \begin{tabular}{@{} ll rrl rrl @{}}
    \toprule
    & & \multicolumn{3}{c}{\textbf{KS distance }} 
      & \multicolumn{3}{c}{\textbf{Total Variation }} \\
    \cmidrule(lr){3-5} \cmidrule(l){6-8} 
    Distribution & $\lambda$ 
      & MLE   & Initial & Pólya Tree  
      & MLE    & Initial  & Pólya Tree\\ 
    \midrule
Uniform & 10 & 0.015 & 0.059 & 0.050 & 0.034 & 0.095 & 0.056 \\  & 50 & 0.015 & 0.045 & 0.050 & 0.033 & 0.086 & 0.056 \\[\smallskipamount]Piecewise Uniform & 10 & 0.015 & 0.037 & 0.249 & 0.066 & 0.086 & 0.500 \\  & 50 & 0.015 & 0.038 &    NA & 0.065 & 0.097 &   NA \\[\smallskipamount]Pareto & 10 & 0.016 & 0.046 & 0.460 & 0.019 & 0.086 & 0.451 \\  & 50 & 0.017 & 0.043 & 0.553 & 0.019 & 0.093 & 0.543 \\[\smallskipamount]Gamma & 10 & 0.015 & 0.053 & 0.235 & 0.023 & 0.104 & 0.386 \\  & 50 & 0.016 & 0.048 & 0.241 & 0.024 & 0.103 & 0.438 \\[\smallskipamount]Beta & 10 & 0.017 & 0.054 & 0.107 & 0.025 & 0.098 & 0.159 \\  & 50 & 0.017 & 0.053 & 0.101 & 0.023 & 0.102 & 0.172 \\[\smallskipamount]
    \bottomrule
  \end{tabular}
  \caption{Kolmogorov-Smirnov (KS) distance and Total variation  distance between each of the three estimators $\hat{F}_{MLE}$, $\hat{F}_{init}$, Pólya Tree estimator(PT) and the true valuation distribution $F$, averaged over 30 replications within each of the 10 simulation settings with larger $\lambda$ and $K=1000$.}
  \label{auction:performance:evaluation:highlambda}
\end{table}

\noindent
where $W$ is the lower branch of Lambert $W$ function that is stably implemented in the {\it pracma} package in {\it R}. We employ this approximation for computing $\hat{F}_{SP}$ in the large $\lambda \tau$ setting. Finally, note that the region where the approximation (\ref{G_lambda_approximation}) works well  is precisely the relevant region for our computations, as $\hat{F}_{SP} (x) = G_{\hat{\lambda}}^{-1}(G_{SP} (x))$, and it turns out that the pre-image under $G_\lambda$ of most non-zero values taken by the empirical CDF $G_{SP}(x)$ is close to $1$ in the large $\lambda \tau$ setting. 

Second, when the underlying expected number of bidders is very large, the joint optimization algorithm can converge slowly, especially for coordinates $\theta_i $ with $i<u_1$. Thus we use a two-stage approach, where for the first step, we fix the $\theta_i $ with $i<u_1$ to the corresponding values obtained from $\hat{F}_{FP}$, and only update the $\lambda$ and $\theta_i $ with $i\geq u_1$ based on conditional maximization in each iteration - until a mild stopping criterion is met. In the second stage, with the final parameter value from the first stage as our initial value, we run the usual coordinate-wise optimization over all coordinates of ${\boldsymbol \theta}$ and $\lambda$. 

{\it Note that these modifications do not deviate from the likelihood principle, we just use a more stable and efficient method to compute the initial value, and selectively optimize over a subset of coordinates for the first few iterations for numerical stability and faster convergence.}

The results for the two settings discussed above, namely with $\lambda \tau = 1000$ and $\lambda \tau = 5000$, are summarized in Table \ref{auction:performance:evaluation:highlambda}. The comparative performance pattern between the proposed $\hat{F}_{MLE}$ and $\hat{F}_{init}$ estimators remains the same as in our original setting with underlying $\lambda \tau = 100$, while the performance of the Pólya tree estimator deteriorates sharply, with numerical issues encountered in some settings leading to {\it NA} values. These additional experiments reinforce the message that the proposed MLE estimator can provide a scalable, more accurate and more stable alternative to existing non-parametric/semi-parametric methods.

\section{Additional Simulation Experiments: Placed bids and waiting times}

\begin{figure}[htbp]
    \centering
        \subfloat[Scatter plot for Unobserved placed bid]{
        \includegraphics[width=0.45\textwidth]{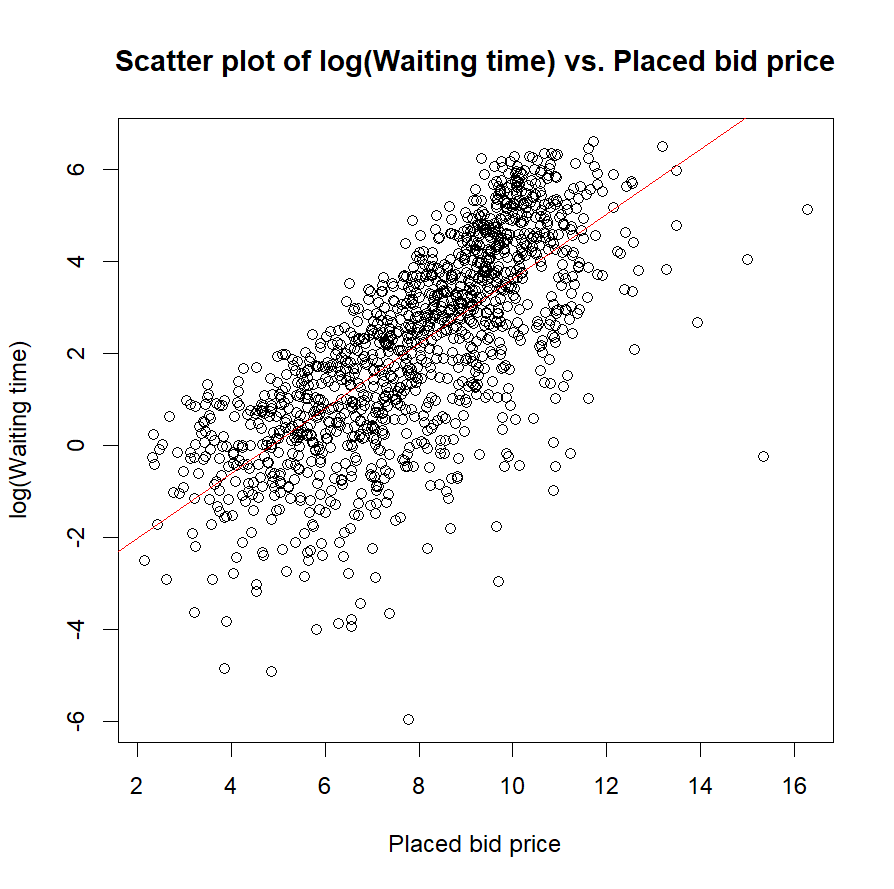}}
    \subfloat[Scatter plot for Observed selling price]{
        \includegraphics[width=0.45\textwidth]{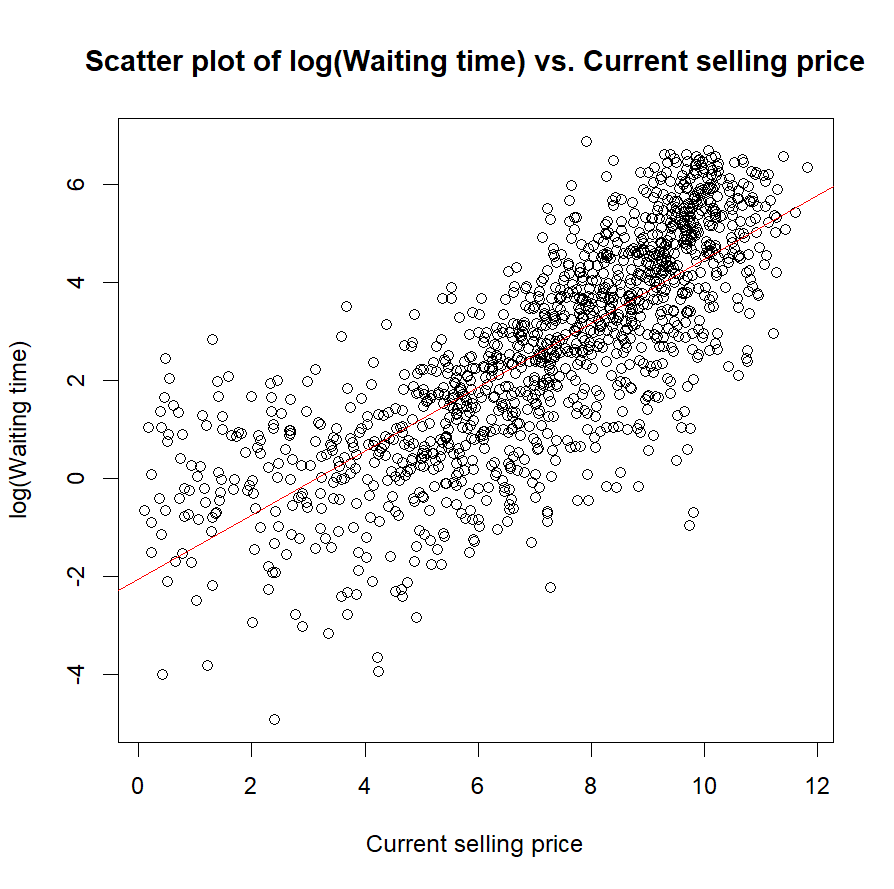}}
    \caption{Scatter plots for waiting time vs bid - (unobserved) placed bid in (a), and (observed) current selling price in (b).}
    \label{fig:wait_time_Poisson.plot}
\end{figure}

\noindent
The purpose of this simulation study is to show that, although the arrival process of potential bidders follows a Poisson process that is independent of the current selling price, the waiting time until the next placed bid is positively correlated with the most recent placed bid—and, consequently, with the current selling price. We simulate 1{,}000 auctions in which potential bidders arrive according to a Poisson process with rate $\lambda = 1$, the auction window length $\tau$ is 100 units, and the underlying valuation distribution $F$ is a Gamma distribution with shape parameter 10 and rate parameter 2. Figure~\ref{fig:wait_time_Poisson.plot} (below) contains plots of - (a) placed bid vs. the logarithm of the waiting time until the next placed bid, and (b) current selling price following a placed bid vs. the logarithm of the waiting time until the next placed bid. Both plots clearly show a strong positive relationship between the relevant quantities. 

\section{Additional Simulation Experiments: Profit‑maximizing price estimation under different production costs}

The simulations in Section~4.3 demonstrate the comparative advantage of the proposed MLE method in estimating both the profit-maximizing price and the corresponding maximum profit (per person). In the baseline experiments, the production cost was fixed at the 25\% quantile of the true valuation distribution. To assess the robustness of the results with respect to the choice of the production cost~$c$, we replicate these simulations (for $K=100$) under three additional cost levels, corresponding to the 10\%, 30\%, and 50\% quantiles of the true valuation distribution. 

\begin{figure}[htbp]
    \centering    \includegraphics[width=0.98\textwidth]{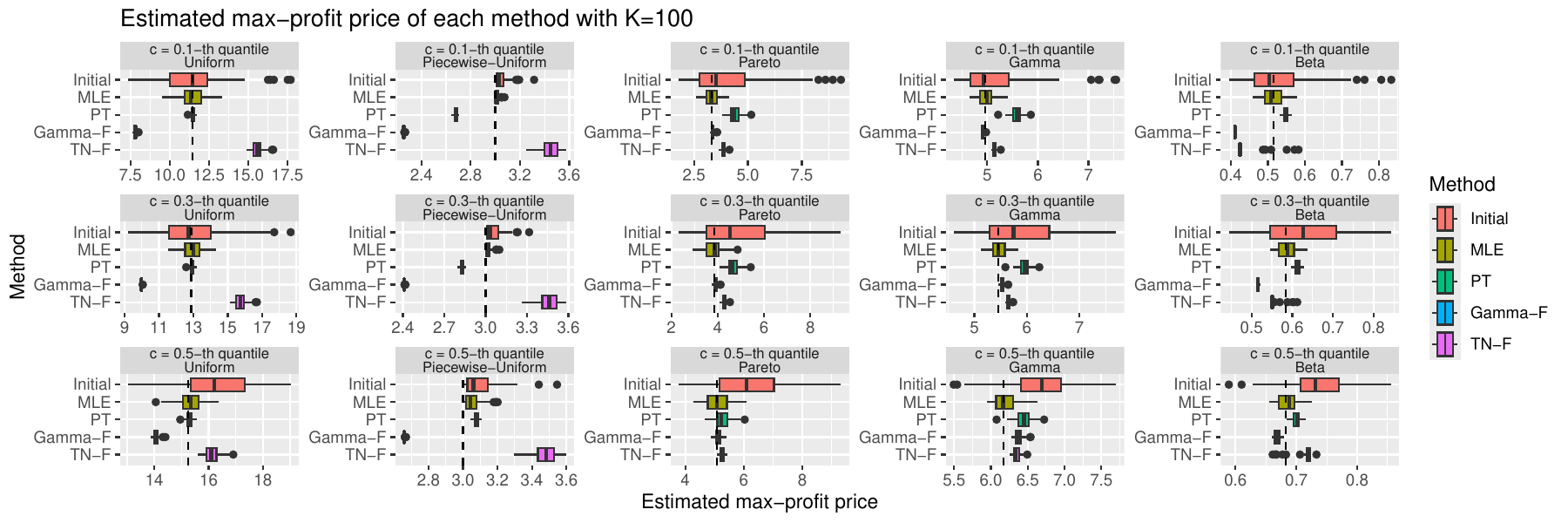}
    \caption{Box plots for the estimated profit maximizing price $x_{\hat F}$ for each of the five estimators, $\hat{F}_{MLE}$, $\hat{F}_{init}$, Pólya tree estimator ({PT}), the Gamma method based on final selling prices ({Gamma-F}), and the Truncated-Normal method based on final selling prices ({TN-F}). Results are based on 100 replications within each of the 10 simulation settings, with different production cost $c$ fixed at different quantile of the true $F$. The first row corresponds to $c$ equals to the 10\% quantile of the true $F$, the second row corresponds to the 30\% quantile, and the third row corresponds to the 50\% quantile. The dashed lines represent the true profit maximizing price $x_F$ in each scenario.}
    \label{app:max_profit price for more}
\end{figure}

As shown in Figure~\ref{app:max_profit price for more}, the MLE method consistently performs best across all cost levels in the Piecewise-Uniform, Gamma, and Beta settings, and performs reasonably well in the Uniform and Pareto settings. By contrast, the initial estimator increasingly overestimates the profit-maximizing price as the cost rises. The P\'olya tree estimator generally overestimates the price across cost levels, with the exception of the Uniform setting; in the Piecewise-Uniform case, it transitions from underestimation to overestimation as the cost increases. The Gamma-F method performs well in the Pareto setting and for low production costs in the Gamma setting, but tends to overestimate the price at higher cost levels. Finally, the TN-F method performs poorly across most settings.

\section{Additional Simulation Experiments: Robustness with respect to choice of arrival process}
\begin{table}[!h]
  \centering
  \small
  
  \begin{tabular}{@{} ll ccccc @{}}
    \toprule
    & & \multicolumn{5}{c}{\textbf{KS distance }} \\
      % & \multicolumn{3}{c}{\textbf{Total Variation }} \\
    \cmidrule(lr){3-7} 
    Distribution & $K$ 
      & $\ \quad $MLE$\ \ $   & Initial & Pólya Tree  
      & Gamma-F & TN-F\\ 
    \midrule
 Uniform & 100  &  0.047 &  0.128 &  0.050 &  0.195 &  0.589 \\ 
   & 1000  &  0.015 &  0.048 &  0.050 &  0.235 &  0.588 \\ 
 [\smallskipamount] Piecewise-Uniform & 100  &  0.043 &  0.099 &  0.230 &  0.327 &  0.586 \\ 
   & 1000  &  0.015 &  0.038 &  0.224 &  0.344 &  0.587 \\ 
 [\smallskipamount] Pareto & 100  &  0.052 &  0.105 &  0.205 &  0.068 &  0.110 \\ 
   & 1000  &  0.017 &  0.041 &  0.163 &  0.020 &  0.100 \\ 
 [\smallskipamount] Gamma & 100  &  0.044 &  0.127 &  0.298 &  0.253 &  0.199 \\ 
   & 1000  &  0.014 &  0.048 &  0.299 &  0.297 &  0.192 \\ 
 [\smallskipamount] Beta & 100  &  0.048 &  0.132 &  0.127 &  0.188 &  0.401 \\ 
   & 1000  &  0.016 &  0.051 &  0.123 &  0.256 &  0.363 \\ 
  \cmidrule(lr){3-7}    & & \multicolumn{5}{c}{\textbf{Total Variation}} \\    \cmidrule(lr){3-7} 
 Uniform & 100  &  0.072 &  0.202 &  0.067 &  0.253 &  0.603 \\ 
   & 1000  &  0.033 &  0.084 &  0.066 &  0.260 &  0.601 \\ 
 [\smallskipamount] Piecewise-Uniform & 100  &  0.087 &  0.149 &  0.467 &  0.453 &  0.598 \\ 
   & 1000  &  0.065 &  0.083 &  0.455 &  0.437 &  0.598 \\ 
 [\smallskipamount] Pareto & 100  &  0.057 &  0.169 &  0.217 &  0.062 &  0.112 \\ 
   & 1000  &  0.020 &  0.064 &  0.174 &  0.02 &  0.102 \\ 
 [\smallskipamount] Gamma & 100  &  0.071 &  0.209 &  0.315 &  0.258 &  0.203 \\ 
   & 1000  &  0.023 &  0.077 &  0.306 &  0.302 &  0.194 \\ 
 [\smallskipamount] Beta & 100  &  0.070 &  0.219 &  0.136 &  0.210 &  0.405 \\ 
   & 1000  &  0.024 &  0.080 &  0.131 &  0.269 &  0.367 \\ 
    \bottomrule
  \end{tabular}
  \caption{Kolmogorov-Smirnov (KS) distance and Total variation  distance between various estimators ($\hat{F}_{MLE}$, $\hat{F}_{init}$, Pólya Tree estimator(PT), the Gamma method based on final selling prices - {Gamma-F}, the Truncated-Normal method based on final selling prices - {TN-F}) and the true valuation distribution $F$, averaged over 100 replications within each of the 10 simulation settings, with Gamma interarrival times.}
  \label{auction:performance:evaluation:gammaarrival}
\end{table}

\begin{table}[!h]
  \centering
  \small
  
  \begin{tabular}{@{} ll ccccc @{}}
    \toprule
    & & \multicolumn{5}{c}{\textbf{KS distance }} \\
      % & \multicolumn{3}{c}{\textbf{Total Variation }} \\
    \cmidrule(lr){3-7} 
    Distribution & $K$ 
      & $\ \quad $MLE$\ \ $   & Initial & Pólya Tree  
      & Gamma-F & TN-F\\ 
    \midrule
 Uniform & 100  &  0.051 &  0.123 &  0.051 &  0.216 &  0.627 \\ 
   & 1000  &  0.016 &  0.054 &  0.059 &  0.254 &  0.626 \\ 
 [\smallskipamount] Piecewise-Uniform & 100  &  0.050 &  0.097 &  0.231 &  0.346 &  0.621 \\ 
   & 1000  &  0.016 &  0.053 &  0.224 &  0.362 &  0.625 \\ 
 [\smallskipamount] Pareto & 100  &  0.053 &  0.107 &  0.219 &  0.065 &  0.115 \\ 
   & 1000  &  0.017 &  0.050 &  0.175 &  0.018 &  0.105 \\ 
 [\smallskipamount] Gamma & 100  &  0.044 &  0.129 &  0.299 &  0.274 &  0.203 \\ 
   & 1000  &  0.015 &  0.052 &  0.300 &  0.309 &  0.205 \\ 
 [\smallskipamount] Beta & 100  &  0.051 &  0.145 &  0.124 &  0.224 &  0.438 \\ 
   & 1000  &  0.017 &  0.057 &  0.126 &  0.292 &  0.400 \\ 
  \cmidrule(lr){3-7}    & & \multicolumn{5}{c}{\textbf{Total Variation}} \\    \cmidrule(lr){3-7} 
 Uniform & 100  &  0.076 &  0.207 &  0.067 &  0.265 &  0.639 \\ 
   & 1000  &  0.034 &  0.090 &  0.078 &  0.274 &  0.638 \\ 
 [\smallskipamount] Piecewise-Uniform & 100  &  0.092 &  0.155 &  0.472 &  0.448 &  0.620 \\ 
   & 1000  &  0.065 &  0.091 &  0.456 &  0.433 &  0.634 \\ 
 [\smallskipamount] Pareto & 100  &  0.059 &  0.167 &  0.23 &  0.06 &  0.116 \\ 
   & 1000  &  0.02 &  0.074 &  0.184 &  0.018 &  0.106 \\ 
 [\smallskipamount] Gamma & 100  &  0.067 &  0.216 &  0.315 &  0.279 &  0.206 \\ 
   & 1000  &  0.023 &  0.082 &  0.310 &  0.314 &  0.206 \\ 
 [\smallskipamount] Beta & 100  &  0.072 &  0.227 &  0.136 &  0.237 &  0.438 \\ 
   & 1000  &  0.024 &  0.088 &  0.135 &  0.302 &  0.400 \\ 
    \bottomrule
  \end{tabular}
  \caption{Kolmogorov-Smirnov (KS) distance and Total variation  distance between various estimators ($\hat{F}_{MLE}$, $\hat{F}_{init}$, Pólya Tree estimator(PT), the Gamma method based on final selling prices - {Gamma-F}, the Truncated-Normal method based on final selling prices - {TN-F}) and the true valuation distribution $F$, averaged over 100 replications within each of the 10 simulation settings, with lognormal interarrival times.}
  \label{auction:performance:evaluation:lognormalarrival}
\end{table}

Note that the proposed methodology uses the assumption that the {\em potential} bidders arrive at the auction according to a Poisson process. To assess robustness of our methodology to misspecification of the arrival process (of potential bidders), we perform two additional simulations where the true bidder arrival process is chosen to be (i) a renewal process with Gamma interarrival times, and (ii) a renewal process with lognormal interarrival times. The results for the Kolmogorov-Smirnov and Total variation distances are presented in Table \ref{auction:performance:evaluation:gammaarrival} and Table \ref{auction:performance:evaluation:lognormalarrival}. The comparative performance of all methods closely mirrors the results reported in Section 4.2 of the main paper. The proposed MLE approach still outperforms the other methods in most of the settings.